%% file: fft_network_1.tex
\documentclass[review,A4,11pt]{elsarticle}
\usepackage{natbib}
\usepackage{amsmath}
\usepackage{amsfonts}
\usepackage{amssymb}
\usepackage[scale=0.85]{geometry}
\usepackage{color}
\usepackage[usenames,dvipsnames,table]{xcolor}
\usepackage{graphicx,import}
\usepackage{epstopdf}
\usepackage{subcaption}
\usepackage{hyperref}
\usepackage{bm}
\usepackage{setspace}
\usepackage{float}
\usepackage{lipsum}
\usepackage{mathtools}
\usepackage{siunitx}
\usepackage{caption}
\usepackage{import}
\usepackage{lineno}

\usepackage{sectsty}
\usepackage{titlesec}
\sectionfont{\fontsize{11pt}{0} \bfseries}
\subsectionfont{\fontsize{11pt}{0} \bfseries}
\subsubsectionfont{\fontsize{11pt}{0} \bfseries}
\titlespacing*{\subsection}{0pt}{12pt}{11pt}
\titlespacing*{\subsubsection}{0pt}{12pt}{11pt}
\titlespacing*{\paragraph}{0pt}{0pt}{11pt}
\definecolor{aqua}{rgb}{0.0,1.0,1.0}

\graphicspath{{./images/}{./images/Cook_Membrane/}{./images/TimoBeam/}{./images/Laminate/}{./images/Circular_Inclusion/}{./images/1D/}}

\usepackage{ifluatex}
\ifluatex
\usepackage{pdftexcmds}
\makeatletter
\let\pdfstrcmp\pdf@strcmp
\let\pdffilemoddate\pdf@filemoddate
\makeatother
\fi
\journal{}
\renewcommand{\d}{\mathrm{d}}
\newcommand{\D}{\mathrm{D}}

\newcommand{\du}{\,\mathrm{d}u}

\newcommand{\dX}{\,\mathrm{d}X}

\newcommand{\dV}{\,\mathrm{d}V}
\newcommand{\dA}{\,\mathrm{d}A}

\newcommand{\bfb}{\mathbf{b}}

\newcommand{\bfj}{\mathbf{j}}
\newcommand{\bfk}{\mathbf{k}}

\newcommand{\bff}{\mathbf{f}}

\newcommand{\bft}{\mathbf{t}}
\newcommand{\bfx}{\mathbf{x}}
\newcommand{\bfy}{\mathbf{y}}

\newcommand{\bfw}{\mathbf{w}}
\newcommand{\bfA}{\mathbf{A}}

\newcommand{\bfC}{\mathbf{C}}

\newcommand{\bfF}{\mathbf{F}}

\newcommand{\bfN}{\mathbf{N}}

\newcommand{\bfP}{\mathbf{P}}

\newcommand{\bfT}{\mathbf{T}}

\newcommand{\bfX}{\mathbf{X}}

\newcommand{\bfW}{\mathbf{W}}

\newcommand{\bfGamma}{\boldsymbol{\Gamma}}

\newcommand{\bfmu}{\boldsymbol{\mu}}
\newcommand{\bfxi}{\boldsymbol{\xi}}

\newcommand{\bftau}{\boldsymbol{\tau}}
\newcommand{\bfvarphi}{\boldsymbol{\varphi}}

\newcommand{\bbC}{\mathbb{C}}

\newcommand{\bbI}{\mathbb{I}}

\newcommand{\bbG}{\mathbb{G}}

\newcommand{\rmD}{\mathrm{D}}

\newcommand{\calB}{\mathcal{B}}

\newcommand{\calF}{\mathcal{F}}

\newcommand{\calH}{\mathcal{H}}

\newcommand{\calO}{\mathcal{O}}

\newcommand{\calR}{\mathcal{R}}
\newcommand{\calT}{\mathcal{T}}

\newcommand{\iunit}{\mathrm{i}\,}

\newcommand{\bfZero}{\mathbf{0}}

\newcommand{\bbCMacro}{\overline{\mathbb{C}}}

\newcommand{\bfvarphiMacro}{\overline{\bfvarphi}}
\newcommand{\bfxMacro}{\overline{\bfx}}
\newcommand{\bfXMacro}{\overline{\bfX}}
\newcommand{\bfFMacro}{\overline{\bfF}}
\newcommand{\bfPMacro}{\overline{\bfP}}

\newcommand{\energyMacro}{\overline{\psi}}

\newcommand{\HDMR}{\text{HDMR}}
\newcommand{\NN}{\text{NN}}

\newcommand{\nablaMacro}{\overline{\nabla}}
\newcommand{\bftMacro}{\overline{\bft}}

\newcommand{\epsilonMacro}{\overline{\epsilon}}

\newcommand{\FMacro}{\overline{F}}

\newcommand{\bfFFluc}{\widetilde{\bfF}}

\newcommand{\bfvarphiFluc}{\widetilde{\bfvarphi}}

\newcommand{\GammaD}{\Gamma_\mathrm{D}}
\newcommand{\GammaN}{\Gamma_\mathrm{N}}

\newcommand{\equivalence}{\quad\Leftrightarrow\quad}

\newcommand{\norm}[1]{\left\lVert#1\right\rVert}

\newcommand{\RVE}{\text{RVE}}

\makeatletter
\newcommand*\bigcdot{\mathpalette\bigcdot@{.5}}
\newcommand*\bigcdot@[2]{\mathbin{\vcenter{\hbox{\scalebox{#2}{$\m@th#1\bullet$}}}}}

\newcommand{\tsr}[1]{\ensuremath\boldsymbol{#1}}

\newcommand{\wb}[1]{\overline{#1}}


\begin{document}
\begin{frontmatter}
\title{A surrogate model for computational homogenization of elastostatics at finite strain using the HDMR-based neural network approximator}
	\author{Vien Minh Nguyen-Thanh\fnref{addressHannover}\corref{}}
	\ead{minh.nguyen@ikm.uni-hannover.de}
	\author{Lu Trong Khiem Nguyen\fnref{addressStuttgart}\corref{nguyen}}
	
	\author{Timon Rabczuk\fnref{addressWeimar}\corref{}}
	\ead{timon.rabczuk@uni-weimar.de}
	\author{Xiaoying Zhuang\corref{zhuang}\fnref{tonducthang1,tonducthang2}}
	\cortext[zhuang]{Corresponding author: \textcolor{blue}{\sffamily xiaoying.zhuang{\fontfamily{ptm}\selectfont @}tdtu.edu.vn}}
	\cortext[nguyen]{Corresponding author: \textcolor{blue}{\sffamily nguyen{\fontfamily{ptm}\selectfont @}mechbau.uni-stuttgart.de}}
	\address[tonducthang1]{Division of Computational Mechanics, Ton Duc Thang University, Ho Chi Minh City, Vietnam}
	\address[tonducthang2]{Faculty of Civil Engineering, Ton Duc Thang University, Ho Chi Minh City, Vietnam}
	\address[addressHannover]{Institute of Continuum Mechanics, Leibniz Universit{\"a}t Hannover, Appelstra{\ss}e 11, 30167 Hannover, Germany}
	\address[addressStuttgart] {Institute of Applied Mechanics, Chair I, University of Stuttgart, Pfaffenwaldring 7, 70569 Stuttgart, Germany}
	\address[addressWeimar] {Institute of Structural Mechanics, Bauhaus-University Weimar, Marienstra{\ss}e 15, 99423 Weimar, Germany}
\begin{abstract}
We propose a surrogate model for two-scale computational homogenization of elastostatics at finite strains. The macroscopic constitutive law is made numerically available via an explicit formulation of the associated macro-energy density. This energy density is constructed by using a neural network architecture that mimics a high-dimensional model representation. The database for training this network is assembled through solving a set of microscopic boundary values problems with the prescribed macroscopic deformation gradients (input data) and subsequently retrieving the corresponding averaged energies (output data). Therefore, the two-scale computational procedure for the nonlinear elasticity can be broken down into two solvers for microscopic and macroscopic equilibrium equations that work separately in two stages, called the offline and online stages. A standard finite element method is employed to solve the equilibrium equation at the macroscale. As for mircoscopic problems, an FFT-based collocation method is applied in tandem with the Newton-Raphson iteration and the conjugate-gradient method. Particularly, we solve the microscopic equilibrium equation in the Lippmann-Schwinger form without resorting to the reference medium and thus avoid the fixed-point iteration that might require quite strict numerical stability condition in the nonlinear regime.
\end{abstract}
\end{frontmatter}
	\section{Introduction}
	Multiscale techniques are important for man-made and natural materials; one such approach is homogenization. Roughly speaking, homogenization is a rigorous version of what is known as averaging.
	It is a powerful tool to study the heterogeneous materials and composites. Based on knowledge of the microstructure of materials, the objective is to understand the response of materials at the macroscale. Homogenization techniques are basically multiscale methods applied to problems where the separation of scales are visible. In the field of homogenization theory for heterogeneous materials, the microscopic length-scale characterizes the fast variation of material properties throughout the continuum body while the macroscopic length-scale corresponds to responses of the entire structure as if the materials were homogeneous.
	
	In solid mechanics, analytical homogenization methods have grown fruitfully in the last half century. They include the Hashin-Strikman variational principles, which are used to estimate the upper and lower bounds of effective modulus tensor for linear elastic composites for quite general class of microstructures; see e.g. \textsc{Hashin\,\&\,Shstrikman}~\cite{Hashin+Shstrikman-1962,Hashin+Shstrikman-1963}, \textsc{Willis}~\cite{Willis-1977,Willis-1981}, \textsc{Avellaneda}~\cite{Avellaneda-1987}, \textsc{Milton}~\cite{Milton-1988}, \textsc{Ponte}~\cite{Ponte-1991}. These contributions aimed at describing the effective response of the materials of rather random structure whose information is only available through statistical relations among composite phases. The extension of the Hashin-Shstrikman (HS) variational principles to the nonlinearity was introduced by \textsc{Willis}~\cite{Willis-1983} and subsequently applied to predict the first bounds for nonlinear composites by \textsc{Talbot}~\cite{Talbot-1985} and for viscous composites by \textsc{Ponte et al.}~\cite{Ponte-1988}. The contribution by \textsc{Ponte\,\&\,Suquet}~\cite{Ponte+Suquet-1998} and the monograph by \textsc{Milton}~\cite{Milton-2002} review excellently the variational techniques of homogenization for heterogeneous materials and various topics of composite materials.
	
	Although the fully- and semi-analytical methods are powerful and able to deliver exact effective response for rather simple microstructure, they fail to provide good bounds when the microstructure is rather complicated or the correlation information between phases is not readily available. Due to these difficulties, the computational homogenization appeared as a promising candidate to move the theory of homogenization forward. One such approach decouples the multiscale problems into two nested problems, namely the microscopic and macroscopic boundary value problems; see \textsc{Hughes et al.}~\cite{Hughes-1998}, \textsc{Miehe et al.}~\cite{Miehe+Schroeder+Schotte-1999}, \textsc{Feyel}~\cite{Feyel-1999} among the first works in this direction. This technique is called two-scale computational homogenization and normally realized in an FE$^2$ procedure. In such procedure, the original multiscale problem is resolved in a bottom-up fashion in that the macroscopic boundary value problem (BVP) is solved using  the finite element method (FEM) and the effective quantities inquired at the macroscale are supplied after resolving many associated microscopic boundary value problems by also FEM. The microscopic BVP is solved within a small region of the whole continuum which is chosen to represent well the microstructure of material. Therefore, it is called representative volume element (RVE). The two-scale solver was made possible thanks to the crucial theoretical contribution of \textsc{Mandel}~\cite{Mandel-1971} and \textsc{Hill}~\cite{Hill-1972} that established a micro-macro transition condition. The latter was coined Hill-Mandel condition or macro-homogeneity condition. In addition, this condition gives rise to appropriate boundary conditions on the RVE problem.
	
	Keeping in mind that FE$^2$ solution method was not efficient in dealing with complex microstructure, \textsc{Moulinec\,\&\,Suquet}~\cite{Moulinec+Suquet-1998} independently published a paper discussing the treatment of microscopic BVPs using the Discrete Fourier Transform (DFT) spectral method. It is worth mentioning that this paper was a follow-up of another work by the same author \textsc{Moulinec\,\&\,Suquet}~\cite{Moulinec+Suquet-1994}. The digital image obtained by Scanning Electron Microscopy can be directly made use resorting to the DFT-based methods as the solution is globally interpolated at the voxels and therefore the meshing is not necessary. The key idea is to recast a differential equation with respect to the gradient field (such as linearized strain and deformation gradient) into an integral equation resorting to a periodic Green's function. This FFT-based solver is considered as an excellent alternative to the FEM with several efforts of improving its speed  (see, e.g., \textsc{Zeman et al.}~\cite{Zeman+Vondrejc+Novak+Marek-2000}) and extensions to scope complex microstructure with high phase contrast and nonlinear materials (see, e.g., \textsc{Michel et al.}~\cite{Michel+Moulinec+Suquet-2001} and \textsc{Brisard\,\&\,Dormieux}~\cite{Brisard+Dormieux-2012}). Recently, \textsc{de Geus et al.}~\cite{Geus+Vondrejc+Zeman+Peerlings+Geers-2017} provided a perspective on FFT-based collocation method as a Galerkin-based method. Accordingly, the solution method for the equilibrium equation at the microscale resorting to a combination of a Newton-Raphson iteration and conjugate-gradient solver did not require the reference medium. This spectral method has been lately applied to computational homogenization for electro-elastically and magneto-elastically coupled materials in \textsc{G\"{o}k\"{u}zum et al.}~\cite{Gokuzum+Nguyen+Keip-2019} and \textsc{Rambausek et al.}~\cite{Rambausek+Gokuzum+Nguyen+Keip-2019}, respectively. All these aforementioned works basically replaced the FE$^2$ solution method in which the FFT-based solver was used for the microscopic BVPs.
	
	A two-scale approach basically computes the gradient and the hessian of macro-energy density numerically at all the quadrature points used in the problem domain of the macroscopic BVP. In principle, the two-scale method makes the macro-energy density numerically available and supplies the macroscopic solver with its derivatives. On the other hand, it is also possible to make the macro-energy density available either in an analytical approach for simple microstructures or in a numerical approximation such as interpolations, regressions or reduced-order models (ROMs). In the latter case, \textsc{Yvonnet\,\&\,He}~\cite{Yvonnet+He-2007} suggested a reduced-order model of multiscale method for studying elastostatics at finite strains in which the ROM therein was based on the proper orthogonal decomposition. In some sense, this work can be classified as data-driven method since it generated the data by solving a sequence of microscopic BVPs and exploited these data in a two-stage fashion to speed up the solution process for the macroscopic BVP. In this manner, the micro-solver for dealing with microscopic BVPs was run concurrently with the macro-solver for macroscopic BVP. The data space of macroscopic strain in their work was assembled rather simply in a uniform pattern but this assembly pattern is generally not a good choice for microstructures exhibiting anisotropicity. In order to overcome this disadvantage, \textsc{Fritzen\,\&\,Kunc}~\cite{Fritzen+Kunc-2018} suggested a ROM that exploited a specific data sampling strategy. This model was proved to be practically better than the former one by \textsc{Yvonnet\,\&\,He}~\cite{Yvonnet+He-2007} when it could handle the composite materials with isotropic and anisotropic microstructures almost equally well.
	
	Although a two-scale computational homogenization reduces the computation effort of a direct numerical simulation, it can still be significantly improved in some cases. For example, when the composites are made of phases of hyperelastic materials, the effective properties of the homogenized material can be described by a homogenized energy density, or macroscopic free energy density (also called macro-energy density above). In fact, in contrast to \textsc{Yvonnet\,\&\,He}~\cite{Yvonnet+He-2007}, \textsc{Fritzen\,\&\,Kunc}~\cite{Fritzen+Kunc-2018} proposed a surrogate model in which the macroscopic free energy density is numerically constructed with the aid of the radial basis functions. The macro-solver was implemented based on this numerically available macro-energy density. In addition, \textsc{Le et al}~\cite{Le+Yvonnet+He-2015} proposed to construct the macro-energy density numerically by a neural network architecture that mimics a high-dimensional model representation and reformulated a two-scale computational problem into a surrogate computational model. It is worth mentioning that this work was inspired by a technique of building numerically a potential energy surface that is an important entity in the community of computational chemistry. As the latter exploits the advantages of both the high-dimensional model representation and the neural-network approximators, the surrogate model naturally counters can handle the energy function of high-dimensional variables while the usage of neural networks renders its implementation rather simple.
	
	In this work we aim at speeding up the computation for a homogenization problem of elasticity at finite strains by combining the FFT-based collocation method for the microscopic BVPs and a surrogate model based on a neural network architecture proposed by \textsc{Manzhos\,\&\,Carrngton}~\cite{Manzhos+Carrington-2006}. The macroscopic BVP is subsequently solved by using a standard finite element procedure. In addition to the reduction of computational cost, the explicit macro-energy density can be not only reused but also improved by enriching the data samples later. Two key points of this contribution are highlighted: 
	\begin{itemize}
		\item[(i)] We point out a connection between the derivations presented in \textsc{Moulinec\,\&\,Suquet}~\cite{Moulinec+Suquet-1998} and in \textsc{de Geus et al}~\cite{Geus+Vondrejc+Zeman+Peerlings+Geers-2017}.
		\item[(ii)] The computational framework in \textsc{Le et al.}~\cite{Le+Yvonnet+He-2015} is extended to scope the elastostatics at finite strain.
	\end{itemize}
	First, the theory of computational homogenization in the context of continuum mechanics is summarized in the next Section. The aforementioned highlights (i) and (ii) are presented in Section~\ref{Section:Methodology-for-microscopic-BVP} and Section~\ref{Section:Surrogate-model-using-NN-approximator}, respectively. Also, we discuss some implementation details which were not addressed in \textsc{Le et al.}~\cite{Le+Yvonnet+He-2015}. Section~\ref{Section:Representative-examples} is devoted to numerical validation of the computational framework via various representative examples. Those include the mathematical problems that admit the analytical solutions as well as the real-world applications. Some concluding remarks are given in the last Section.

	\section{Computational homogenization for finite strain problems}
	\subsection{Fundamental equations}
	We summarize the theory of first-order computational homogenization for nonlinear elasticity. The starting point is the variational formulation describing the response of a continuum body of elastic heterogeneous materials. In order to facilitate the presentation, we briefly go through the fundamental continuum mechanics. In this work, we adopt a part of notation as well as the presentation of \textsc{Miehe et al.}~\cite{Miehe+Vallicotti+Teichtmeister-2016}.
	
	\subsubsection{Deformation of continuum body of heterogeneous materials}
	Deformation of a continuum body $\calB$ can be characterized by the primary field
	\begin{equation*}
	\bfvarphi = \begin{cases}
	\calB \times \calT &\hspace{-6pt} \rightarrow \mathbb{R}^{3} \\
	(\bfX, t) &\hspace{-6pt} \mapsto \bfvarphi(\bfX, t)
	\end{cases}
	\end{equation*}
	that maps reference coordinates $\bfX \in \calB$ onto points $\bfx = \bfvarphi(\bfX, t)$ of the current configuration $\bfvarphi(\calB, t)$. When the initial time is fixed, This field is called the \emph{deformation mapping} and the \emph{deformation gradient} $\bfF$ is defined as its gradient, that is $\bfF = \nabla\bfvarphi$.
	
	For practical considerations, when the work done by the inertial force is negligible compared to that done by the total stress stored in the system, a static analysis can be used. In this work, we scope the body on which the external force is slowly applied and therefore a classic first-order homogenization theory can be exploited.
	
	We start with the variational formulation: The true motion $\bfvarphi \in H^{1}_{\GammaD}(\calB)$ of the continuum body $\calB$ is the stationary point of the following potential
	\begin{equation}\label{original-variational-problem}
	\Pi(\bfvarphi) =  \int_{\calB} \psi (\bfX,\bfF)\dV - \int_{\calB} \bff\bigcdot\bfvarphi\dV - \int_{\GammaN}\bftMacro \bigcdot\bfvarphi\dA,
	\end{equation}
	where $\GammaD$ is a part of the boundary of $\calB$ where the deformation mapping is prescribed, $\psi(\bfX, \bfF)$ is the free Helmholtz energy, $\bff$ is the external force per unit volume, $\bftMacro$ is the traction force exerted on the body at the boundary $\GammaN$. In this formulation, the dependence of the free energy on the spatial coordinate $\bfX$ indicates the heterogeneity of the body. In addition, the function space
	\begin{equation*}
	H_{\GammaD}^{1} = \big\{\bfvarphi \in H_{\GammaD}^{1} \,|\, \bfvarphi(\bfX) = \bfvarphiMacro(\bfX) \quad\forall\bfX \in \GammaD \big\}
	\end{equation*}
	is the space of all vector-valued functions in the Sobolev space $H^{1}(\calB)$ that are prescribed by the given function $\bfvarphi_0(\bfx)$ defined on the boundary $\GammaD$. The variational equation derived from the stationary statement $\delta\Pi = 0$ is the equilibrium equation and the boundary conditions
	\begin{equation*}
	\nabla \bigcdot \bfP^{T} = \bff \quad \text{in }\calB, \quad
	\bfvarphi(\bfX) = \bfvarphiMacro \quad \text{on }\GammaD, \quad
	\frac{\partial\psi}{\partial\bfF}\bigcdot \bfN = \bftMacro\quad \text{on }\GammaN,
	\end{equation*}
	where $\bfP = \partial\psi/\partial\bfF$ is called the first Piola-Kirchhoff stress and $\bfN = \bfN(\bfX)$ is the outward unit vector normal to $\GammaN$ measured in the reference coordinates
	
	\begin{figure}[htb]
		\centering
		\def\svgwidth{0.7\textwidth}
		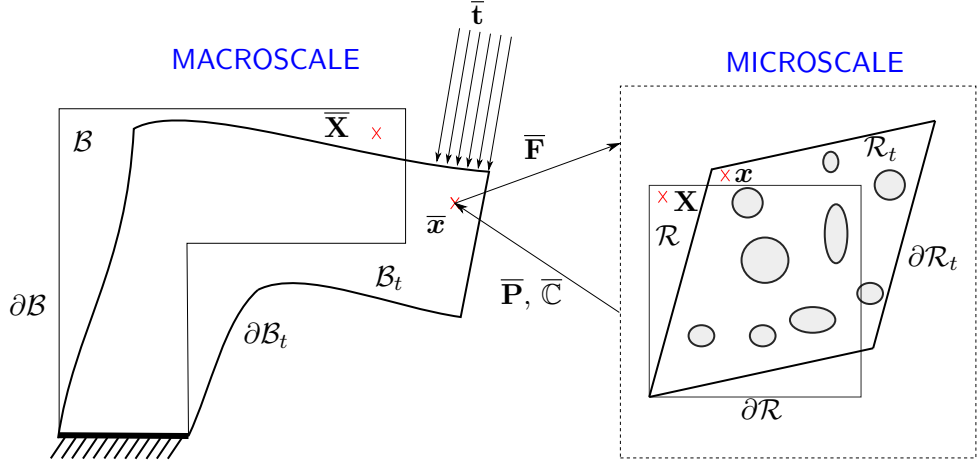
		\caption{\textit{A continuum body of heterogeneous materials.} The heterogeneities in the body around the material point $\bfXMacro$ can be ``averaged out'' to understand the mechanical response of the body at $\bfXMacro$ as if it is a homogeneous body by looking at a representative volume element enclosing the material point $\bfXMacro$. In this way, we could realize the entire heterogeneous body as the homogeneous one by looking at infinite number of RVEs surrounding all material points.}
		\label{Fig: Continuum Body}
	\end{figure}
	
	\subsubsection{First-order computational homogenization}
	Following the theory of first-order computational homogenization, the variational problem defined by \eqref{original-variational-problem} can be effectively replaced by a sequence of nested two-scale boundary value problems: microscopic problem and macroscopic problem. 
	
	First, one of the most fundamental assumptions in this theory is the decomposition of the primary field $\bfvarphi$ about one certain coordinate $\bfXMacro$ imitating a Taylor expansion as follows
	\begin{equation*}
	\bfvarphi(\bfX,t) = \bfvarphiMacro_{|\bfXMacro}(t) + \nablaMacro\big(\bfvarphiMacro_{|\bfXMacro}\big)(t) \bigcdot \left(\bfX - \bfXMacro\right) + \bfvarphiFluc(\bfX,t), \quad \bfvarphiMacro_{|\bfXMacro}\in \calO(\epsilon^0), \overline{\nabla}(\bfvarphiMacro_{\bfXMacro})\in \calO(\epsilon^0), \bfvarphiFluc\in\calO(\epsilon^1),
	\end{equation*}
	where the subscript $\bfXMacro$ indicates that the macroscopic field $\bfvarphiMacro$ is associated with such ``macroscopic'' coordinate and $\overline{\nabla}$ denotes the derivative with respect to $\bfxMacro$. In this expansion, $\calO$ denotes the standard big-$O$ used in the approximation theory (see, e.g., \textsc{Trefethen}~\cite{Trefethen-2013}). In a static analysis, we may remove the the term $\bfvarphiMacro_{\bfXMacro}(t)$ in the subsequent derivation as this term does not contribute to the macroscopic variational problem. We obtain
	\begin{equation}\label{deformation-decomposition}
	\bfvarphi(\bfX,t) = \bfFMacro(t)\bigcdot\left(\bfX - \bfXMacro\right) + \bfvarphiFluc(\bfX, t), \quad \bfFMacro = \nablaMacro\bfvarphiMacro.
	\end{equation}
	This expansion is valid in a neighborhood domain of $\bfXMacro$ called the representative volume element and denoted by $\calR(\bfXMacro)$; see Fig.~\ref{Fig: Continuum Body}. Without ambiguity, we drop out the macroscopic coordinate $\bfXMacro$ associated with the corresponding $\RVE$. It is immediately seen that
	\begin{equation}\label{deformation-gradient-decomposition}
	\bfF := \nabla\bfvarphi = \bfFMacro + \nabla\bfvarphiFluc \quad\Rightarrow\quad \frac{1}{|\calR|}\int_{\calR}\bfF\dV = \bfFMacro + \frac{1}{|\calR|}\int_{\calR}\nabla\bfvarphiFluc\dV.
	\end{equation}
	
	In a two-scale computational homogenization, the macroscopic constitutive response is obtained by an appropriate \emph{upscaling} of the microscopic quantities resulting from the RVE calculation. This is usually achieved by enforcing the so-called \emph{Hill-Mandel} or \emph{macro-homogeneity} condition. This condition states that the microscopic volume average of the variation of work performed on the RVE $\calR(\bfXMacro)$ associated with $\bfX$ is equal to the local variation of work at $\bfXMacro$ on the microscale. This condition essentially guarantees an energetic consistency of the first-order homogenization theory. It is derived from the definition of the macroscopic free energy density (macro-energy density)
	\begin{equation}\label{macroscopic-energy-density}
	\energyMacro(\bfFMacro) = \inf\limits_{\bfvarphi\in \calH\left(\bfFMacro\right)}\frac{1}{|\calR|}\int_{\calR} \psi (\bfX, \nabla\bfvarphi)\dV,
	\end{equation}
	where $\calH(\bfFMacro)$ is the function space of all vector fields that satisfy the decomposition \eqref{deformation-decomposition} and certain appropriate boundary conditions on the RVE boundary $\partial\calR$. This macro-energy density is clearly not well-defined unless a boundary condition for $\bfvarphi$ is defined, which leaves different possibilities for various averaging operators. Let us assume that $\bfvarphi$ is a stationary point of the above minimization problem, then we can write
	\begin{equation*}
	\energyMacro(\bfFMacro) = \frac{1}{|\calR|}\int_{\calR} \psi \dV = \big\langle \psi \big\rangle_{\calR}
	\end{equation*}
	and derive \emph{macro-homogeneity condition}, or the \emph{Hill-Mandel condition} (see \textsc{Mandel}~\cite{Mandel-1971} and \textsc{Hill}~\cite{Hill-1972})
	\begin{equation}\label{macro-homogeneity-condition}
	\frac{\partial \energyMacro}{\partial\bfFMacro} : \delta\bfFMacro = \bigg\langle \frac{\partial \psi}{\partial \bfF}: \delta \bfF\bigg\rangle_{\calR}.
	\end{equation}
	Substitution of equation \eqref{deformation-decomposition} into the right-hand side of this relation, we obtain
	\begin{equation}\label{macro-homogeneity-RHS}
	\bigg\langle \frac{\partial \psi}{\partial \bfF}: \delta \bfF\bigg\rangle_{\calR} = \frac{1}{|\calR|}\int_{\calR} \frac{\partial\psi}{\partial\bfF} : \delta \bfF\dV = \frac{1}{|\calR|}\int_{\calR}\bfP\dV : \delta\bfFMacro + \frac{1}{|\calR|}\int_{\calR}\bfP : \delta\bfFFluc\dV,
	\end{equation}
	where we have used the denotation
	\begin{equation*}
	\bfFFluc = \nabla\bfvarphiFluc \quad \Rightarrow\quad \delta\bfFFluc = \nabla\delta\bfvarphiFluc.
	\end{equation*}
	Using the Gauss theorem, the last term can be transformed to the area integral
	\begin{equation}\label{Gauss-theorem-for-fluctuation-work}
	\frac{1}{|\calR|}\int_{\calR}\bfP : \delta\bfFFluc\dV = \frac{1}{|\calR|}\int_{\partial\calR}\big(\bfP \bigcdot\bfN\big)\delta\bfvarphiFluc\dV - \frac{1}{|\calR|}\int_{\calR}\big(\bfP\bigcdot\nabla \big)\delta\bfvarphiFluc\dV = \frac{1}{|\calR|}\int_{\partial\calR}\big(\bfP \bigcdot\bfN\big)\delta\bfvarphiFluc\dV
	\end{equation}
	in which we have used the assumption that $\bfvarphi$ is the solution of the minimization problem \eqref{macroscopic-energy-density}, that is the microscopic equilibrium equation $\bfP\bigcdot\nabla = \bfZero$ holds.
	Combining equations \eqref{macro-homogeneity-condition}--\eqref{Gauss-theorem-for-fluctuation-work}, we arrive at
	\begin{equation*}
	\bfPMacro : \delta\bfFMacro = \bigg(\frac{1}{|\calR|}\int_{\calR}\bfP\dV\bigg) : \delta\bfFMacro + \frac{1}{|\calR|}\int_{\partial\calR}\big(\bfP\bigcdot\bfN)\,\delta\bfvarphiFluc\dV,
	\end{equation*}
	where the macroscopic stress is defined as $\bfPMacro = \partial\energyMacro/\partial\bfFMacro$. If we choose the boundary condition on $\bfvarphi$ such that the area integral in the latter condition vanishes, this will lead to the relation
	\begin{equation}\label{macroscopic-stress}
	\bfPMacro = \frac{1}{|\calR|}\int_{\calR}\bfP\dV \equivalence \frac{\partial\energyMacro}{\partial\bfFMacro} = \frac{1}{|\calR|}\int_{\calR}\frac{\partial \psi}{\partial \bfF}\dV.
	\end{equation}
	The macrohomogeneity condition reduces to the following condition
	\begin{equation*}
	\frac{1}{|\calR|}\int_{\calR}\big(\bfP\bigcdot\bfN\big)\delta\bfvarphiFluc\dV = \bfZero,
	\end{equation*}
	which can be fulfilled in various ways. Among several approaches, the following three are normally chosen
	\begin{itemize}
		\item Dirichlet boundary condition: $\bfvarphiFluc = \bfZero$ on $\partial\calR$.
		\item Neumann boundary condition: $\bfP \bigcdot\bfN = \bfPMacro\bigcdot\bfN$ on $\partial\calR$.
		\item Periodic boundary condition: $\bfvarphiFluc$ is periodic on $\partial\calR$ and $\bfP\bigcdot \bfN$ is anti-periodic on $\partial\calR$.
	\end{itemize}
	
	In this work, we study heterogeneous materials with periodic microstructure and it is indeed numerically verified in \cite{Terada+Hori+Kyoya+Kikuchi-2000} that the periodic condition provides better prediction on the macroscale response. In addition, we will use the FFT-based collocation method for solving the microscopic problems. Therefore, we adopt in this work the periodic boundary condition for microscopic problems. That is, the functional space $\calH(\bfFMacro)$ is now defined as
	\begin{equation*}
	\calH(\bfFMacro) = \big\{\bfvarphi = \bfFMacro\bigcdot(\bfX - \bfXMacro) + \bfvarphiFluc \in H^{1}(\calR)\,| \quad \bfvarphiFluc \text{ is periodic on } \partial\calR \big\}.
	\end{equation*}
	The fluctuation fields $\bfvarphiFluc$ belong to the function space $\calH_{\#}$ consisting of all vector-valued functions in $H^{1}(\calR)$ that are periodic of $\calR$. Accordingly, it is immediately implied from equation \eqref{deformation-gradient-decomposition} that
	\begin{equation}\label{macroscopic-deformation-gradient}
	\bfFMacro = \frac{1}{|\calR|}\int_{\calR}\bfF\dV \equivalence \frac{1}{|\calR|}\int_{\calR}\bfFFluc\dV = \bfZero.
	\end{equation}
	
	Up to the first approximation in the micro-macro length-scale ratio $\epsilon$, the macro-energy density will be used as a replacement in the original variational problem \eqref{original-variational-problem}. In this way, the fast varying properties of the heterogeneous materials are ``averaged out'', leaving a homogenized elastic body characterized by the homogenized free energy density. Indeed, this density is a function of $\bfFMacro = \nablaMacro$ which is in turn determined through the macroscopic deformation mapping $\bfvarphiMacro$.
	
	The presented theory can be transferred to a two-scale computational homogenization approach in the following
	\begin{itemize}
		\item With the input macroscopic $\bfFMacro$, we solve the microscopic boundary value problem
		\begin{equation}\label{microscopic-boundary-value-problem}
		\energyMacro(\overline{\bfF}) = \inf\limits_{\bfvarphiFluc \in \calH_{\#}} \frac{1}{|\calR|}\int_{\calR} \psi(\bfX, \bfFMacro + \nabla\bfvarphiFluc)\dV
		\end{equation}
		for the fluctuation field with the periodic boundary condition. According to the definition \eqref{macroscopic-energy-density} we can determine the macro-energy density.
		\item With the macro-energy density obtained above, we solve the macroscopic boundary value problem: Find the macroscopic deformation mapping $\bfvarphiMacro\in H^1_{\GammaD}(\calB)$ that minimizes the following potential
		\begin{equation}\label{macroscopic-boundary-value-problem}
		\overline{\Pi}(\bfvarphiMacro) = \int_{\calB}\energyMacro(\bfFMacro)\dV - \int_{\calB}\bff\bigcdot\bfvarphiMacro\dV - \int_{\GammaN}\bftMacro \bigcdot \bfvarphiMacro\dA.
		\end{equation}
	\end{itemize}
	
	\section{Methodology for microscopic boundary value problem}\label{Section:Methodology-for-microscopic-BVP}
	\subsection{Review of multi-dimensional discrete Fourier transform}
	In order to facilitate the derivation of the FFT-based collocation method for multi-dimensional boundary value problems, we use herein two sets of running indices. The Roman letters indicate the indices associated with the collocation nodes used in the discrete Fourier transform, while the Greek letters indicate those associated with the problem dimension.
	
	Let us consider a periodic function $f$ defined on domain $\mathfrak{G}^{h}=\Pi_{\alpha=1}^d(-L_\alpha, L_\alpha)$ of $d$-dimension. Consider a uniform mesh of grid points
	\begin{equation*}
	\bfX_{\bfj} = \big( (X_1)_{j_1}, \ldots, (X_d)_{j_d}\big), \quad \bfj = (j_1,\ldots, j_d),
	\end{equation*}
	with $(X_{\alpha})_{j_{\alpha}}$ for $\alpha = 1,\ldots, d$ defined as follows
	\begin{equation*}
	(X_{\alpha})_{j_\alpha} = -L_{\alpha} + h_{\alpha}\bigg(j_{\alpha} - \frac{1}{2}\bigg), \quad j_{\alpha} = 1, \ldots, N_\alpha, \quad h_\alpha = \frac{2 L}{N_\alpha}.
	\end{equation*}
	We define a periodic \emph{grid function} $v$ as a restriction of the periodic function on this uniform mesh. The value of $v$ at node $\bfX_{\bfj}$ can be denoted as $v_{\bfj} = f(\bfX_\bfj)$. In this presentation, we restrict ourselves to the odd numbers of grid points. This assumption means all $N_{\alpha}$ for $\alpha = 1,\ldots,d$ are odd numbers and $\lfloor N_{\alpha} \rfloor = (N_{\alpha} - 1)/2$ denotes the floor rounding of $N_{\alpha}/2$.
	
	Multi-dimensional discrete Fourier transform (DFT) of a multi-dimensional array $v_{j_1\ldots j_d}$ that is a grid function of $d$ discrete variables $X_{j_\alpha}$, $\alpha = 1, \ldots, d$, is defined by
	\begin{equation}\label{forward-DFT-full}
	\widehat{v}_{k_1\ldots k_d} =  \sum\limits_{j_1=1}^{N_1}\exp[-\iunit \xi_1(k_1)(X_1)_{j_1}\big]\times\cdots\times \sum\limits_{j_d = 1}^{N_d}\exp[-\iunit \xi_d(k_d)(X_d)_{j_d}\big] v_{j_1\ldots j_d},\quad -\lfloor N_\alpha/2\rfloor \leq k_\alpha\leq \lfloor N_\alpha/2 \rfloor.
	\end{equation}
	In this definition, the \emph{scaled wavenumbers} $\xi_{\alpha} = (\pi/ L_\alpha)k_\alpha$ for $\alpha = 1,\ldots, d$, are defined by scaling the integer wavenumbers $-\lfloor N_\alpha/2\rfloor \leq k_\alpha \in \mathbb{Z} \leq \lfloor N_\alpha/2 \rfloor$ by the wavelength ratio $\pi/L_\alpha$. Using the denotations
	\begin{equation*}
	\bfk = (k_1,\ldots, k_d), \quad \bfxi(\bfk) = \bigg(\frac{\pi}{L_1}k_1,\ldots, \frac{\pi}{L_d}k_d\bigg),\quad \bfj = (j_1,\ldots, j_d), \quad \bfN = (N_1,\ldots, N_d),
	\end{equation*}
	we will write the full expression ``formally'' and compactly as
	\begin{equation}\label{forward-DFT}
	\widehat{v}_{\bfk} = \sum\limits_{\bfj = 1}^{\bfN} \exp\big[-\iunit \bfxi(\bfk) \bigcdot\bfX_{\bfj}\big]v_{\bfj}, \quad -\lfloor \bfN/2 \rfloor \leq \bfk \leq \lfloor \bfN/2 \rfloor,
	\end{equation}
	in which the summation indicates that all the indices $j_\alpha$ run from $1$ to $N_{\alpha}$. The inverse multi-dimensional discrete Fourier transform is per definition given by
	\begin{equation}\label{inverse-DFT}
	v_{\bfj} = \frac{1}{|\bfN|} \sum\limits_{\bfk = -\lfloor\bfN/2\rfloor}^{\lfloor\bfN/2\rfloor} \exp\big[\iunit \bfxi(\bfk)\bigcdot\bfX_{\bfj}\big]\widehat{v}_{\bfk}, \quad |\bfN| = \prod\limits_{\alpha=1}^{d}N_{\alpha}.
	\end{equation}
	It can be proved that definition \eqref{inverse-DFT} leads to the true inverse formula of the ``forward'' discrete Fourier transform \eqref{forward-DFT}. It is important to keep in mind that the spectral methods are based on the global interpolations. In case of periodic data, the appropriate interpolant is naturally constructed in virtue of the invserse discrete Fourier transform. To this end, we replace coordinate of the collocation points $\bfX_{\bfj}$ with the generic coordinate $\bfX$ to obtain the interpolant
	\begin{equation*}
	p(\bfX) = \frac{1}{|\bfN|}\sum\limits_{\bfk = -\lfloor\bfN/2 \rfloor}^{\lfloor \bfN/2 \rfloor} \exp\big[\iunit \bfxi(\bfk)\bigcdot\bfX\big]\widehat{v}_{\bfk}.
	\end{equation*}
	
	\subsection{Numerical method for microscopic boundary value problem}
	We outline the numerical strategy for solving the microscopic boundary value problem by using the FFT-based collocation method. Although the outcome of our formulation is in line with the results obtained in \textsc{Moulinec\,\&\,Suquet}~\cite{Moulinec+Suquet-1998} and \textsc{de Geus et al.}~\cite{Geus+Vondrejc+Zeman+Peerlings+Geers-2017}, we pursue a different track of presentation in that it will be shown that their results are connected. 
	
	It was mentioned in \textsc{Michel et al.}~\cite{Michel+Moulinec+Suquet-2001} that the equilibrium equation in the Lippmann-Schwinger form can be derived by using the Green operator defined through \emph{a priori} chosen reference medium tensor. On the other hand, the equilibrium of the same form for heterogeneous materials undergoing large deformation was derived by \textsc{de Geus et al.}~\cite{Geus+Vondrejc+Zeman+Peerlings+Geers-2017} using the Green operator defined independent from a reference medium. In fact, the derivation followed by \textsc{de Geus et al}~\cite{Geus+Vondrejc+Zeman+Peerlings+Geers-2017} can be well summarized by the four properties listed in \textsc{Section 2.4},~\textsc{Michel et al.}~\cite{Michel+Moulinec+Suquet-2001} and this paper attempts to demonstrate this connection.
	
	\subsubsection{Derivation of micro-equilibrium with the polarization field}
	\paragraph{Index notation} We make a remark regarding the index notation used in this contribution. First, all the formulation using index notation adopt the Einstein convention in which the double-duplicated index implies the summation over it. Second, we use small Roman index, such as $i$, $j$, \ldots to for the derivative with respect to the reference coordinate $X_i$. This practice is a bit different from some literature but it is beneficial in the current paper. In addition, there is no ambiguity in doing so.
	
	The starting point is the balance equation at the microscale
	\begin{equation*}
	\nabla\bigcdot \bfP^{T} = \bfZero \quad\Leftrightarrow\quad P_{ij,j} = 0.
	\end{equation*}
	Now, the reference medium $\bbC^0$ and the polarization field $\bftau$ are introduced according to
	\begin{equation}\label{polarization-definition}
	P_{ij} = C_{ijkl}^{0} F_{kl} - \tau_{ij} \quad\Rightarrow\quad \widehat{P}_{ij} = C_{ijkl}^0 \widehat{F}_{kl} - \widehat{\tau}_{ij}.
	\end{equation}
	Applying the Fourier transform to the balance equation and using the definition $\bfF = \nabla\bfvarphi$, we obtain
	\begin{equation*}
	\iunit{} \widehat{P}_{ij}\xi_j = 0 \equivalence \iunit\big(C_{ijkl}^0 \iunit\widehat{\varphi}_k\xi_l - \widehat{\tau}_{ij} \big)\xi_j = 0 \equivalence C_{ijkl}^0\widehat{\varphi}_k\xi_l\xi_j = -\iunit \widehat{\tau}_{ij}\xi_j.
	\end{equation*}
	We define the acoustic tensor $A_{ik}(\bfxi) = C_{ijkl}^0 \xi_l\xi_j$ and resolve this equation for $\widehat{\bfvarphi}$ to obtain
	\begin{equation*}
	\widehat{\varphi}_k = -\iunit A^{-1}_{ki} \widehat{\tau}_{ij}\xi_j \quad \forall\bfxi \neq \bfZero.
	\end{equation*} 
	Note that in this derivation, we have assumed that the second-order tensor $\bfF$ is compatible in the sense $\nabla\times \bfF = \nabla \times \nabla\bfvarphi = \bfZero$. Keeping in mind that $\bftau = \bftau(\bfF)$, we can relate $\bfF$ to the Piola-Kirchhoff stress $\bfP$ by using the above relation as follows
	\begin{equation}\label{Lippmann-Schwinger-nonzero-wavenumber}
	\widehat{F}_{ij} = \iunit\widehat{\varphi}_i\xi_j = \Gamma^{0}_{ijkl}\widehat{\tau}_{kl},\quad \Gamma^0_{ijkl} = -A_{ik}^{-1}\,\xi_l\xi_j \quad \forall\bfxi \neq\bfZero.
	\end{equation}
	This relation provides a tool for computing $\widehat{\bfF}$ at all nonzero wavenumbers in the Fourier space. At zero wavenumbers, we have from the definition of the Fourier transform
	\begin{equation}\label{Lippmann-Schwinger-zero-wavenumber}
	\widehat{\bfF}(\bfxi = \bfZero) = \frac{1}{|\calR|}\int_{\calR} \bfF\dV = \bfFMacro.
	\end{equation}
	According to two equations \eqref{Lippmann-Schwinger-nonzero-wavenumber} and \eqref{Lippmann-Schwinger-zero-wavenumber}, we may define the Green operator $\bfGamma^0$ in the Fourier space as
	\begin{equation*}
	\widehat{\Gamma}^0_{ijkl} = \begin{cases}
	-A_{ik}^{-1}\xi_l \xi_j & \forall\bfxi\neq \bfZero, \\
	0 & \phantom{\forall}\bfxi = \bfZero,
	\end{cases}
	\end{equation*}
	and arrive at
	\begin{equation}\label{fixed-point-formula}
	\bfF = \bfFMacro + \bfGamma^0 \ast \bftau(\bfF)\quad\Rightarrow\quad \bfF = \bfFMacro + \bfGamma^0 \ast \big(\bbC^0 : \bfF - \bfP \big).
	\end{equation}
	
	If the entire derivation is repeated for the zero-valued stress field $\bfP^{0} \equiv \bfZero$, or the following obviously true equation $P^{0}_{ij,j} = 0$, we obtain
	\begin{equation}\label{compatibiliy-condition-in-fixed-point-formula}
	\bfF = \overline{\bfF} + \bfGamma^{0}\ast \big(\bbC^{0} : \bfF\big).
	\end{equation}
	Combining two equations \eqref{fixed-point-formula}-\eqref{compatibiliy-condition-in-fixed-point-formula}, we end up with the equilibrium equation at the microscale 
	\begin{equation}\label{microscale-strong-form}
	\bfGamma^{0} \ast \bfP = \bfZero.
	\end{equation}
	
	\subsubsection{Derivation of micro-equilibrium with compatibility-enforcing operator}
	We start with the stationary condition of the variational principle \eqref{microscopic-boundary-value-problem} as follows
	\begin{equation}\label{microsocpic-variational-equation}
	\int_{\calR} \bfP : \delta\bfF \dV = 0 \quad \forall \delta\bfF \in V_{c}(\overline{\bfF}),
	\end{equation}
	where $V_{c}(\overline{\bfF})$ is the space of tensor-valued functions that are compatible in the sense it can be derived from a deformation mapping field and that are averaged over the volume of the RVE to give a constant tensor $\overline{\bfF}$. In mathematical terms, the volume-averaged condition is given by \eqref{macroscopic-deformation-gradient} and the compatibility condition reads $\nabla\times\bfF = \bfZero$.
	
	Let $\bfW$ be a second-order tensor-valued function, then $\bfGamma^{0}\ast \bfW$ is a compatible field and its volume average vanishes, that is
	\begin{equation}\label{compatibility-enforcement-operator}
	\nabla\times \big(\bfGamma^{0}\ast \bfW\big) = \bfZero,\quad \big\langle\bfGamma^{0}\ast \bfW \big\rangle_{\calR} = \bfZero.
	\end{equation}
	The second condition is simply a consequence of the specific definition of $\widehat{\bfGamma}^{0}$ at the zero wavenumbers, $\widehat{\bfGamma}^{0}(\bfZero) = \bfZero$. In the index notation, the first equation reads
	\begin{equation*}
	\epsilon_{ijk}(\Gamma^{0}_{mkrs} \ast W_{rs})_{,j} = 0 \quad\Rightarrow\quad \epsilon_{ijk} \widehat{\Gamma}^0_{mkrs} \widehat{W}_{rs} \iunit \xi_j = 0 \quad\Rightarrow\quad -\epsilon_{ijk} \big(A_{mr}^{-1} \xi_s \widehat{W}_{rs}\big)\xi_k\xi_j = 0.
	\end{equation*}
	It is immediately seen that $\epsilon_{ijk} \xi_j \xi_k = 0$ for all $i$ and hence the last equation holds true for all $i, m = \overline{1,d}$.
	
	Using the decomposition $\bfF = \overline{\bfF} + \bfFFluc$ and keeping in mind that $\bfP = \bfP (\overline{\bfF} + \bfFFluc)$, the variational equation \eqref{microsocpic-variational-equation} can be rewritten as
	\begin{equation*}
	\int_{\calR} \bfP : \delta\bfFFluc \dV = \bfZero \quad \forall\bfFFluc \in V_c(\bfZero).
	\end{equation*}
	The trial function space $V_c(\bfZero\big)$ can be relaxed by considering the equivalent variational equation
	\begin{equation}\label{microscale-weak-form}
	\int_{\calR}\bfP : \delta \big(\bfGamma^{0} \ast\bfW \big) \dV = 0 \quad\Leftrightarrow\quad \int_{\calR} \big(\bfGamma^{0} \ast \bfP \big) : \delta\bfW \dV = 0\quad  \forall\delta\bfW.
	\end{equation}
	As the latter equation must hold true for arbitrary tensor-valued function $\delta\bfW$, the equilibrium equation at the microscale \eqref{microscale-strong-form} can be equivalently derived from this equation. In fact, equations \eqref{microscale-strong-form} and \eqref{microscale-weak-form} are respectively the strong form and weak form of the microscopic boundary value problem \eqref{microscopic-boundary-value-problem}.
	
	\subsubsection{Compatibility projection operator and connection to the existing works}
	Even though the equilibrium equation \eqref{microscale-strong-form} has exactly the same form derived in \textsc{Geus et al.}~\cite{Geus+Vondrejc+Zeman+Peerlings+Geers-2017}, the main difference is that $\bfGamma^0$ is not a projection onto the function space $V_c(\bfZero)$. Indeed, we will point out that the Green function $\bbG = \bbC^{0} : \bfGamma^{0}$ is a projection operator onto the function space $V_c(\bfZero)$. That means 
	\begin{itemize}
		\item[(i)]$\bbG$ acting on any arbitrary tensor-valued function $\bfW$ produces an element in $V(\bfZero)$
		\begin{equation}\label{compatibility-enforcement-property}
		\nabla \times \big(\bbG \ast \bfW\big) = \bfZero, \quad \big\langle \bbG \ast \bfW \big\rangle_{\calR} = \bfZero \quad \forall \bfW.
		\end{equation}
		\item[(ii)] $\bbG$ acting on any element $\bfW$ in $V(\bfZero)$ gives itself
		\begin{equation}\label{compatibility-projection-property}
		\bbG\ast\big[\bbG \ast \bfW\big] = \bbG \ast \bfW \quad \forall \bfW.
		\end{equation}
	\end{itemize}
	The first property (i) means that $\bbG$ projects $\bfW$ onto $V_c(\bfZero)$: $\bbG[\bfW]\in V_c(\bfZero)$ while the second (ii)  means $\bbG$ is the idempotent: $\bbG[\bbG] = \bbG$.
	
	First, multiplying equation \eqref{compatibility-enforcement-operator} by $\bbC^{0}$ on the left-hand side, we obtain
	\begin{equation*}
	\nabla \times \big[\big(\bbC^{0} : \bfGamma^{0}\big) \ast \bfW \big] = \bfZero, \quad \big\langle \big(\bbC^{0} : \bfGamma^{0}\big)\ast \bfW \big\rangle_{\calR} = \bfZero, 
	\end{equation*}
	which is identical to equation \eqref{compatibility-enforcement-property} by identifying $\bbG = \bbC^{0} : \bfGamma^{0}$. Second, let us consider a periodic tensor-valued function $\bfW$ with the property $\langle\bfW \rangle_{\calR} = \bfZero$, then equation\eqref{compatibiliy-condition-in-fixed-point-formula} reduces to
	\begin{equation*}
	\bfW = \bfGamma^{0} \ast \big(\bbC^{0} : \bfW \big) \quad\Rightarrow \quad \bfW = \big(\bbC^{0} : \bfGamma^{0}\big) \ast \bfW.
	\end{equation*}
	This equation implies
	\begin{equation*}
	\big(\bbC^{0} : \bfGamma^{0}\big)\ast \bfW = \big(\bbC^{0}:\bfGamma{^0}\big)\ast \big[\big(\bbC^{0} : \bfGamma \big)\ast \bfW\big],
	\end{equation*}
	which is basically nothing else but equation \eqref{compatibility-projection-property}.
	
	According to this analysis, various projections can be constructed by choosing different reference elasticity tensor $\bbC^{0}$. The Green projection operator derived in \textsc{Geus et al.}~\cite{Geus+Vondrejc+Zeman+Peerlings+Geers-2017} is obtained by setting the reference medium $\bbC^{0}$ to the ``minus'' fourth-order identity tensor: $\bbC^{0} : \bfW = \bbI : \bfW = \bfW$. In that case, we have $\bbG^{0} = \bbI : \bfGamma^{0} = \bfGamma^{0}$ and at the same time
	\begin{equation*}
	C^{0}_{ijkl} = \delta_{ik}\delta_{jl} \quad\Rightarrow\quad A_{ik} = \delta_{ik} \delta_{jl}\xi_j\xi_l  = -\norm{\bfxi}^2\delta_{ik} \quad\Rightarrow\quad \widehat{\Gamma}^{0}_{ijkl} = -A^{-1}_{ik}\xi_j \xi_l = -\frac{\delta_{ik}}{\norm{\bfxi}^2}\xi_j \xi_l.
	\end{equation*}
	Thus, among many possibilities of choosing the Green projection operator, we may choose as
	\begin{equation}\label{orthogonal-projection-operator}
	G_{ijkl} = \begin{cases}
	\displaystyle \frac{\delta_{ik}}{\norm{\bfxi}^2} \xi_j \xi_l & \forall \bfxi \neq \bfZero, \\
	0 & \phantom{\forall} \bfxi = \bfZero.
	\end{cases}
	\end{equation}
	As the right-hand side of equilibrium equation is zero, the minus sign does not effect in the final result and is not taken into account. At the same time, equilibrium equation at the microscale is equivalent with
	\begin{equation}\label{microscale-equilibrium-equation}
	\bbG \ast \bfP = \bfZero.
	\end{equation}
	
	\paragraph{\textbf{\sffamily Remark}} The above analysis leads to not only the existing results obtained in \textsc{Michel et al.}~\cite{Michel+Moulinec+Suquet-2001} and \textsc{Geus et al.}~\cite{Geus+Vondrejc+Zeman+Peerlings+Geers-2017} but also draw a connection between the two routes of derivation. At the same time, it also reveals that there are many possibilities of choosing projection operators other than \eqref{orthogonal-projection-operator}, each of which is obtained by fixing the reference tensor $\bbC^{0}$.

	\subsubsection{Numerical method for microscopic equilibrium equation and macroscopic tangent moduli}
	
	We will solve the equilibrium equation \eqref{orthogonal-projection-operator}-\eqref{microscale-equilibrium-equation} by the Newton method; see also \textsc{Geus et al}~\cite{Geus+Vondrejc+Zeman+Peerlings+Geers-2017}. Note also that a detail numerical procedure for system accounted for magneto-elastostatics can be found in \textsc{Rambausek et al.}~\cite{Rambausek+Gokuzum+Nguyen+Keip-2019}.
	
	In this contribution, we compare the solutions obtained by a two-scale approach and the proposed surrogate model by means of the resultant macroscopic stress $\bfPMacro$ and tangent moduli $\overline{\bbC}$. For such comparison, we need to compute these quantities as the secondary variables resulted from the solution $\bfF$ of the microscopic BVP.
	
	\paragraph{\textbf{\sffamily Numerical procedure for microscopic equilibrium}} A linearization \eqref{microscale-equilibrium-equation} with respect to $\bfFFluc^{[n]}$ at step $n$ gives
	\begin{equation*}
	\bbG \ast \bigg[ \frac{\partial \bfP}{\partial \bfF} \big( \bfF^{[n]}\big) : \Delta \bfFFluc^{[n]}\bigg] = - \bbG \ast \bfP^{[n]},
	\end{equation*}
	in which we have used $\bfF^{[n]} = \bfFMacro + \bfFFluc^{[n]}$ and $\bfP^{[n]} = \partial_{\bfF} \psi \big (\bfFMacro + \bfFFluc^{[n]}\big)$. Since $\bbG$ is numerically explicit in the Fourier space, this equation is evaluated in the Fourier space and then mapped back to the physical space. To sum up, we obtain the iterative scheme
	\begin{equation}\label{Newton-Raphson-for-micro-equilibrium}
	\begin{aligned}
	\calF^{-1}\bigg\{ \widehat{\bbG} : \calF\bigg[\frac{\partial\bfP}{\partial\bfF}(\bfF^{[n]}\big) : \Delta \bfFFluc^{[n]}\bigg] \bigg\} &= -\calF^{-1}\bigg\{\widehat{\bbG} : \calF\big(\bfP^{[n]} \big) \bigg\}, \\
	\bfFFluc^{[n+1]} &= \bfFFluc^{[n]} + \Delta\bfFFluc^{[n+1]},
	\end{aligned}
	\end{equation}
	where $\calF$ and $\calF^{-1}$ denote the discrete Fourier transform (DFT) and its inverse transform, also called the inverse DFT. The initial guess $\bfFFluc^{[0]}$ can be chosen so that its volume average vanishes, that is $\big\langle\bfFFluc^{[0]}\big\rangle_{\calR} = \bfZero$. Accordingly, we need to apply the following boundary condition at each iteration given by \eqref{Newton-Raphson-for-micro-equilibrium}
	\begin{equation*}
	\big\langle\bfFFluc^{[n]}\big\rangle_{\calR} = \bfZero \quad\Leftrightarrow\quad \big\langle\Delta\bfFFluc^{[n]}\big\rangle_{\calR} = \bfZero \quad \forall n.
	\end{equation*}
	Enforcement of this condition is conducted by setting the Green operator to be zero at the zero wavenumbers as defined in equation \eqref{orthogonal-projection-operator}.
	
	\paragraph{\textbf{\sffamily Computation of macroscopic tangent moduli}} It will be argued in the next section that the microscopic boundary value problems essentially characterize the constitutive law at the macroscale (see also next Section~\ref{Sec:Approximation-macro-constitutive-law}). Keeping this spirit in mind, we must be able to compute the macroscopic tangent moduli resorting to the information at the microscale. To this end, we pursue the strategy in \textsc{Rambausek et al.}~\cite{Rambausek+Gokuzum+Nguyen+Keip-2019} and briefly derive the formulation for macroscopic tangent moduli. According to equation~\eqref{macroscopic-stress}, we can compute
	\begin{equation}\label{macroscopic-tangent-moduli}
	\overline{\bbC} = \frac{\partial\bfPMacro}{\partial\bfFMacro} = \frac{1}{|\calR|}\int_{\calR}\bigg[ \frac{\partial\bfP}{\partial\bfF}: \bigg(\bbI + \frac{\partial\bfFFluc}{\partial\bfFMacro}\bigg) \bigg]\dV = \bigg\langle \frac{\partial \bfP}{\partial \bfF} : \bigg(\bbI + \frac{\partial\bfFFluc}{\partial\bfFMacro} \bigg) \bigg\rangle_{\calR},
	\end{equation}
	where $\bbI$ is the fourth-order identity tensor. In this equation, the deformation gradient $\bfF$ must be obtained as the solution of the microscopic BVP. Thus, it remains to determine the sensitivity of the fluctuation field $\bfFFluc$ with respect to the macroscopic field $\bfFMacro$, which is also called fluctuation sensitivity (see \textsc{Miehe et al.}~\cite{Miehe+Schroeder+Schotte-1999}). Such fluctuation sensitivity will be revealed by taking derivative of equation \eqref{microscale-equilibrium-equation} with respect to $\bfFMacro$ with the help of decomposition $\bfF = \bfFMacro + \bfFFluc$. In doing so, we arrive at
	\begin{equation}\label{fluctuation-sensitivity-equation}
	\bbG \ast \bigg[\frac{\partial \bfP}{\partial \bfF} : \bigg(\bbI + \frac{\partial\bfFFluc}{\partial\bfFMacro} \bigg)\bigg] = \bfZero \quad \Leftrightarrow \quad \bbG \ast \bigg(\frac{\partial\bfP}{\partial\bfF} : \frac{\partial\bfFFluc}{\partial\bfFMacro}\bigg) = -\bbG \ast \frac{\partial\bfP}{\partial\bfF},
	\end{equation}
	where the derivative $\partial\bfP/\partial \bfF$ can be determined since $\bfF$ is understood as given by the solution of equation~\eqref{microscale-equilibrium-equation}. It is not surprising that this is a linear system for the fluctuation sensitivity as it is resulted from a linearization of equation \eqref{microscale-equilibrium-equation}. This equation can be solved by the conjugate gradient (CG) method acting on the form
	\begin{equation*}
	\calF^{-1}\bigg\{\widehat{\bbG} : \calF\bigg(\frac{\partial\bfP}{\partial \bfF}: \frac{\partial\bfFFluc}{\partial\bfFMacro}\bigg)\bigg\} = -\calF^{-1}\bigg(\widehat{\bbG} : \calF\bigg[\frac{\partial\bfP}{\partial\bfF}\bigg]\bigg).
	\end{equation*}
	The CG method is used for solving equation~\eqref{fluctuation-sensitivity-equation} because of two reasons. (i) It is not easy to construct the matrix form of equation \eqref{microscale-equilibrium-equation} as $\bbG$ is only available in the Fourier space. (ii) The left-hand side of this equation is conveniently computed in a componentwise manner and the CG method can be applied to each component of equation.

	Since the latter equation is a secondary derivation from the equilibrium equation \eqref{microscale-equilibrium-equation}, it is essential to pass the average condition associated with \eqref{microscale-equilibrium-equation} on to this equation. Taking derivative of $\big\langle\bfFFluc \big\rangle = \bfZero$ with respect to $\bfFMacro$,  the average condition on the fluctuation sensitivity is revealed as follows
	\begin{equation*}
	\bigg\langle \frac{\partial\bfFFluc}{\partial\bfFMacro} \bigg\rangle_{\calR} = \bfZero.
	\end{equation*}
	
	\section{Surrogate model using approximator of macro-energy density based on neural networks}
	\label{Section:Surrogate-model-using-NN-approximator}
	
	\subsection{Rationale towards determination of the macroscopic constitutive law with a feedforward neural network}\label{Sec:Approximation-macro-constitutive-law}
	
	The theory of first-order periodic homogenization basically provides a macroscopic constitutive law in a numerical basis. In a mathematical formulation, this statement is reflected by two variational problems \eqref{microscopic-boundary-value-problem} and \eqref{macroscopic-boundary-value-problem}. The first variational problem defines the macroscopic energy density while excluding the given externally applied condition $\bff$ and $\bfT$, the second problem is characterized by this energy density. A numerical solution of the minimization problem \eqref{macroscopic-boundary-value-problem} based on gradient-based techniques such as Newton method and requires the evaluation of the derivatives of $\energyMacro$. In a finite element procedure, such derivatives are only inquired at the quadrature points of the entire problem domain. Therefore, the approximation solution can be achieved in a two-scale computation procedure in that the inquiries of the solver for the macroscopic BVP \eqref{macroscopic-boundary-value-problem} at one quadrature point is resolved by solving the corresponding microscopic BVP with the inputs obtained from the solver at the macroscale. 
	
	When the heterogeneous materials under consideration can be, up to the first-order approximation, characterized by only single representative volume element, the homogenized material behavior is represented by one single ``continuous'' macro-energy density. This will be a basic assumption underlying the application of the proposed method. At the first sight, this condition seems to be strict but it has proved to be successful in many engineering applications and real-world problems so far. In fact, most of the existing works use only one RVE for the two-scale computation process.
	
	While a two-scale computational can guarantee a convergence state of the solution, it is computationally demanding. We articulate in the following that in practice, gaining extra little accuracy is not as important as computational speed and saving computing resources while maintaining a certain level of accuracy. The two-scale computational procedure we are pursuing is the not the unique way to achieve the homogenized solution. Indeed, the most fundamental assumption is that we may replace the original problem by the homogenized problem characterized by \eqref{microscopic-boundary-value-problem} and \eqref{macroscopic-boundary-value-problem}. This results in the macro-homogeneity condition \eqref{macro-homogeneity-condition} that should be solved in couple with \eqref{microscopic-boundary-value-problem}-\eqref{macroscopic-boundary-value-problem} in order to link the micro- and macroscopic quantities. These three equations constitutes of an under-determined system. The macro-homogeneity condition can be fulfilled \emph{a priori} by applying the either one of three boundary conditions in the microscopic BVP or combination of those types different boundary sections. In doing so, we have plenty of possibilities of solving the microscopic BVP \eqref{microscopic-boundary-value-problem}, one of which is associated to one way of computing macroscopic energy density. According to this analysis, it is fair to interpret that there are certain amount of noise in the sampling data in the form $(\bfFMacro, \energyMacro)$. This articulation is also a strong reinforcement in why we may use neural networks as a sort of interpolation of the macro-energy density.
	
\begin{figure}[htb]
		\centering
		\includegraphics[width=0.75\textwidth]{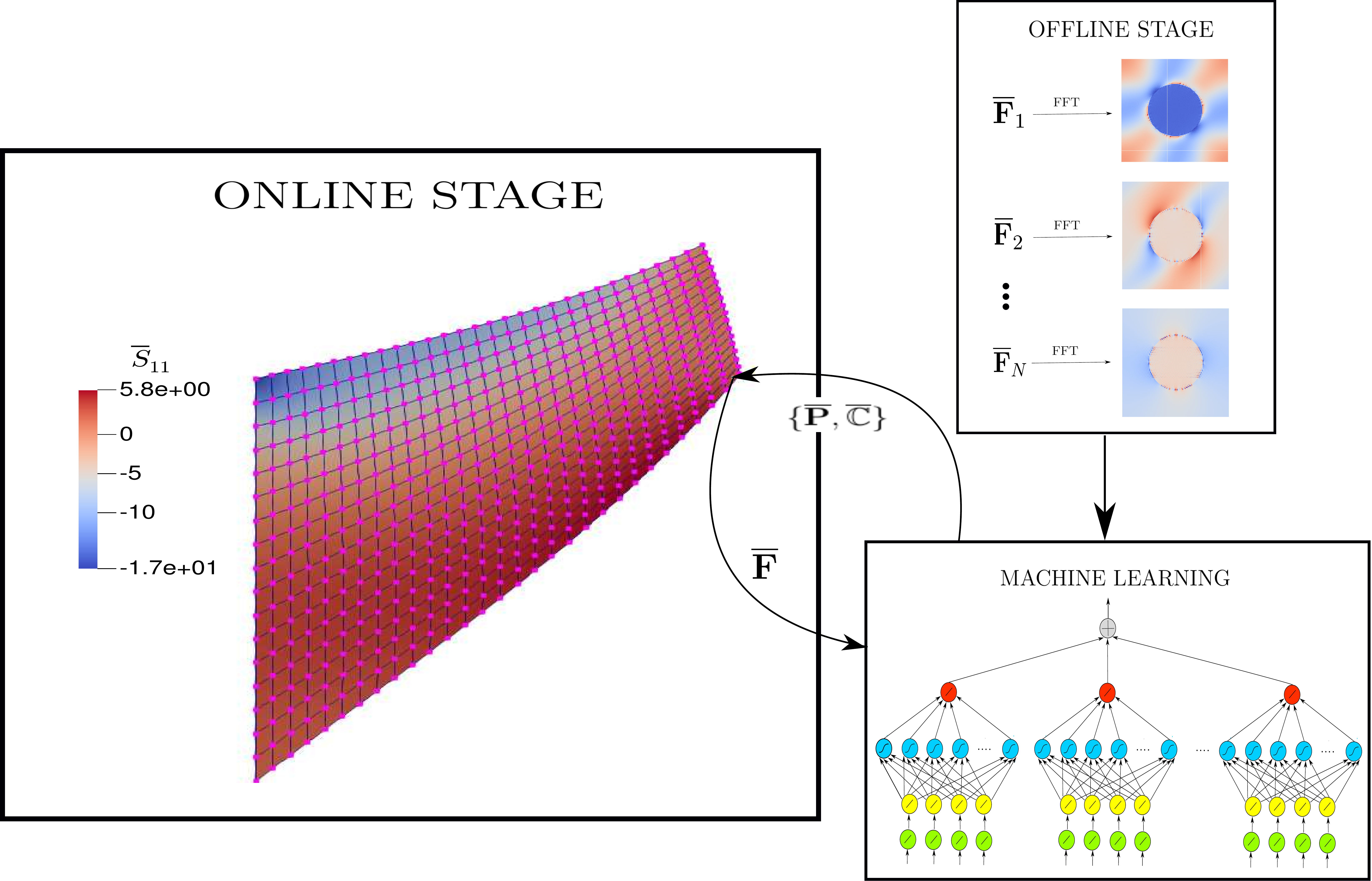}
		\caption{\textit{Surrogate model for computational homogenization by means of approximator of macro-energy density.}}
		\label{Fig:Surrogate-model-based-on-NN}
	\end{figure}

	\paragraph{\textbf{\sffamily Interpretation and terminology}} In the current approach, the macro-energy density will be made numerically available by means of an approximator that is constructed based on neural networks and input-output data in the form $(\bfFMacro, \energyMacro)$ . The process of collecting the data is conducted by solving many microscopic boundary value problems with given values $\bfFMacro$ as the inputs and computing the resultant macro-energy density \eqref{macroscopic-energy-density} as the outputs. Such a process is called \emph{offline stage} and while solving the macroscopic boundary value problem \eqref{macroscopic-boundary-value-problem} using the approximate macro-energy density is called \emph{online stage}. In combination, these two stages constitutes a surrogate model whose overall picture could be recapitulated in Fig.~\ref{Fig:Surrogate-model-based-on-NN}.

	\subsection{HDMR-based neural networks}
	The neural networks employed in this work stems from the contribution of \textsc{Manzhos\,\&\,Carrington}~\cite{Manzhos+Carrington-2006} which embedded partially the structure of high-dimensional model representation (HDMR) into neural network (\NN) fits. In fact, this work was aimed at building multidimensional potential energy surfaces which frequently arise in the field of computational chemistry. An efficient implementation of this kind of neural networks can be found in \textsc{Manzhos et al.}~\cite{Manzhos+Yamashita+Carrington-2009}.
	
	\subsubsection{High-dimensional model representation}
	A function $f$ of muli-variable $\bfx \in [0,1]^{D}$ can be approximated by the expansion (cf. \textsc{Manzhos\,\&\,Carrington}~\cite{Manzhos+Carrington-2006})
	\begin{equation}\label{HDMR-expansion}
	f^{\HDMR}(\bfx) = f_0 + \sum\limits_{i=1}^{D} \underbrace{f_{i}^{(1)}(x_i)}_{\text{component\newline function}} + \underbrace{\sum\limits_{1\leq i < j \leq D} f_{ij}^{(2)}(x_i, x_j)}_{\text{mode term}} \quad +\quad \cdots + f_{12\ldots D}^{(D)}(x_1, \ldots, x_D),
	\end{equation}
	which is called a high-dimensional model representation. This expansion consists of a sum of mode terms, each of which is in turn a sum of component functions $f_{i_1,\ldots, i_n}(x_{i_1},\ldots, x_{i_n})$. In general, such HDMR approximation is achieved by a least-squared optimization problem in which we minimize the following error functional
	\begin{equation}\label{least-square-for-HDMR}
	\min\int_{\mathbb{R}^{D}} \big[ f(\bfx) - f^{\HDMR} \big] \d\bfmu(\bfx),
	\end{equation}
	where $\d\bfmu$ stands for a predefined measure (see, e.g., \textsc{Stein\,\&\,Shakarchi}~\cite{Stein+Shakarchi-2005}). Several methods have been proposed by choosing different measures $\d\bfmu$ and the associated strategies were developed for determining the component functions.
	
	As long as the functional form of the component functions characterized by controlling parameters and the measures $\d\bfmu$ are known \emph{a priori}, the minimization problem \eqref{least-square-for-HDMR} can be solved by a suitable gradient-based method. Among all possibilities, using neural networks as component functions is attractive because the functions they representing are universal approximators and efficient methods for computing the network weights are available.
	
	\subsubsection{Neural networks based on the structure of high-dimensional representation model}
	We summarize here the neural networks introduced by \textsc{Manzhos et al.}~\cite{Manzhos+Yamashita+Carrington-2009}. In addition, we adapt the presented theory to our specific application. A theory of artificial neural networks can be found in the standard text by \textsc{Goodfellow et al.}~\cite{Goodfellow+Bengio+Courville-2016}.
	
	We recall that a multilayer perceptron (MLP) belongs to the class of feedforward neural network and thus consists of (i) an input layer, (ii) a number of hidden layers and (iii) an output layer. The construction of \HDMR{} expansion below will be based on the sum of many MLPs. As for our application, the input layer corresponds to the components of the macroscopic deformation gradient and the output layer to the macro-energy density. Except for the input nodes, each node is a neuron characterized by an associated activation function and its weight. First, we postulate the Ansatz of the the HDMR function $f^{\HDMR}$ in the form
	\begin{equation}\label{expansion-in-neural-network}
	f^{\HDMR}(x_1, \ldots, x_D) = \sum\limits_{i = 1}^{L} g_i^{\NN}(\bfx),
	\end{equation}
	where each $g_i^{\NN}(\bfx)$ is a neural network and there are $L$ component functions in this expansion. As compared to \eqref{HDMR-expansion}, the expansion into multiple mode terms have been imitated at this step. Next, we perform dimensionality reduction in the arguments of $g_i^{\NN}$ in such a way that
	\begin{equation}\label{g-neural-network}
	g_i^{\NN}(x_1, \ldots, x_D) = f_i^{\NN}(y_1^{i}, y_2^{i}, \ldots, y_d^{i}),
	\end{equation}
	where $d<D$ is the reduced dimension and the reduced coordinate $\bfy^{i}$ is obtained from the linear mappings
	\begin{equation}\label{dimensionality-reduction}
	\bfy^{i} = \bfA^{i}\bigcdot \bfx + \bfb^{i}.
	\end{equation}
	In general, the dimension of $\bfy^{i}$ for all component index $i$ is smaller than the dimension of $\bfx$, so $\bfA^{i}$ is a matrix of size $(d\times D)$. At this step, the partial spirit of dimensionality reduction in the component functions of the HDMR expansion \eqref{HDMR-expansion} has been copied. Now that we employ the feedforward neural network with one hidden single-layer to represent $f_{i}^{\NN}$ as follows
	\begin{equation}\label{feedforward-neural-network}
	f_i^{\NN}(y_1^i, \ldots, y_d^{i}) = \sum\limits_{n=1}^{N^{i}} c_n^{i} \sigma(\bfw_n^{i} \bigcdot \bfy^{i} + v_n^{i}) + v_0^{i},
	\end{equation}
	where equation $\sigma$ is the activation function for the the hidden layer, $N^{i}$ is the number of hidden neurons corresponding to the $i$-th component function. The specific activation function will be presented later for a direct relevance.
	
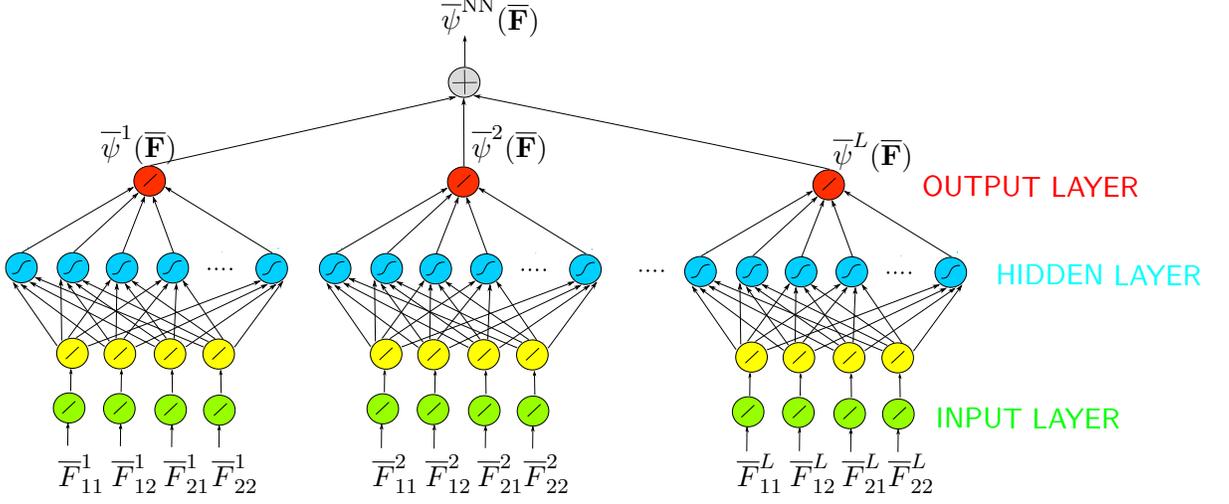
\begin{figure}[htb]
		\centering{
			\def\svgwidth{0.75\textwidth}
			\input{./images/NeuronNetwork.pdf_tex}
		}\vspace{12pt}
		\caption{\textit{Architecture of the HDMR-based neural network function.} A function $f$ of multidimensional variable $(x_1,\ldots, x_D)$ is approximated by a summation of $L$ component functions. One component is a neural network with two hidden layers which uses in order the linear activation functions and $\tanh$ activation functions. It should be interpreted that $\FMacro_{ij}^{k} = \FMacro_{ij}$ for all the component functions.}
		\label{Fig:Architecture-of-HDMR-neural-network}
	\end{figure}

	Combing equations \eqref{expansion-in-neural-network}--\eqref{feedforward-neural-network}, we arrive at
	\begin{equation}\label{HDMR-expansion-neural-network-function}
	f^{\HDMR}(x_1, \ldots, x_D) = \sum\limits_{i = 1}^{L} \Bigg\{ \sum\limits_{n=1}^{N^{i}} c_n^{i} \sigma^{i}\big( \bfw_n^{i} \bigcdot \big[\bfA^{i}\bigcdot \bfx + \bfb^{i}\big] + v_n^{i}\big) + v_0^{i}\Bigg\}.
	\end{equation}
	In doing so, we actually use $L$ neural networks with two hidden functions, first of which uses linear activation function while the second uses the nonlinear activation function $\sigma$. In Fig.~\ref{Fig:Architecture-of-HDMR-neural-network}, the architecture of the sum of neural networks \eqref{HDMR-expansion-neural-network-function} is presented.
	The larger number of component functions $L$ is used, the more terms we may reflect in \eqref{HDMR-expansion} up to the reduced dimension $d$. Note that the function form \eqref{HDMR-expansion-neural-network-function} already reflects the \HDMR{} expansion \eqref{HDMR-expansion} even though $d$ is chosen to be identical for all component functions. In fact, if we restrict the some rows of $\bfA^{i}$ to be zero, then corresponding zero values of $\bfy^{i}$ are not effected. Nevertheless, it is not necessarily the case in minimizing the functional \eqref{least-square-for-HDMR}.
	
	Note that the open-source program provided by \textsc{Manzhos et al.}~\cite{Manzhos+Yamashita+Carrington-2009} permits more alternatives. For example, the reduced dimension $d$ can be chosen differently for component functions. For more detail, the readers are referred to the original work.
	
	\subsubsection{Training of the neural networks for the approximation of macro-energy density}
	An appropriate measure $\d\bfmu$ can be chosen so that the function defined in \eqref{least-square-for-HDMR} can be reduced to the arithmetic average of $f - f^{\HDMR}$ evaluated at the input data. In our case, such input data are the sampling data of the macroscopic deformation gradient $\bfFMacro$ and the output data are the corresponding macro-energy density as the outcome of solving the microscopic BVP. So we need to minimize
	\begin{equation}\label{minimization-problem-for-macro-energy-density}
	\bigg\{\frac{1}{| \rmD |} \sum\limits_{\bfFMacro \in \mathrm{D}}\Big[ \energyMacro_{|\bfFMacro} - \energyMacro^{\NN}(\bfFMacro) \Big]^2 \bigg\}^{1/2} \longrightarrow \min
	\end{equation}
	with respect to the neuron weights, where $|\rmD|$ denotes the number of data in $\rmD$. In this formulation $\energyMacro^{\NN}$ is the approximation of the macro-energy density that we are ultimately seeking. Its structure is described as in equation \eqref{HDMR-expansion-neural-network-function} with $\bfx$ being replaced with $\bfFMacro$.
	
	It should be emphasized here that the HDMR-based function \eqref{HDMR-expansion-neural-network-function} using the neural networks as each component function is not of standard feedforward neural networks. The weights associated with hidden neurons for each component function are interrelated. Therefore, standard methods for seeking the neuron weights such as back propagation method is not directly applicable. Although one could apparently solve this minimization \eqref{minimization-problem-for-macro-energy-density} by a gradient method, \textsc{Manzhos et al.}~\cite{Manzhos+Yamashita+Carrington-2009} has provided an efficient algorithm to obtain the neuron weights by reusing the training network algorithm for each component functions. Our contribution utilizes the program published therein.
	
	The outcome of the training process is the approximator of the macro-energy density given by
	\begin{equation}\label{HDMR-macro-energy-density}
	\energyMacro^{\NN}(\overline{\bfF}\big) = \sum\limits_{i=1}^{L} \Bigg\{ \sum\limits_{n = 1}^{N^{i}}c_n^{i} \sigma^{i}\big(\bfw_{n}^{i} \bigcdot\big[\bfA^{i}\bigcdot \bfFMacro + \bfb^{i}\big] + v_n^{i} \big) + v_0^{i}\Bigg\}.
	\end{equation}
	The architecture of this neural network macro-energy density is illustrated in Fig.~\ref{Fig:Architecture-of-HDMR-neural-network}.
	
	\subsection{Analytical derivatives of neural networks}
	\subsubsection{Computation of macroscopic stress and tangent moduli}
	The finite element method is used in tandem with the Newton-Raphson method to solve the resultant macroscopic boundary value problem. We briefly derive the formulation for the Newton scheme for the macroscopic boundary value problem. The variational equation derived from the minimization problem \eqref{macroscopic-boundary-value-problem} with $\energyMacro$ replaced by $\energyMacro^{\NN}$ is given by
	\begin{equation*}
	J(\bfvarphiMacro, \delta\bfvarphiMacro) := \big\langle \D\Pi(\bfvarphiMacro), \delta\bfvarphiMacro \big\rangle = \int_{\calB}\nabla\delta\bfvarphiMacro :  \frac{\partial \energyMacro^{\NN}}{\partial \bfFMacro}\dV - \int_{\calB}\bff \bigcdot\delta\bfvarphiMacro\dV - \int_{\GammaN} \bfT \bigcdot \delta\bfvarphiMacro\dA = 0 \quad \forall\delta\bfvarphiMacro.
	\end{equation*}
	Linearization of the left-hand side of this equation with respect to $\bfvarphiMacro$ givess
	\begin{equation*}
	\big\langle\D J(\bfvarphiMacro, \delta\bfvarphiMacro), \Delta\bfvarphiMacro\big\rangle = \int_{\calB} \delta\bfFMacro : \frac{\partial^2\energyMacro^{\NN}}{\partial\bfFMacro\partial\bfFMacro} : \Delta\bfFMacro \dV.
	\end{equation*}
	Then, the Newton-Raphson iteration is formulated as
	\begin{equation}\label{Newton-iteration-for-macroproblem}
	\begin{aligned}
	\big\langle \D J(\bfvarphi^{[k], \delta\bfvarphiMacro}, \Delta\bfvarphiMacro^{[k]} \big\rangle & = \big\langle \bff, \delta\bfvarphiMacro \big\rangle + \big\langle \bftMacro,\delta\bfvarphiMacro \big\rangle_{\GammaN} - \Big\langle \bfPMacro^{\NN,[k]}, \delta\bfFMacro \Big\rangle \qquad \forall\delta\bfvarphiMacro, \\
	\bfvarphiMacro^{[k+1]} &= \bfvarphiMacro^{[k]} + \Delta\bfvarphiMacro^{[k]},
	\end{aligned}
	\end{equation}
	where we have used the following inner product denotations
	\begin{equation*}
	\Big\langle \bfPMacro^{\NN,[k]}, \delta \bfFMacro \Big\rangle = \int_{\calB}\partial_{\bfFMacro}\energyMacro^{\NN}\big(\bfFMacro^{[k]}\big) : \nabla\delta\bfvarphiMacro\dV, \quad \big\langle \bff, \delta\bfvarphiMacro\big\rangle = \int_{\calB} \bff\bigcdot\delta\bfvarphiMacro \dV, \quad \big\langle \bfT, \delta\bfvarphiMacro \big\rangle_{\GammaN} = \int_{\GammaN}\bfT \bigcdot \delta\bfvarphiMacro \dA. 
	\end{equation*}
	
	At each iteration of the algorithm \eqref{Newton-iteration-for-macroproblem}, we need to compute the stress and the tangent stiffness as follows
	\begin{equation*}
	\bfPMacro^{\NN} = \frac{\partial\energyMacro^{\NN}}{\partial \overline{\bfF}}, \quad \overline{\bbC}^{\NN} = \frac{\partial^2 \energyMacro^{\NN}}{\partial\bfFMacro \partial\bfFMacro}.
	\end{equation*}
	In our contribution, we exclusively use the $\tanh$ as the activation function, i.e. $\sigma^{i}(x)= \sigma(x) = \tanh(x)$ for all activation functions, which implies
	\begin{equation*}
	\sigma^{\prime}(x) = 1 - \sigma(x)^2, \quad \sigma^{\prime\prime}(x) = 2\sigma(x)\big[\sigma(x)^2 - 1\big].
	\end{equation*}
	For artificial neural networks, $\mathrm{tansig}$ is normally used instead of $\tanh$ and implemented in such a way that it speeds up the training process even though they are mathematically identical.
	
	Taking the first and second derivatives of the explicit expression \eqref{HDMR-macro-energy-density}, we arrive at
	\begin{equation}\label{neural-network-stress-tangent}
	\begin{aligned}
	\bfPMacro^{\NN} &= \sum\limits_{i=1}^{L}\bigg\{ \sum\limits_{n=1}^{N^{i}}c_n^{i} \bfw_n^{i} \bigcdot \bfA^{i} \big[1 - \sigma(q^{i})^2 \big] \bigg\},\quad \text{with} \quad q^{i} = \bfw_n^{i} \bigcdot \big[\bfA^{i} \bigcdot\bfFMacro + \bfb^{i}\big] + v_n^{i}, \\
	\overline{\bbC}^{\NN} &= \sum\limits_{i=1}^L \bigg\{ \sum\limits_{n=1}^{N^{i}} 2 c_n^{i} (\bfw_n^{i} \bigcdot\bfA^{i}\big)^{T}\bigcdot\big(\bfw_n^{i} \bigcdot \bfA^{i}\big) \big[\sigma(q^{i})^2 - 1 \big]\sigma(q^{i}) \bigg\}.
	\end{aligned}
	\end{equation}
	These two expressions are the approximators of $\bfPMacro$ and $\overline{\bbC}$ that will be used in the macrosopic solver.
	
	\subsubsection{Feature normalization and recover to the physical values}
	In this part, we draw special attention to one crucial practice in implementation for those who wish to use the package provided in \textsc{Manzhos et al.}~\cite{Manzhos+Yamashita+Carrington-2009}. As standardized in MATLAB package and many existing libraries, the training process of neural networks accept the normalized features, including input and outputs, to accelerate the convergence of the algorithm. This step of feature normalization must be taken into account for our implementation.
	
	We borrow again the representation \eqref{HDMR-expansion-neural-network-function} to illustrate the feature normalization step. Assume that the range of the input coordinates and the output coordinates are respectively given by
	\begin{equation}\label{data-range}
	\begin{aligned}
	\Omega_r^{\text{input}} &= [(x_r)_{\max}, (x_r)_{\max}], \quad & (x_r)_{\min} &= \min\{(x_r)_{j}\}_{j=1}^{N_{\text{data}}},\quad & (x_r)_{\max} &= \max\{ (x_r)_{j}\}_{j=1}^{N_{\text{data}}}, \\
	\Omega^{\text{output}} &= [f_{\min}, f_\text{max}], \quad & f_{\min} &= \min\{ f_j \}_{j=1}^{N_\text{data}}, \quad & f_\text{max} &= \max\{ f_j \}_{j=1}^{N_\text{data}},
	\end{aligned}
	\end{equation}
	in which $(x_r)_{\min}$, $(x_r)_{\max}$ means the minimum and maximum of the input data in direction $r$, with $1\leq r \leq D$, and $f_{\min}$, $f_{\max}$ are the minimum and maximum of the output data. The input and output data are both normalized to the range $[-1,1]$ by the following scaling
	\begin{equation}\label{data-scaling}
	(\xi_r)_j = 2\frac{(x_r)_j - (x_r)_{\min}}{(x_r)_{\max} - (x_r)_{\min}} - 1, \quad g_j = 2\frac{f_j - f_{\min}}{f_{\max} - f_{\min}} - 1, \quad \quad \forall r = 1,\ldots, D,
	\end{equation}
	where $\bfxi_j$ and $g_j$ are the scaled data. Following definitions of the min- and max-quantities given in \eqref{data-range}, it is plain that $(\bfx_r)_j \in [-1,1]$ and $g_j \in [-1,1]$.
	
	Let us assume that $f^{\HDMR} = f^{\HDMR}(\bfx)$ of the neural network function trained from the original data $\rmD = \{\bfx_j, f_j\}_{j=1}^{N_\text{data}}$ and similarly $g^{\HDMR} = g^{\HDMR}(\bfxi)$ the outcome corresponding to the normalized data $\rmD_\text{norm} = \{\bfxi_j, g_j\}$. These two neural network functions are related to each other by the transformation which is the \emph{continuous} version of data scaling \eqref{data-scaling} as follows
	\begin{equation*}
	f^{\HDMR}(\bfx) = \frac{\Delta f}{2}\big[g^{\HDMR}(\bfxi(\bfx)) + 1\big] + f_{\min}, \quad \xi_r = 2\frac{x_r - (x_r)_{\min}}{\Delta x_r}-1,
	\end{equation*}
	where $\Delta f = f_{\max} - f_{\min}$ and $\Delta x_r = (x_r)_{\max} - (x_r)_{\min}$.
	
	Now, note that the package given in \textsc{Manzhos et al.}~\cite{Manzhos+Yamashita+Carrington-2009} accepts the normalized data $\rmD_\text{norm}$ and generate $g^{\HDMR}$ in terms of the associated neural network weights. Therefore, in order to compute the derivative of $f^{\HDMR}$ with respect to the \emph{original} coordinate $\bfx$, we rely on the chain rule
	\begin{equation*}
	\frac{\partial}{\partial x_r}f^{\HDMR} =  \frac{\Delta f}{2}\bigg[\sum\limits_{s = 1}^{D}\frac{\partial g^{\HDMR}}{\partial \xi_s} \frac{2}{\Delta x_s}\delta_{s r}\bigg] = \frac{\Delta f}{\Delta x_r} \frac{\partial}{\partial \xi_r} g^{\HDMR} \qquad \forall r = 1, \ldots, D.
	\end{equation*}
	This first derivative serves later for the computation of the macroscopic stress \eqref{neural-network-stress-tangent}$_1$. Similarly, the second derivative
	\begin{equation*}
	\frac{\partial^2}{\partial x_r\partial x_s}f^{\HDMR} = \frac{\Delta f}{\Delta x_r \Delta x_s} \frac{\partial^2}{\partial \xi_r \partial \xi_s} g^{\HDMR} \qquad \forall r, s = 1,\ldots D
	\end{equation*}
	is employed for the computation of the macroscopic tangent moduli \eqref{neural-network-stress-tangent}$_2$.

	\section{Representative numerical examples}
	\label{Section:Representative-examples}
	The examples are carefully chosen in the order of difficulty and at the same time demonstrate the robustness of the proposed computational framework. In order to earn confidence in the reliability of the neural network macro-energy density in action, we showcase in all examples, except one as it is not relevant, the full-field solution as a mean of comparison.
	
	We start with a one-dimensional toy problem where the heterogeneity is idealized and mathematically characterized by the energy density with oscillating material parameters. Second and third examples are aimed at justifying the approximate stress field and tangent moduli given by \eqref{neural-network-stress-tangent} as compared to the corresponding numerical quantities obtained through the FFT-based solver for microscopic BVP. The last two numerical examples delve into the real-world applications where three types of solutions are constructed for comparison: (i) numerical solution by a surrogate modeling (ii) numerical solution by a two-scale FE-FFT computation approach, (iii) full-field solution.
	
	Prior to going to details of all below examples, we record here the information regarding the neural network architectures used thereof in Table~\ref{Table:Architecture-of-neural-networks}. The meanings of variables in this table are also briefly recalled in Table~\ref{Table:Meanings-of-variables-in-architecture}.
	\begin{table}[htb]
		\begin{center}
			{\def\arraystretch{1.4}
				\begin{tabular}{|l|l|}
					\hline
					$D$ & Original dimension of the variable $\bfFMacro$ as the input argument of $\energyMacro$ \\
					$d$ & Reduced dimension; refer to Eq.~\eqref{g-neural-network} \\
					$L$ & Number of component functions; refer to Eq.~\eqref{HDMR-macro-energy-density} \\
					$N$ & Number of neurons in the second hidden layer of the component function; refer to Eq.~\eqref{feedforward-neural-network} \\
					$N_\text{data}$ & Number of data for training the network \eqref{HDMR-expansion-neural-network-function}; refer to Eq.~\eqref{least-square-for-HDMR} \\
					\hline
			\end{tabular}}
		\end{center}
		\caption{\textit{Meanings of the parameters that describe the architecture of neural network \eqref{HDMR-macro-energy-density}}.}
		\label{Table:Meanings-of-variables-in-architecture}
	\end{table}
	\begin{table}[htb]
		\begin{center}
			\def\arraystretch{1.4}
			\begin{tabular}{|c|c|c|c|c|c|c|}
				\hline
				\textbf{Examples} & $D$ & $d$  & $L$ & $N$ & $N_\text{data}$ & Database size \\ \hline
				\textsf{Example 5.1} & $1$ & $1$ & $2$ & $5$ & $10^{3}$ & $10\times 10^{3}$ \\ \hline
				\textsf{Example 5.2} & $4$ & $4$ & $15$ & $20$ & $50\times 10^{3}$ & $200\times 10^{3}$ \\ \hline
				\textsf{Example 5.3} & $4$ & $4$ & $15$ & $20$ & $30\times 10^{3}$ & $200\times 10^{3}$ \\
				\hline  
			\end{tabular}
		\end{center}
		\caption{\textit{Record of architectures of neural networks used in different numerical examples}}
		\label{Table:Architecture-of-neural-networks}
	\end{table}
	
	\subsection{A mathematical one-dimensional toy problem}
	\begin{figure}
		\centering
		\begin{subfigure}{0.47\textwidth}
			\centering{ \includegraphics[width=0.7\textwidth]{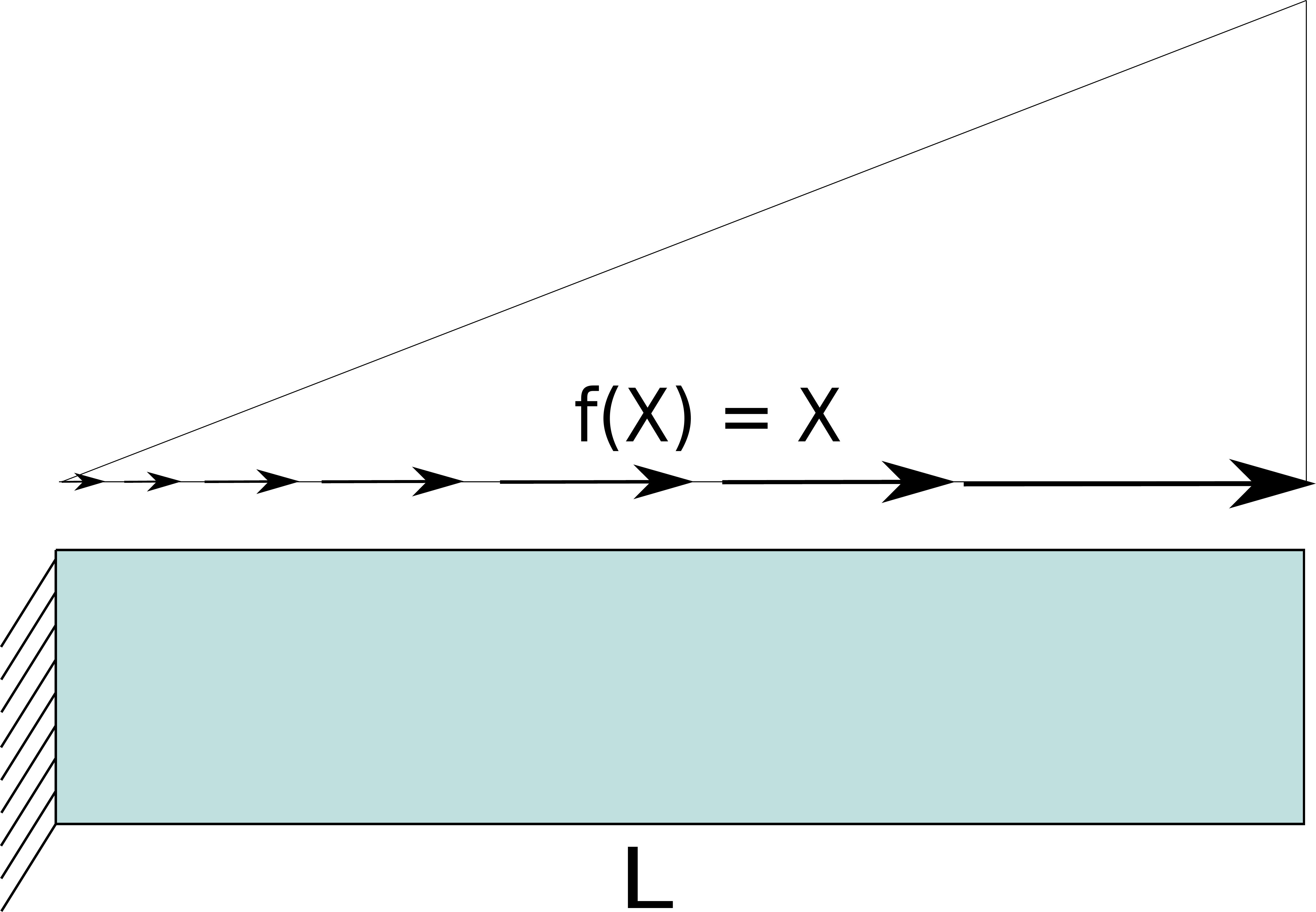}}
		\end{subfigure}
		\begin{subfigure}{0.47\textwidth}
			\centering{ \includegraphics[width=1.0\textwidth]{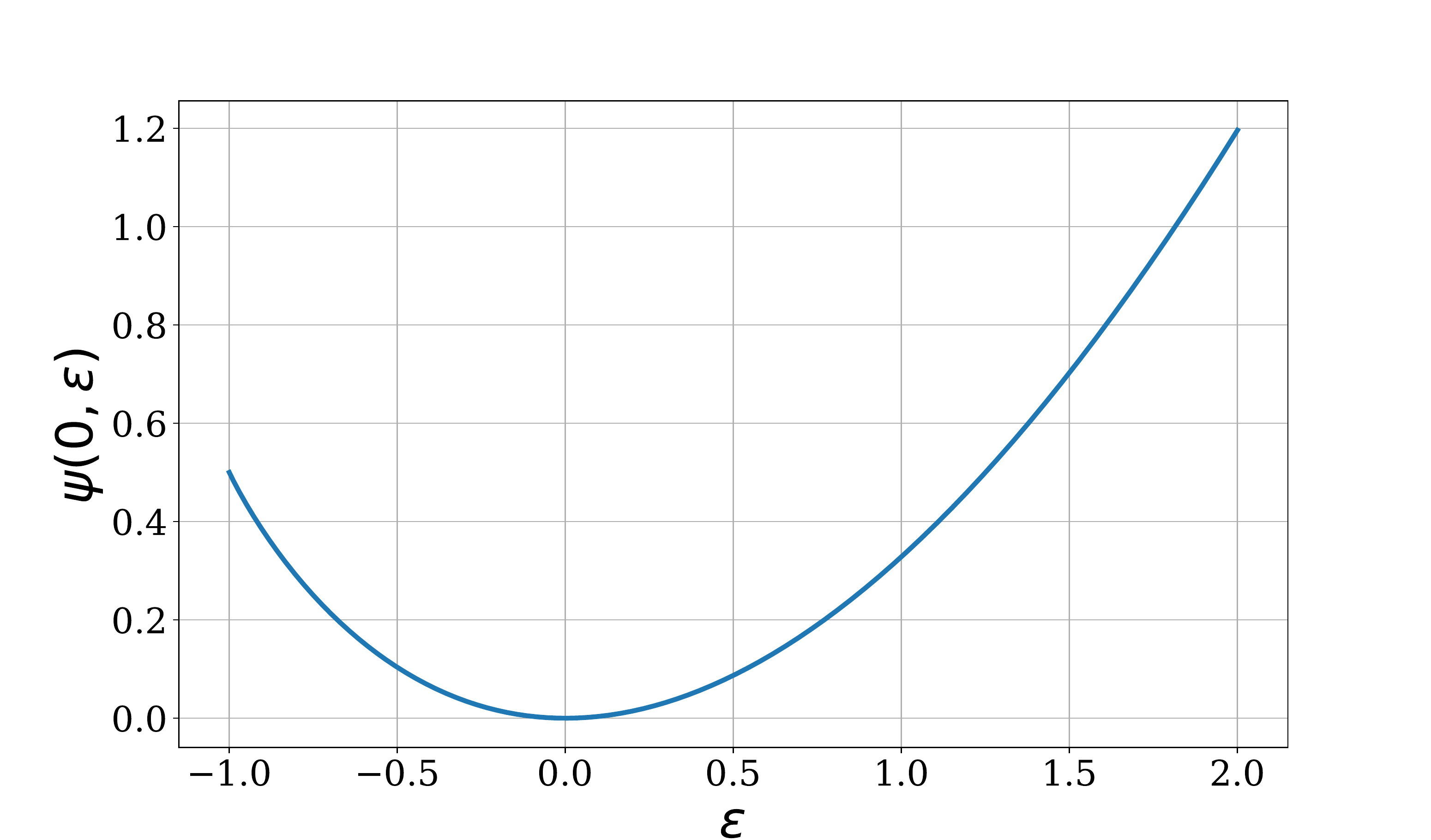} }
		\end{subfigure}
		\caption{\textit{Mathematical toy problem.} (\emph{left}) Problem setting of the mathematical heterogeneous bar subject to a traction. (\emph{right}) With $\mu(0) = 3/2$, the energy function $\psi(0,\epsilon)= (1+\epsilon)^{3/2} - 3/2\epsilon - 1$ is plotted against the $\epsilon$-coordinate.}
		\label{Fig:Mathematical-Bar-Problem}
	\end{figure}
\begin{figure}
		\centering
		\begin{subfigure}{0.47\textwidth}
			\centering
			\includegraphics[width=1.0\textwidth]{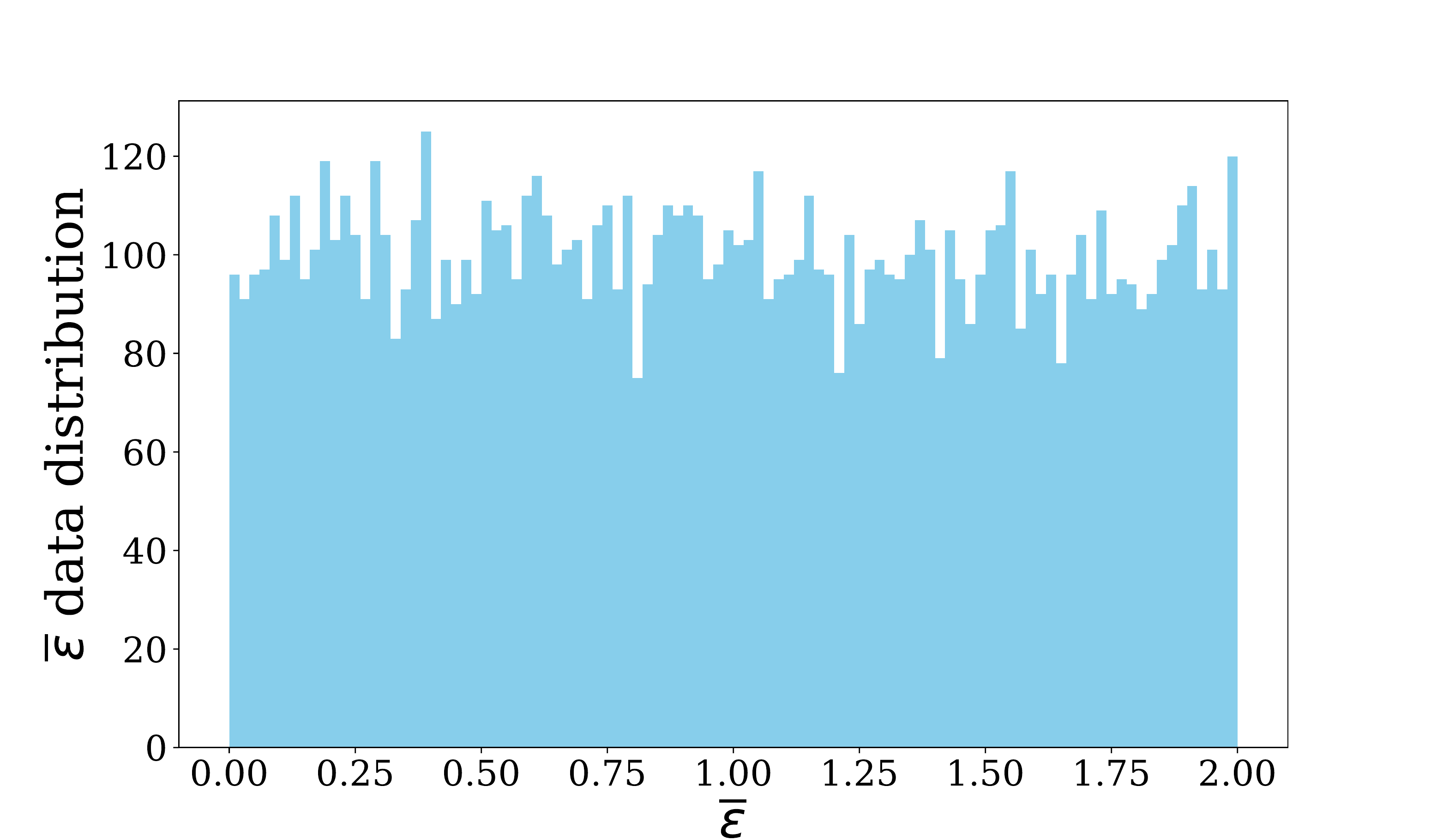}
		\end{subfigure}
		\begin{subfigure}{0.47\textwidth}
			\centering
			\includegraphics[width=1.0\textwidth]{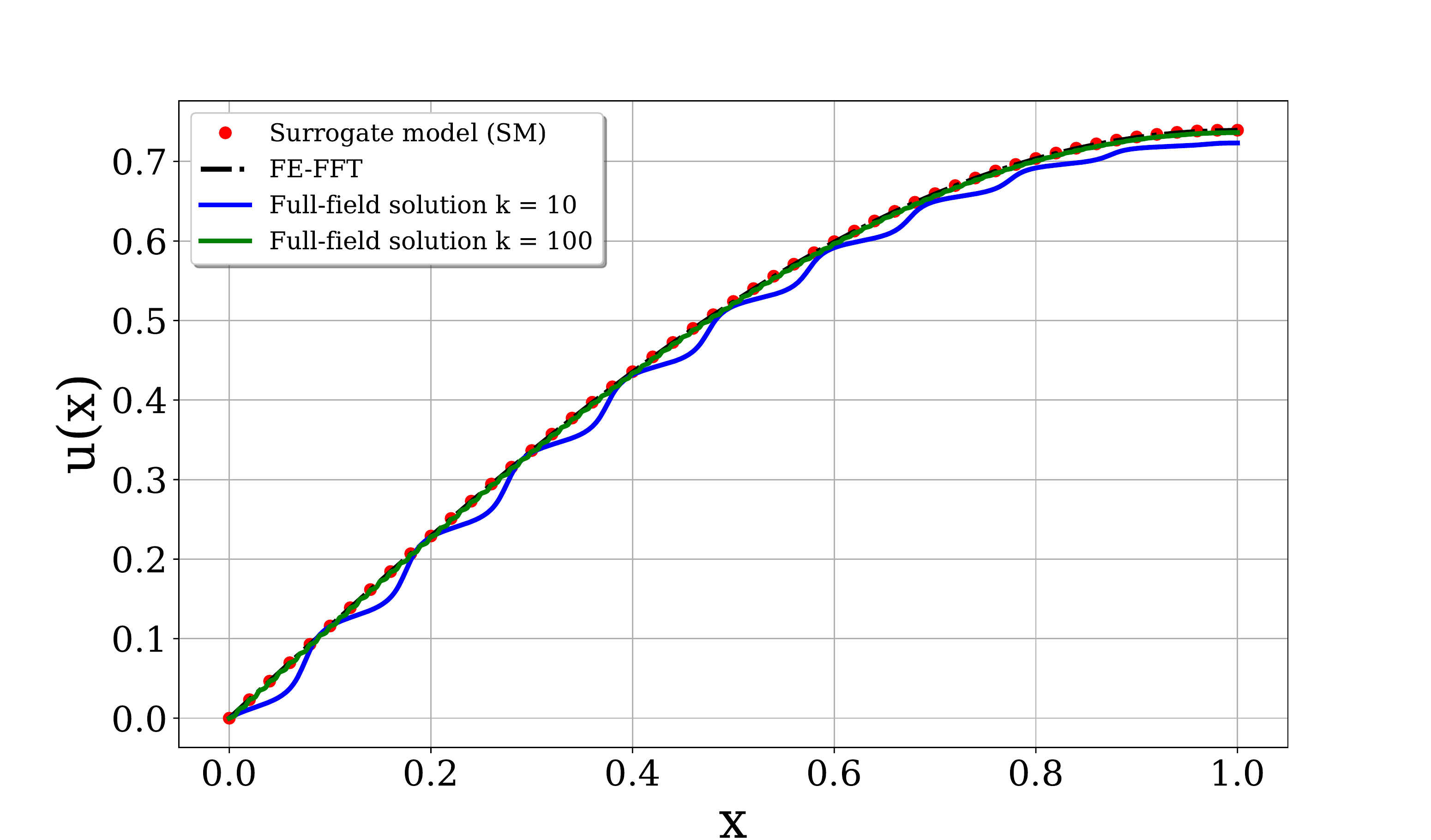}
		\end{subfigure}
		\caption{\textit{Comparison between homogenization solution and full-field solution.} (\emph{left}) The $100$ macroscopic strain data $\epsilonMacro$ is uniformly distributed in the range $[0, 2]$, which can be seen by its histogram plot.  (\emph{right}) The homogenized solution is obtained by using the neural network macro-energy density with $1000$ sampling data (red dots) and by the two-scale FE-FFT method (black dashed line). The full-field solution is obtained by using standard FEM with a high number of elements. }
		\label{Fig:One-Dimensional-Solution-Comparison}
	\end{figure}

	Let us start with a simple bar problem described by the following minimization problem (see Fig.~\ref{Fig:Mathematical-Bar-Problem}
	\begin{equation*}
	\delta \bigg\{ \int_{0}^{L} \psi(X,\epsilon)\dX - \int_{0}^{L}f(X) u(X)\dX - \big[\overline{t}_0\,u(X)\big]_{X\rightarrow L} \bigg\} = 0, \qquad \psi(X,\epsilon) = \mu(X) \bigg[\frac{2}{3}(1 + \epsilon)^{3/2} - \epsilon - \frac{2}{3}\bigg]
	\end{equation*}
	with the essential boundary condition $u(0) = 0$. Note that $\psi(X, \epsilon = 0) = 0$. In this formulation, $X$ is the reference coordinate, $\epsilon = \du/\dX$ denotes the gradient of the displacement field, $t_0$ is the traction force applied to the bar at $X = L$ and the mathematical parameter $\mu=\mu(X)$ representing the inhomogeneities is given by
	\begin{equation*}
	\mu(X) = \frac{3}{2} + \sin(2\pi k X), \quad k \in \mathbb{Z}^{+}.
	\end{equation*}
	This boundary value problem in the strong form reads
	\begin{equation*}
	\frac{\d}{\dX} \frac{\partial\psi}{\partial\epsilon} + f(X) = 0,
	\end{equation*}
	where the boundary conditions are translated to
	\begin{equation*}
	u(0) = 0, \quad \frac{\partial\psi}{\partial\epsilon}\bigg(\frac{\du}{\dX}(L)\bigg) = t_0.
	\end{equation*}
	The RVE is depicted by $\mu(\xi)$ on one wavelength $1/k$, that is $\mu(\xi)$, $\xi \in (X - 1/2k, X+ 1/2k)$. The microscopic BVPs are solved for $10^{4}$ input macroscopic strain data $\epsilonMacro$ that are uniformly distributed in the range $[0,2.0]$ to compute the macro-energy density that are employed as output data. Then, we select randomly only $10\%$ for the training process. The corresponding approximate energy density $\overline{\psi}^{\NN}$ is obtained by using two component functions and each has five neurons in its second hidden layer (see Row \textsf{Example 5.1} of Table~\ref{Table:Architecture-of-neural-networks}) . In Fig.~\ref{Fig:One-Dimensional-Solution-Comparison}, the homogenized solutions obtained by the surrogate model and a two-scale computational approach are compared with the full-field solutions using different wavenumber values $k$. It can be seen that when $k \rightarrow\infty$, the full-field solution converges to the homogenized solution. In addition, the surrogate-model solution agrees excellently with the two-scale solution computed by the FE-FFT method.
	
	\subsection{Surrogate model for a laminate microstructure}
	\label{Sec:Laminate-RVE-Problem}
	\begin{figure}[htb]
		\centering{
			\def\svgwidth{0.3\textwidth}
			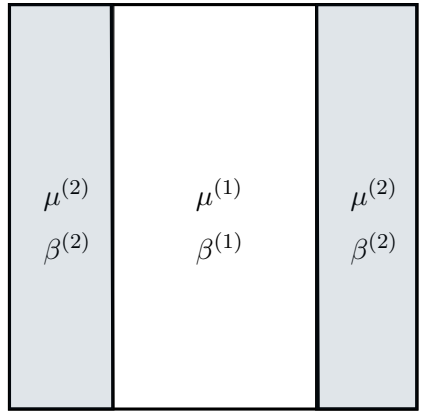
		}
		\caption{\textit{An RVE of laminate structure under plane strain condition.} The laminate consists of two phases of Neo-Hookean materials. Each phase takes up $50$\% volume of the entire RVE.}
	\end{figure}
	\paragraph{\textbf{\sffamily Problem setting}}  In this section, we delve into a two-dimensional homogenization problem that accepts the analytical solution. Particularly, we analyse a two-dimensional laminate RVE in the plane strain condition. This RVE comprises two phases of Neo-Hookean materials
	\begin{equation*}
	\psi(\bfF) = \frac{\mu}{2}\big[\mathrm{trace}(\bfF^{T}\bigcdot\bfF\big) - 2\big] + \frac{\mu}{\beta}\big[\det(\bfF)^{-\beta} - 1\big], \quad i = 1,2.
	\end{equation*}
	where $\mu$ is the shear modulus and $\beta$ is given in term of Poisson ratio $\nu$. The material parameter $\beta$ is determined via the Poisson ratio according to $\beta = 2\nu/(1-\nu)$. The gradient and hessian of $\psi(\bfF)$ are the stress tensor and tangent moduli that are explicitly computed as
	\begin{equation*}
	\begin{aligned}
	P_{ij} &= \frac{\partial \psi}{F_{ij}} = \mu F_{ij} - \mu \det(\bfF)^{-\beta} F_{ji}^{-1}, \\
	C_{ijkl} &= \frac{\partial^2 \psi}{\partial F_{kl}\partial F_{ij}} = \mu \delta_{ik} \delta_{jl} + \mu \det(\bfF)^{-\beta}(\beta F_{ji}^{-1} F_{lk}^{-1} + F_{jk}^{-1} F_{li}^{-1}\big),
	\end{aligned}
	\end{equation*}
	where $\bfF^{-1}$ is the inverse tensor of $\bfF$. We denote by $(\ast)^{(i)}$ the material parameter $(\ast)$ associated with phase $i$, $i = 1,2$ and consider in our example
	\begin{equation*}
	\mu^{(1)} = 100, \quad \mu^{(2)} = 1000,\quad \beta^{(1)} = \beta^{(2)} = 1.
	\end{equation*}
	
	\paragraph{\textbf{\sffamily Analytical solution}} The microscopic boundary value problem corresponding to the laminate structure above admits analytical solution. For self-contained reading, we summarize the main steps given in the work \textsc{Go\"{u}z\"{u}m et al.}~\cite{Gokuzum+Nguyen+Keip-2019} which dealt with the laminate RVE problem with the electro-mechanically coupled materials. We will derive here a system of algebraic equations for determining the gradient deformation fields $\bfF^{(i)}$ distributed at two phases which are resulted from the application of a generic macroscopic deformation gradient $\bfFMacro$. First, we recall the two following differential equations must be fulfilled throughout the \RVE
	\begin{equation*}
	\mathrm{div}\bfP^{T} = \bfZero, \quad \nabla\times\bfF = 0 \quad\Leftrightarrow\quad P_{ij,j} = 0, \quad \epsilon_{imn} \partial_{X_m} F_{jn} = 0.
	\end{equation*}
	The first equation is the microscopic equilibrium equation while the second is the compatibility condition. We specialize these equations to the two-dimensional case as follows
	\begin{equation*}
	\begin{aligned}
	P_{11,1} + P_{12,2} = 0, \quad F_{12,1} - F_{11,2} = 0, \\
	P_{21,1} + P_{22,2} = 0, \quad F_{22,1} - F_{21,2} = 0.
	\end{aligned}
	\end{equation*}
	It can be deduced from the assembly of the laminate phases, the variables appearing in the last equations are independent from $X_2$. Taking this fact into account, we arrive at
	\begin{equation*}
	\begin{aligned}
	P_{11,1} &= 0, \\
	P_{21,1} &= 0,
	\end{aligned}\qquad 
	\begin{aligned}
	F_{12,1} &= 0, \\
	F_{22,1} &= 0.
	\end{aligned}
	\end{equation*}
	which implies $P_{11}$, $P_{21}$, $F_{12}$, $F_{22}$ are all independent of $X_1$, $X_2$. In short, these fields are constant throughout the whole RVE domain. It is also due to the laminate structure, all variables $P_{12}$, $P_{21}$ and $F_{11}$, $F_{21}$ are constant within one phase. Accordingly, it is possible to reuse the notation $F_{ij}^{(i)}$ and $P_{ij}^{(i)}$ to denote the scalar values which the components of deformation gradient $\bfF$ and Piola-Kirchhoff stress $\bfP$ take on throughout the phase $(i)$.
	
	At this point, we have 8 unknown $F_{ij}^{(i)}$ and four equations
	\begin{equation*}
	\begin{aligned}
	F_{12}^{(1)} = F_{12}^{(2)}, \\
	F_{22}^{(1)} = F_{22}^{(2)},
	\end{aligned}\quad
	\begin{aligned}
	P_{11}^{(1)} = P_{11}^{(2)}, \\
	P_{21}^{(1)} = P_{21}^{(2)}.
	\end{aligned}
	\end{equation*}
	The other four equations come from the essential boundary condition of the microscopic BVP. They are the average condition
	\begin{equation*}
	\frac{1}{|\calR|}\int_{\calR}F_{ij}\dV = \overline{F}_{ij}\quad\Leftrightarrow\quad \frac{1}{2}\big( F_{ij}^{(1)} + F_{ij}^{(2)} \big) = \overline{F}_{ij}.
	\end{equation*}
	Keeping in mind that $\bfP = \partial_{\bfF}\psi$ given in terms of $\bfF$, we have just obtained the eight equations for determining the eight unknown $F_{ij}^{(i)}$ as follows
	\begin{equation}\label{Laminate-analytical-formula}
	\begin{aligned}
	F_{12}^{(1)} = F_{22}^{(2)} = \overline{F}_{22}, \quad \frac{1}{2}\big[ F_{11}^{(1)} + F_{11}^{(2)} \big] = \overline{F}_{11}, \quad \frac{\partial\psi^{(1)}}{\partial F_{11}} = \frac{\partial \psi^{(2)}}{\partial F_{11}}, \\
	F_{22}^{(1)} = F_{12}^{(2)} = \overline{F}_{12}, \quad \frac{1}{2}\big[ F_{21}^{(1)} + F_{21}^{(2)} \big] = \overline{F}_{21}, \quad \frac{\partial \psi^{(1)}}{\partial F_{21}} = \frac{\partial \psi^{(2)}}{\partial F_{21}}.
	\end{aligned}
	\end{equation}
	As long as the variable $\bfFMacro$ is given, $\bfFMacro^{(i)}$ for each phase can be computed with very high accuracy. As in our comparison, we address the solutions obtained from this nonlinear system of algebraic equations as analytical solutions even though they can be obtained only in a numerical basis. Upon these solutions, we can compute the macro-energy density according to
	\begin{equation}\label{exact-macro-energy-density}
	\energyMacro^{(e)} = \frac{1}{|\calR|}\int_{\calR}\psi\dV = \frac{1}{2}\Big[\psi\big(\bfF^{(1)} \big) + \psi\big(\bfF^{(2)}\big) \Big].
	\end{equation}
	The exact macroscopic stress is easily computed as
	\begin{equation}\label{exact-stress}
	\bfPMacro^{(e)} = \frac{1}{2}\bigg[\frac{\partial\psi}{\partial\bfF}(\bfF^{(1)}) + \frac{\partial\psi}{\partial \bfF}(\bfF^{(2)})\bigg] = \frac{1}{2}\big[\bfP^{(1)} + \bfP^{(2)}\big].
	\end{equation} 
	We use the \emph{central difference} formula with the extremely small perturbation $\epsilon$ to compute the corresponding ``exact'' stress and tangent moduli
	\begin{equation}\label{exact-tangent-moduli}
	\overline{C}_{ijkl}^{(e)} \approx \frac{\overline{P}_{ij}^{(+)} - \overline{P}_{ij}^{(-)}}{2 \epsilon}, \quad \overline{P}_{ij}^{(\pm)} = \overline{P}_{ij}(\bfFMacro)_{\big|\FMacro_{kl} \rightarrow \FMacro_{kl} \pm \epsilon},
	\end{equation}
	where $P_{ij}^{(\pm)}$ are computed according to equation~\eqref{exact-stress} by using the input $\bfFMacro$ with $\FMacro_{kl}$ being replaced by $\FMacro_{kl} \pm \epsilon$. Exact quantities \eqref{exact-macro-energy-density}--\eqref{exact-tangent-moduli} will be used to compared with its counterparts obtained by using pure numerical solver FFT approach and neural network surrogate model. 
	
	\begin{figure}[htb]
		\centering{
			\begin{subfigure}{0.31\textwidth}
				\centering{ \includegraphics[width=1.05\textwidth]{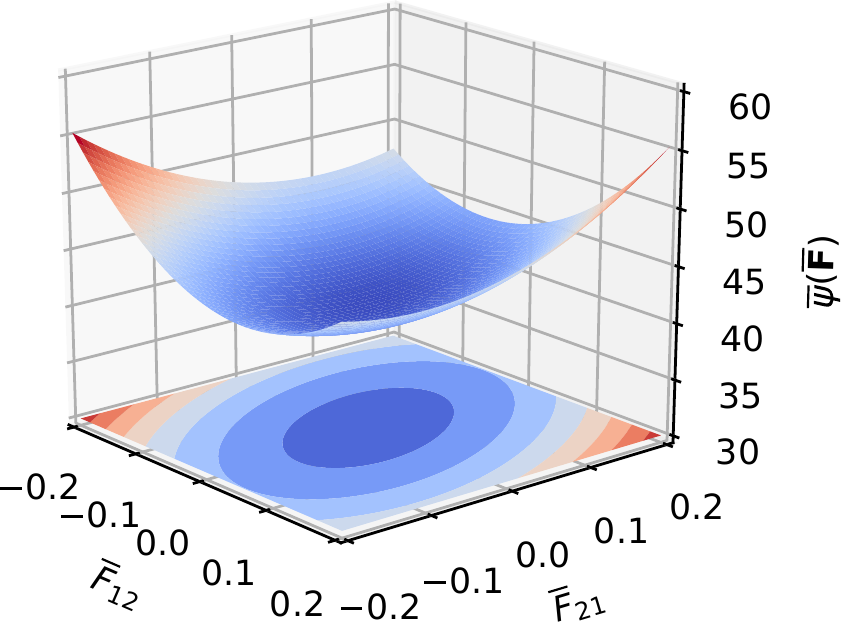} }
			\end{subfigure}\hspace{12pt}
			\begin{subfigure}{0.31\textwidth}
				\centering{ \includegraphics[width=1.05\textwidth]{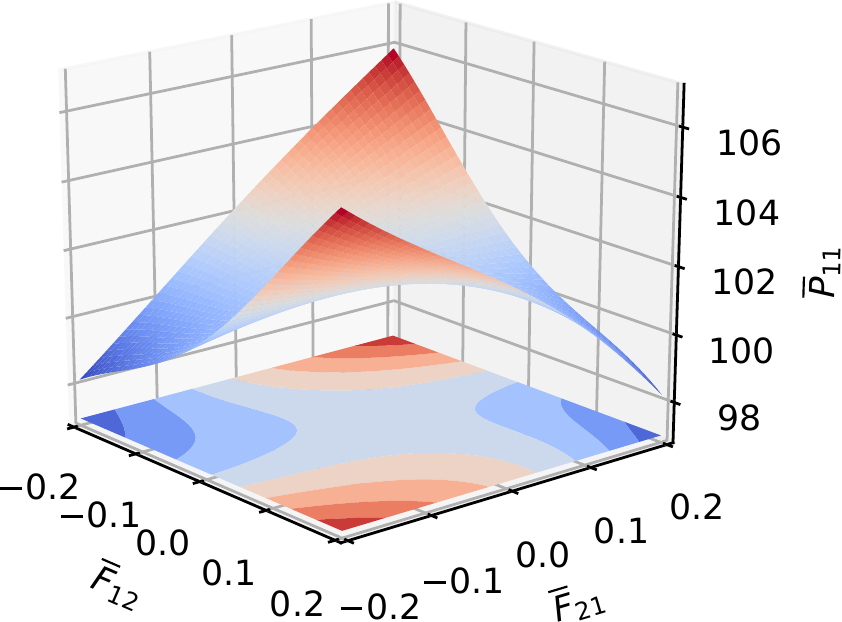} }
			\end{subfigure}\hspace{12pt}
			\begin{subfigure}{0.31\textwidth}
				\centering{ \includegraphics[width=1.05\textwidth]{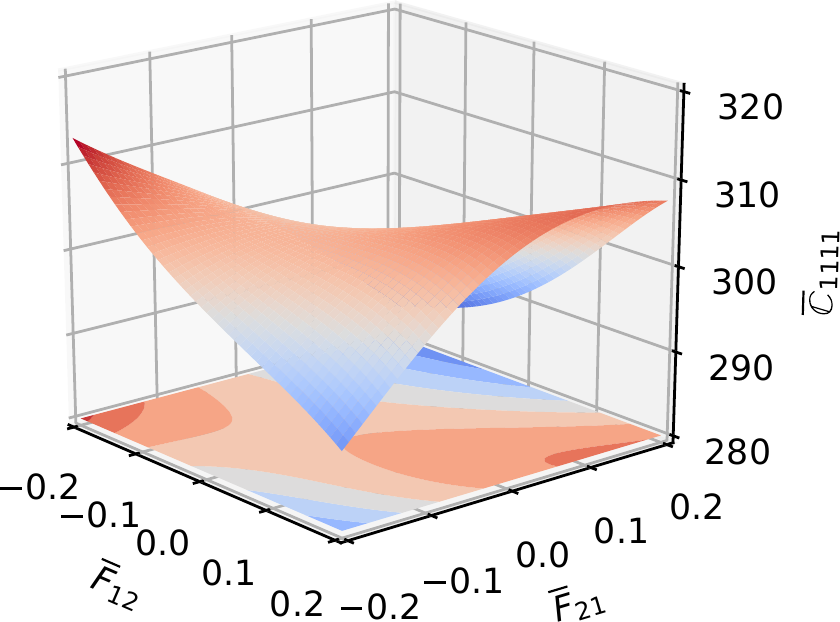} }
		\end{subfigure}}
		\centering{
			\begin{subfigure}{0.31\textwidth}
				\centering{ \includegraphics[width=1.05\textwidth]{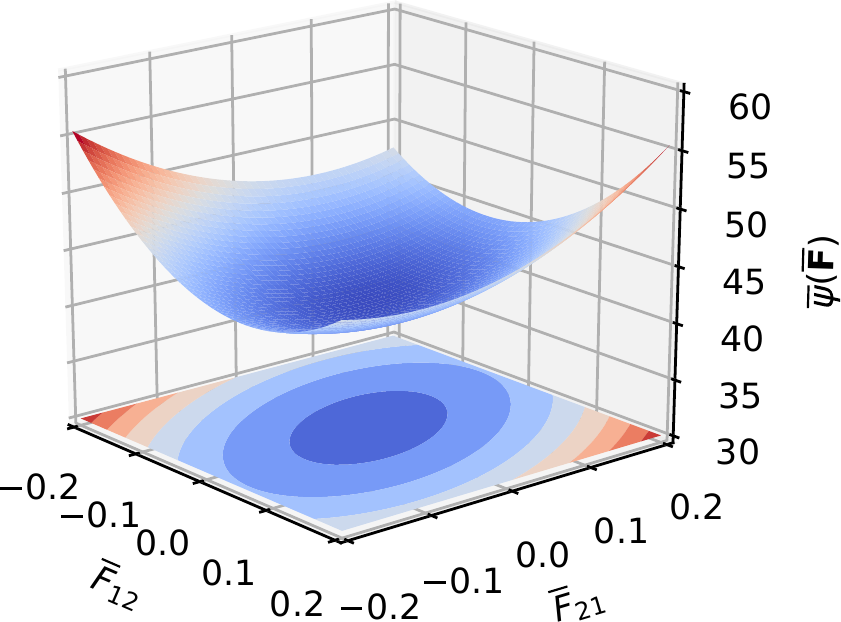} }
			\end{subfigure}\hspace{12pt}
			\begin{subfigure}{0.31\textwidth}
				\centering{ \includegraphics[width=1.05\textwidth]{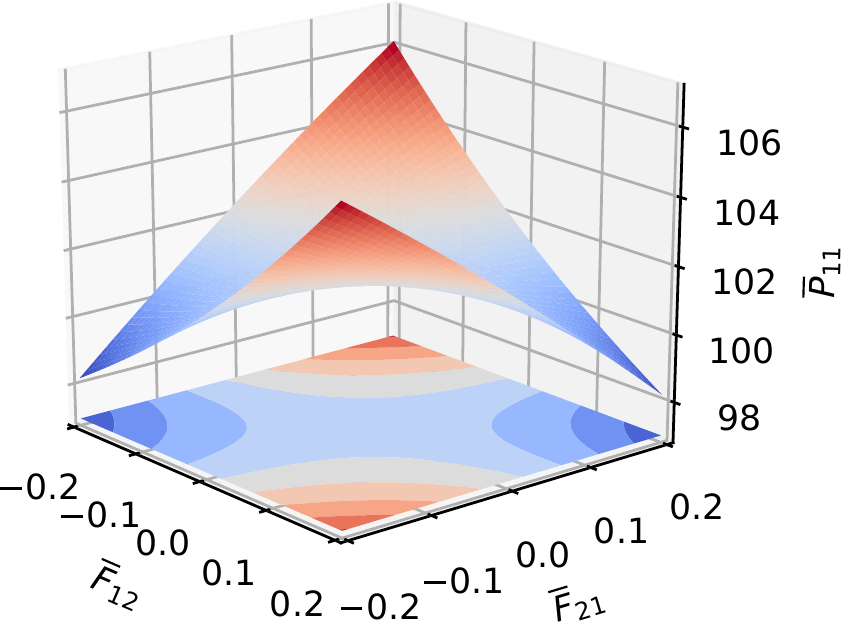} }
			\end{subfigure}\hspace{12pt}
			\begin{subfigure}{0.31\textwidth}
				\centering{ \includegraphics[width=1.05\textwidth]{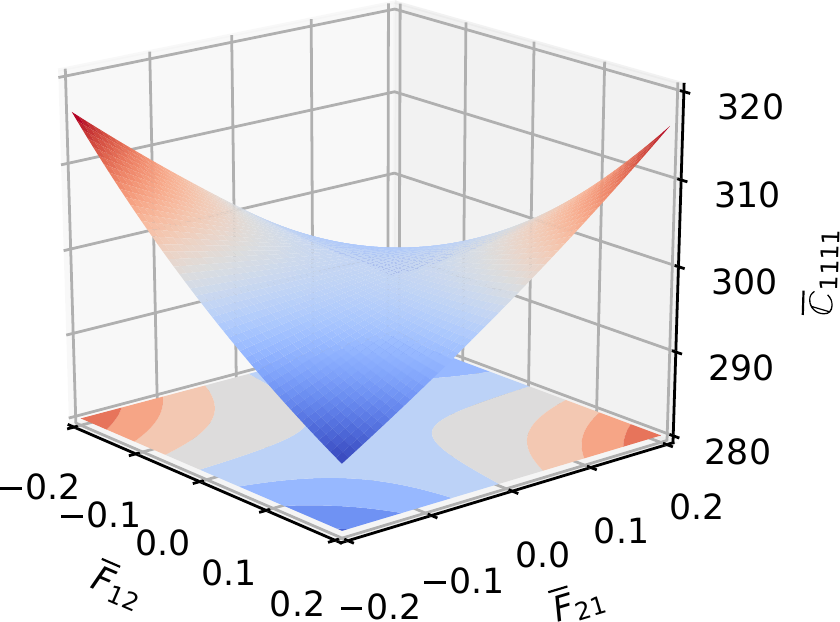} }
			\end{subfigure}
		}
		\centering{
			\begin{subfigure}{0.31\textwidth}
				\centering{ \includegraphics[width=1.05\textwidth]{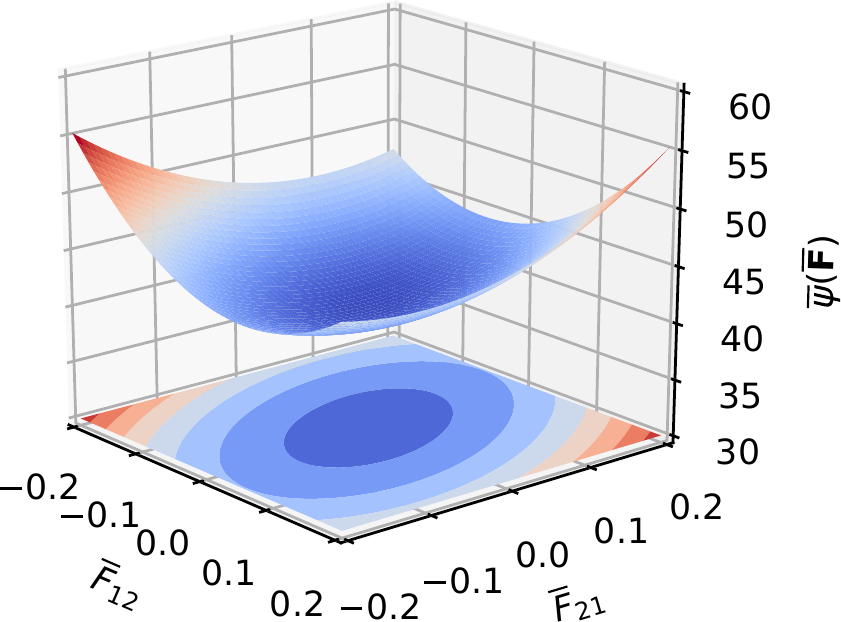} }
			\end{subfigure}\hspace{12pt}
			\begin{subfigure}{0.31\textwidth}
				\centering{ \includegraphics[width=1.05\textwidth]{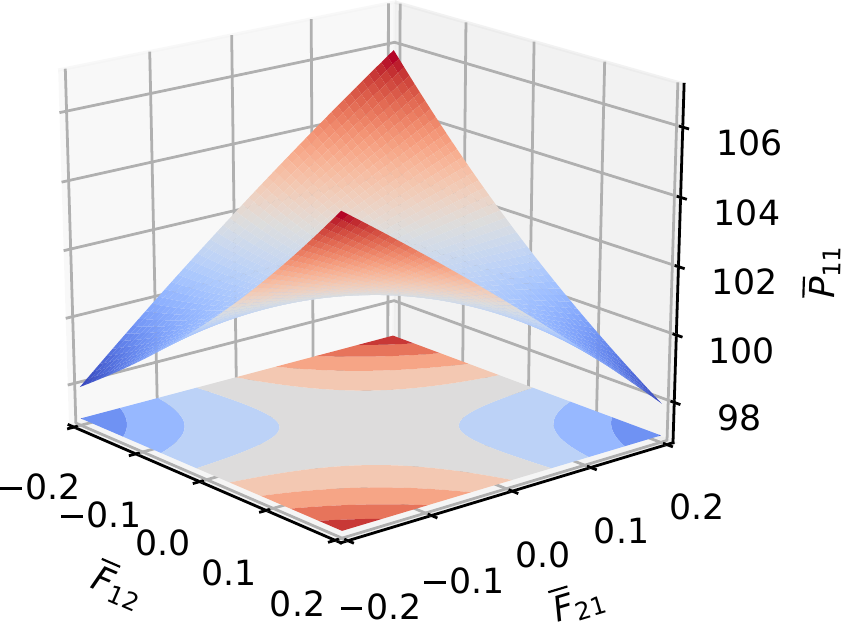} }
			\end{subfigure}\hspace{12pt}
			\begin{subfigure}{0.31\textwidth}
				\centering{ \includegraphics[width=1.05\textwidth]{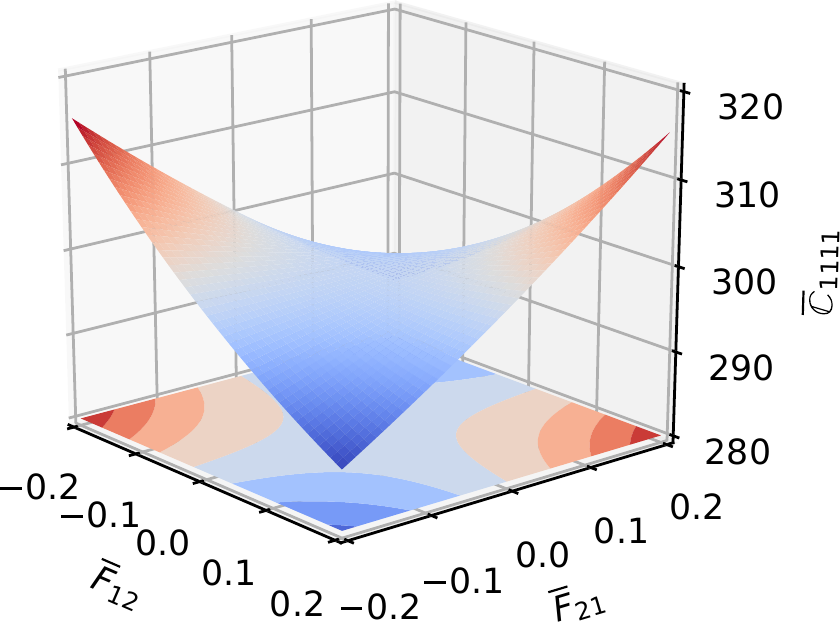} }
		\end{subfigure}}
		\caption{\textit{Comparison between surrogate-model solution, high-fidelity solution and analytical solution}. The macro-scopic energy density (\emph{left column}), the stress component $\overline{P}_{11}$ (\emph{middle column}) and the tangent moduli component $\overline{C}_{1111}$ (\emph{right column)} are ordered according to the following navigation: (\emph{top row}) surrogate-model solution, (\emph{middle row}) high-fidelity solution and (\emph{bottom row}) analytical solution. The plots are generated by fixing $\FMacro_{11} = \FMacro_{22} = 1.2$ and varying $\FMacro_{12}$, $\FMacro_{21}$ in the interval $[-0.2,0.2]$.}
		\label{Fig:Laminate-solution-comparison}
	\end{figure}
	\paragraph{\textbf{\sffamily Numerical results}} In Fig.~\ref{Fig:Laminate-solution-comparison}, we show the macro-energy density, the stress component $\overline{P}_{11}$ and the tangent moduli component $\overline{C}_{1111}$ which are in turn computed according to the solutions obtained respectively by the surrogate model, the high-fidelity solution directly obtained by FFT-based method and the analytical solution \eqref{Laminate-analytical-formula}. As for this comparison, we assemble $N_\text{data} = 50\times 10^{3}$ data points which are extracted from a database of $200\times 10^{3}$ data points $\overline{F}_{ij}$ uniformly distributed in the range
	\begin{equation*}
	\begin{bmatrix}
	\FMacro_{11} & \FMacro_{12} \\
	\FMacro_{21} & \FMacro_{22}
	\end{bmatrix} :: \begin{bmatrix}
	\phantom{-}0.7 \rightarrow 1.3 & -0.3\rightarrow 0.3 \\
	-0.3\rightarrow 0.3 & \phantom{-}0.7\rightarrow 1.3
	\end{bmatrix}.
	\end{equation*} We recall that macroscopic deformation gradients play the role of input data and the result macro-energy density the role of output data. In addition, the architecture for this training is given by $L = 15$ component functions and $N = 20$ as shown in Row~\textsf{Example 5.2} of Table~\ref{Table:Architecture-of-neural-networks}.
	
	\begin{figure}
		\centering
		\begin{subfigure}{0.47\textwidth}
			\centering
			\includegraphics[width=0.95\textwidth]{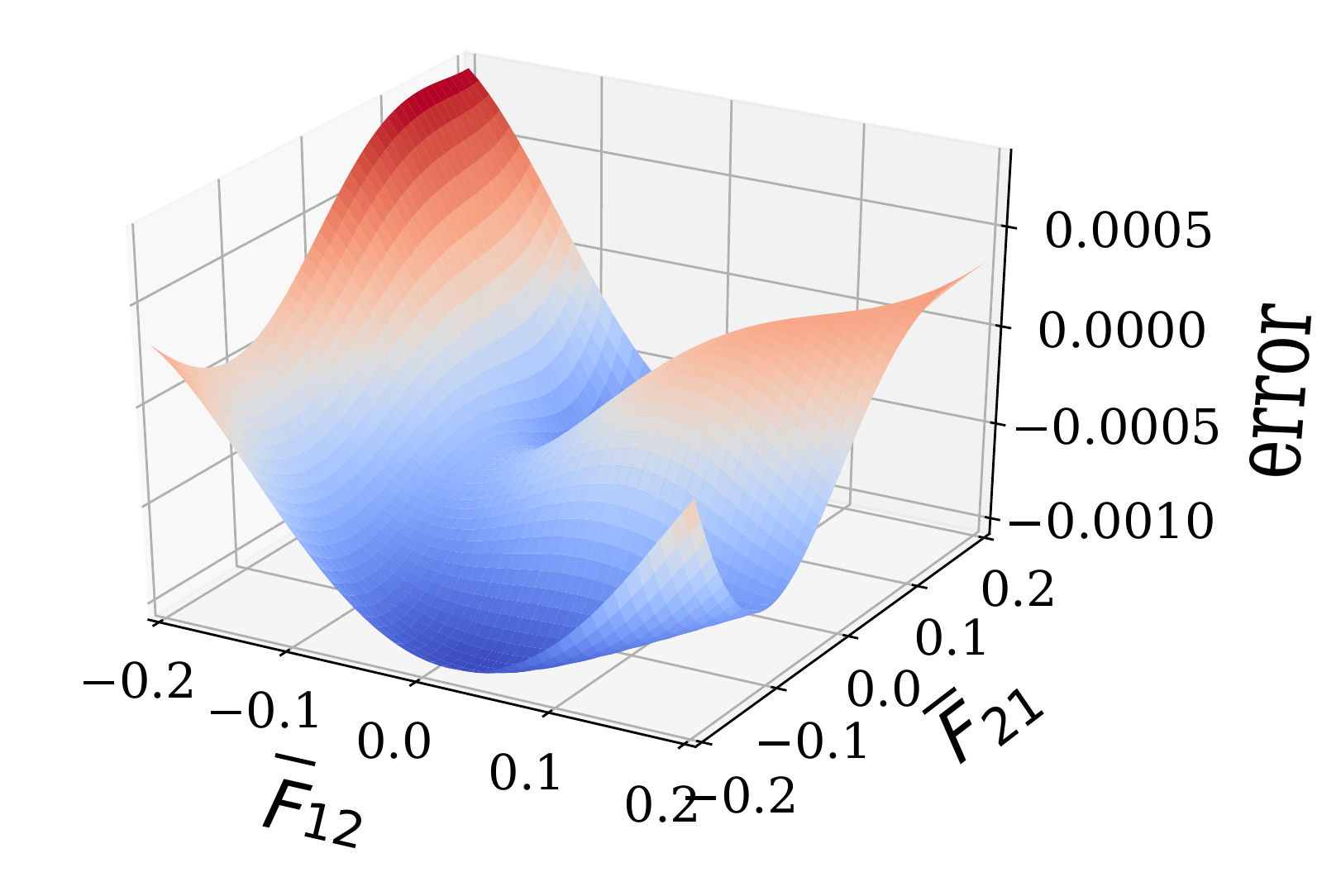}
		\end{subfigure}
		\begin{subfigure}{0.47\textwidth}
			\centering
			\includegraphics[width=0.95\textwidth]{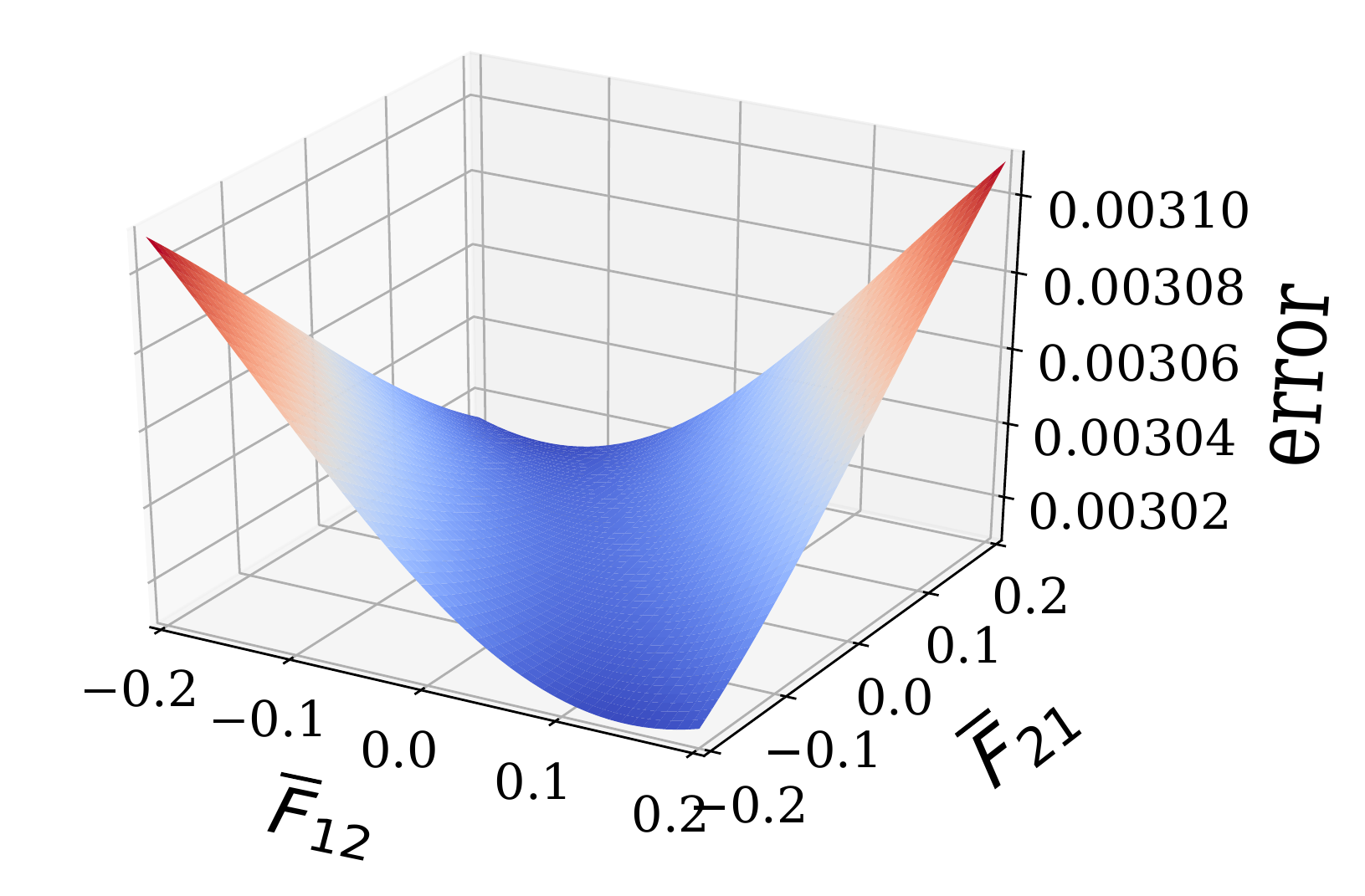}
		\end{subfigure}
		\caption{\textit{Relative error in macro-energy density.} The relative error in macro-energy density constructed by using the surrogate model (\emph{left}) and the FFT solution (\emph{right}) with respect to the exact one are computed by dividing the corresponding absolute error by the exact macro-energy density.}
		\label{Fig:macro-energy-density-comparison}
	\end{figure}
	Although we could notice the differences in stress and tangent moduli components, it is yet difficult to see any differences in the macro-energy density. Indeed, it can be observed from Fig.~\ref{Fig:macro-energy-density-comparison} that the relative error in energy provided by the surrogate-model solution and the FFT-based solution as compared to the exact solution are very small. We highlight here our argument regarding the obvious differences in the stress field and tangent moduli. At the first sight, it might lead to the impression that the method generates high approximation error in the solution. However, it is not entirely true that the final response of the exact mechanical response differs from the approximate response using the surrogate model because the minimum point of the approximate macro-energy density is close to exact one. This reasoning is reflected in the first column of Fig.~\ref{Fig:Laminate-solution-comparison} and also in subsequent numerical examples.
	
	\begin{figure}[htb]
		\centering
		\begin{subfigure}{0.49\textwidth}
			\centering
			\includegraphics[width=1.05\textwidth,angle=0]{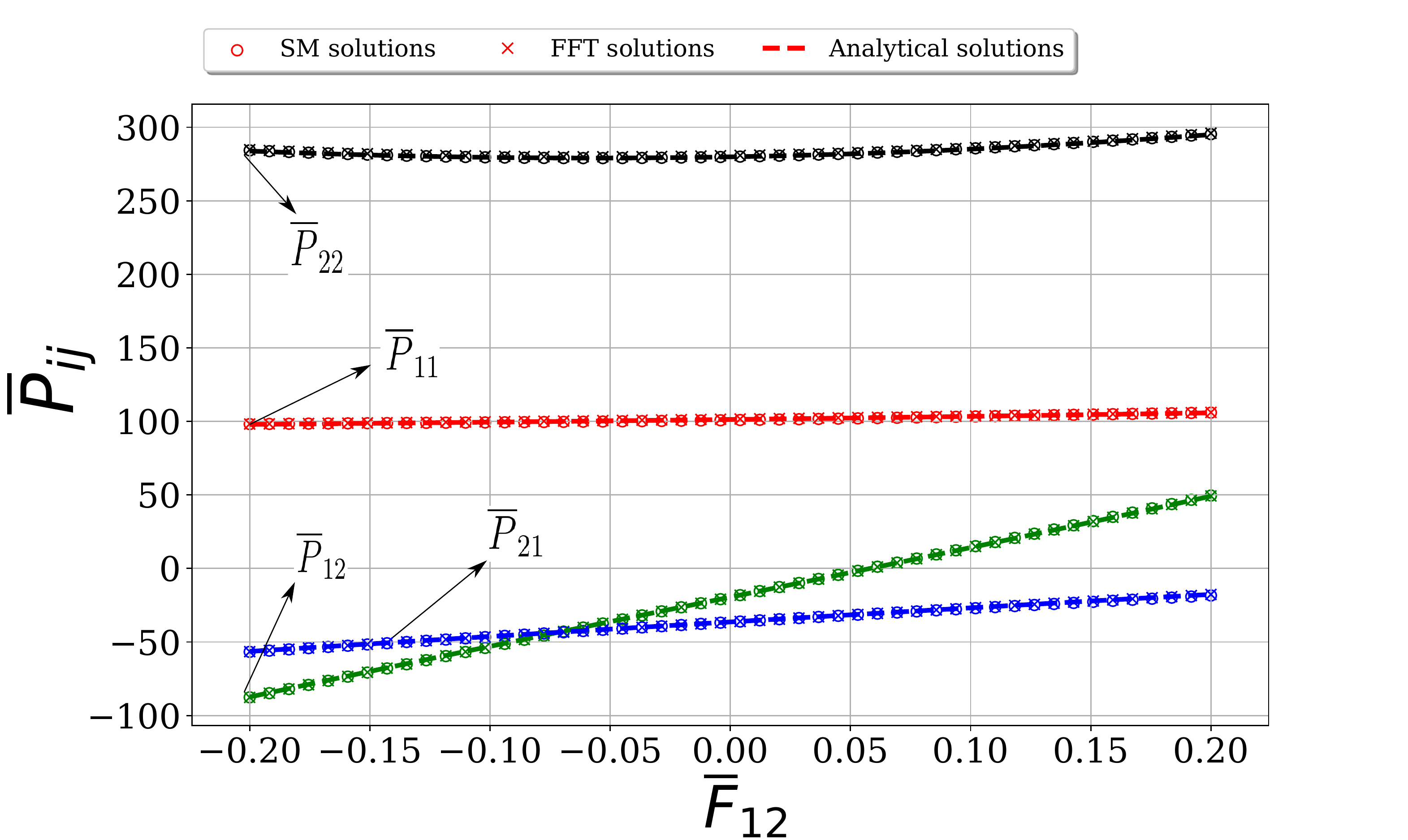}
		\end{subfigure}
		\begin{subfigure}{0.49\textwidth}
			\centering
			\includegraphics[width=1.05\textwidth,angle=0]{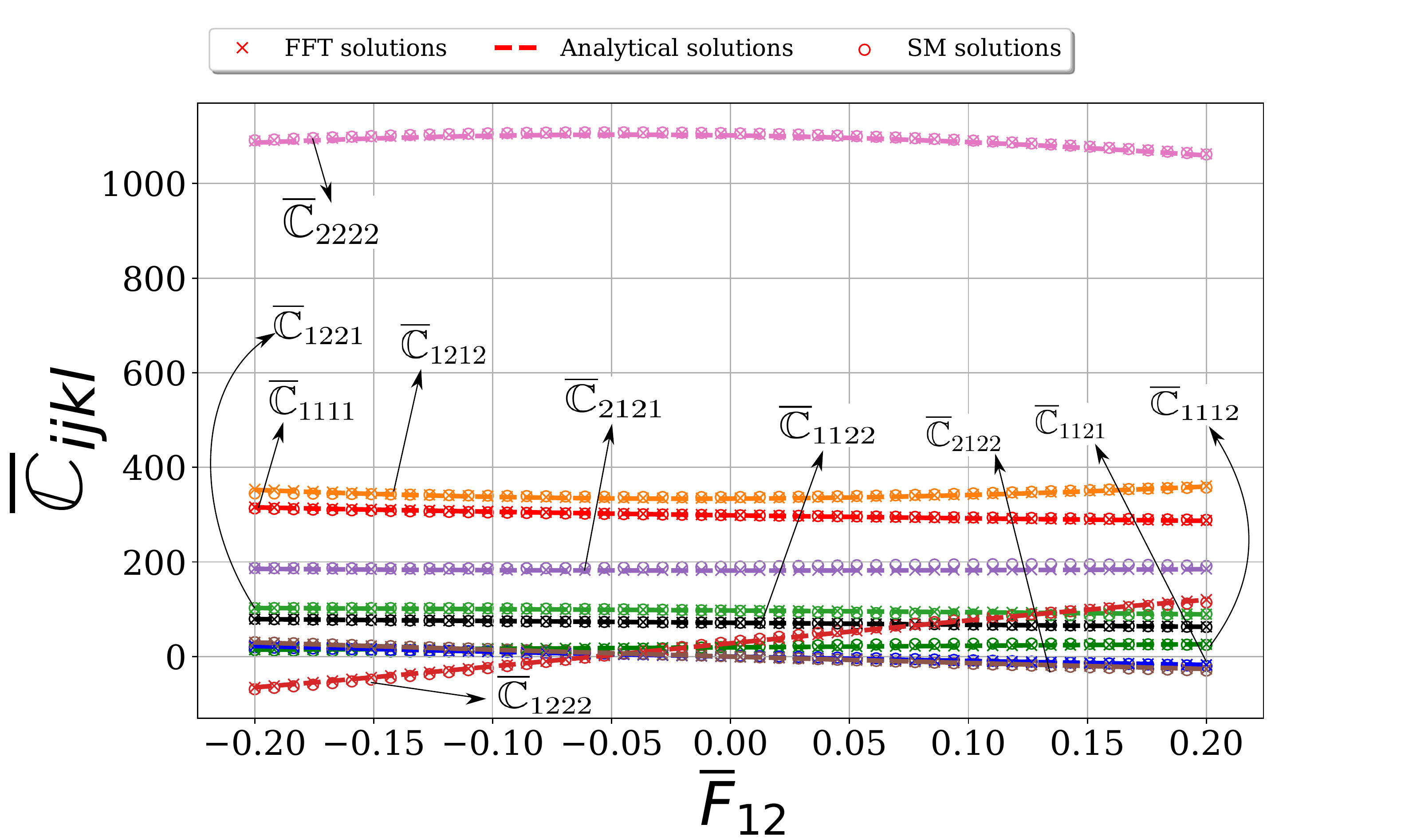}
		\end{subfigure}
		\caption{\textit{Macroscopic stress} $\bfPMacro$ \textit{and tangent moduli} $\bbCMacro$. The macroscopic stress (\emph{left}) and tangent moduli (\emph{right}) produced by surrogate model are computed by using $50 \times 10^3$ data points for training $\energyMacro^{NN}$ and these effective entities are compared with the numerical solutions (FFT solutions) and the analytical solutions. These tensor quantities are plotted against varying component $\FMacro_{12} \in [-0.2,0.2]$ while other three components $\FMacro_{11} = 1.2= \FMacro_{22} = 1.2$ and $\FMacro_{21} = -0.2$ are kept fixed.}
		\label{Fig:Stress+Tangent-Moduli-Comparison-Laminate}
	\end{figure}
	In order to earn more confidence in this computational framework, we conduct another set of numerical experiments. We study the components of macroscopic stress and tangent moduli
	by fixing the components $\FMacro_{11} = \FMacro_{22} = 1.2$, $\FMacro_{21} = -0.2$ letting $\FMacro_{12}$ vary arbitrarily in the range $[-0.2, 0.2]$. In Fig.~\ref{Fig:Stress+Tangent-Moduli-Comparison-Laminate}, the multiple components of $\bfPMacro^{\NN}$ and $\bbCMacro^{\NN}$ are presented by using $N_{\text{data}} = 50\times 10^3$ training data for construction of $\energyMacro^{\NN}$ (see Row \textsf{Example 5.4} of Table~\ref{Table:Architecture-of-neural-networks}). At the same, their counterparts obtained by using two-scale formulas \eqref{macroscopic-stress} and \eqref{macroscopic-tangent-moduli} so-called high-fidelity solution which only FFT is used to solve the BVP in microscale are shown in the same coordinate systems for comparison. The comparison shows excellent results obtained by not only the FFT-based solver for microscopic BVP but also the neural network approximation. Besides, it reveals that the surrogate model captures the expected anisotropy property of the homogenized material extremely well. This leads us to the next numerical study where analytical solution of the microscopic BVP is not available.
	
	\subsection{Surrogate model for a microstructure of circular inclusion}
	\label{Sec:RVE-with-circular-inclusion}
\begin{figure}[htb]
		\centering{
			\def\svgwidth{0.3\textwidth}
			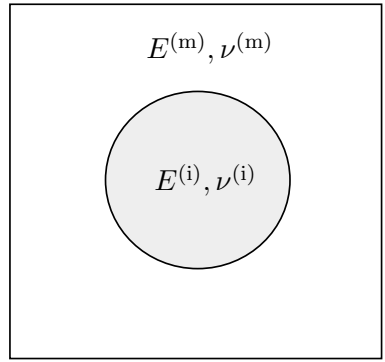 
		}
		\caption{\textit{AN RVE with circular inclusion.} The RVE consists of two phases: (i) circular inclusion and (m) the matrix material surrounding the inclusion. Two phases are of the Neo-Hookean materials that are characterized by the energy density \eqref{Neo-Hookean-materials} and Young modulus and Poisson ratio. The material parameters are specialized to each phase.}
		\label{Fig:RVE-with-circular-inclusion}
	\end{figure}

	\paragraph{\textbf{\sffamily Problem setting}} In this numerical experiment, we study a two-dimensional RVE with circular inclusion in a plane strain analysis. Such RVE is the two-dimensional reduction of the cylindrical inclusion of a three-dimensional RVE with quite large thickness in the direction of cylinder axis. We denote the quantities associated with the inclusion and matrix phases by the superscripts $(\text{i})$ and $(\text{m})$, respectively. The circular inclusion takes up a volume fraction $f^{(\text{i})} = 20\%$ and hence its radius is determined according to $\pi \big[R^{(\text{i})}\big]^2 = f^{(\text{i})} L_1 L_2$, where $L_1$ and $L_2$ are the sizes of the RVE (see Fig.~\ref{Fig:RVE-with-circular-inclusion}).
	
	The constituted phases are of the Neo-Hookean materials characterized by the energy density (cf. \textsc{Yvonnet et al.}~\cite{Yvonnet+Monteiro+He-2013})
	\begin{equation}\label{Neo-Hookean-materials}
	\psi(\bfC) = \frac{1}{2}\lambda \big[\log (J)\big]^2 - \mu \log(J) + \frac{1}{2}\mu\big[\mathrm{trace}(\bfC) - 2\big],
	\end{equation} 
	where $\bfC = \bfF^{T}\bigcdot\bfF$ is the right Cauchy stress tensor, $J = \det(\bfF)$ $\lambda$ and $\mu$ are the Lame parameters given in terms of Young modulus $E$ and Poisson ratio as follows
	\begin{equation*}
	\lambda = \frac{E \nu}{(1+\nu)(1-2\nu)}, \quad \mu = \frac{E}{2(1+\nu)}.
	\end{equation*}
	As for our example, we choose the following material parameters
	\begin{equation}\label{matrix-inclusion-material-parameters}
	E^{(\text{m})} = 100~\text{MPa}, \quad \nu^{(\text{m})} = 0.4, \quad E^{(\text{i})} = 1000~\text{MPa}, \quad  \nu^{(\text{i})} = 0.3.
	\end{equation}
	The stress tensor $\bfP = \partial\psi/\partial\bfF$ and tangent moduli $\bbC = \partial^2\psi/\partial\bfF \partial\bfF$ are derived according to
	\begin{equation*}
	\begin{aligned}
	P_{ij} &= \mu F_{ij} + \big[\lambda \log(J) - \mu\big]F_{ji}^{-1}, \\
	C_{ijkl} &= \lambda F_{ji}^{-1} F_{lk}^{-1} - \big[\lambda\log(J) - \mu\big] F_{jk}^{-1} F_{li}^{-1} + \mu \delta_{ik} \delta_{jl}.
	\end{aligned}
	\end{equation*}
	
	\begin{figure}[htb]
		\centering{
			\begin{subfigure}{0.31\textwidth}
				\centering{ \includegraphics[width=1.05\textwidth]{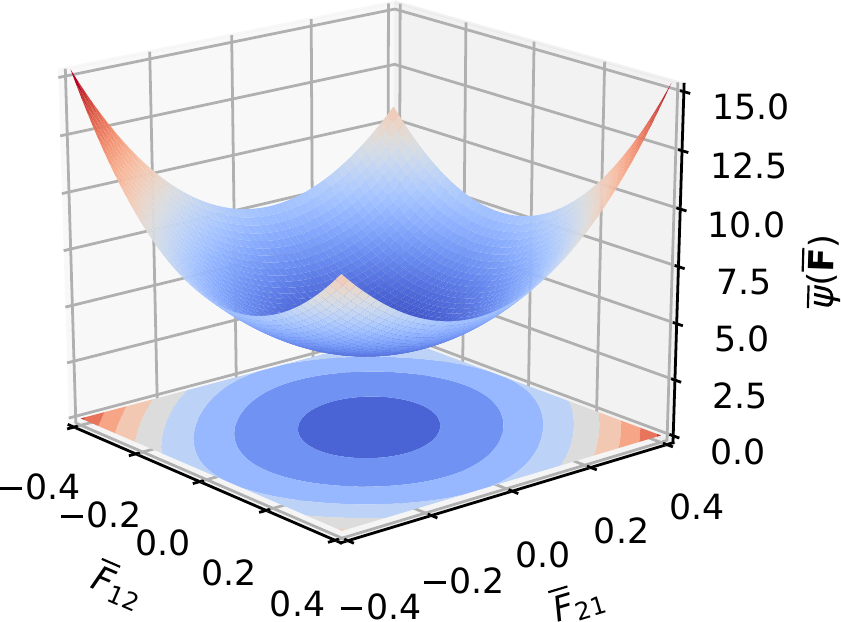} }
			\end{subfigure}\hspace{12pt}
			\begin{subfigure}{0.31\textwidth}
				\centering{ \includegraphics[width=1.05\textwidth]{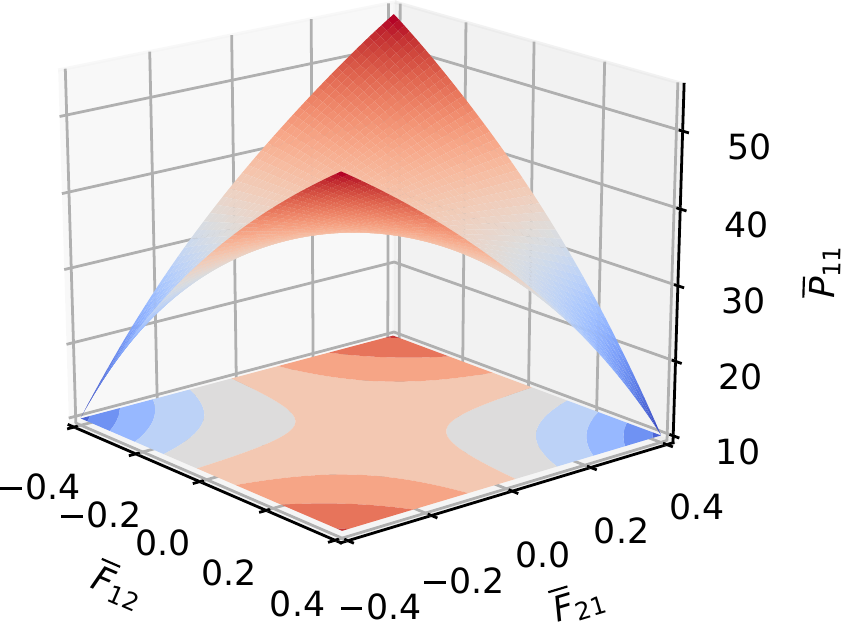} }
			\end{subfigure}\hspace{12pt}
			\begin{subfigure}{0.31\textwidth}
				\centering{ \includegraphics[width=1.05\textwidth]{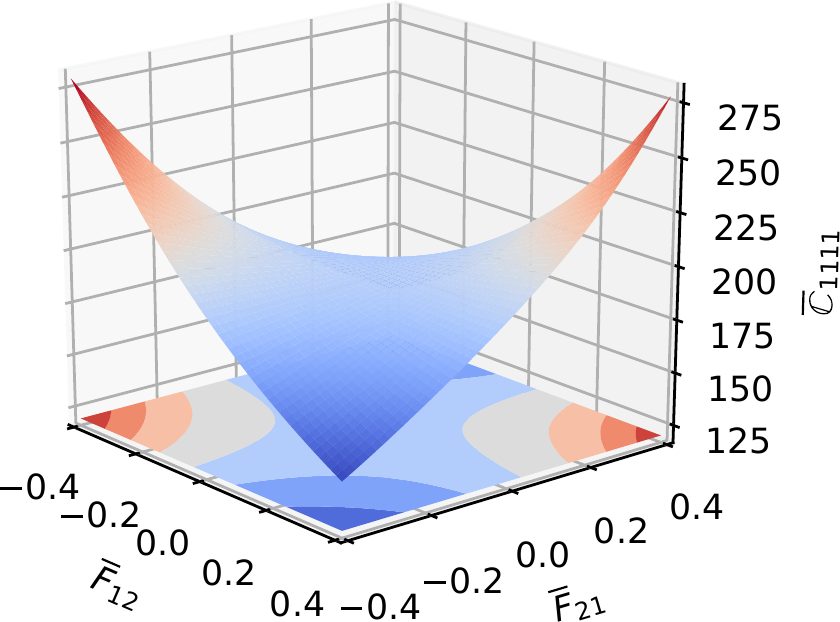} }
			\end{subfigure}
		}
		
		\centering{
			\begin{subfigure}{0.31\textwidth}
				\centering{ \includegraphics[width=1.05\textwidth]{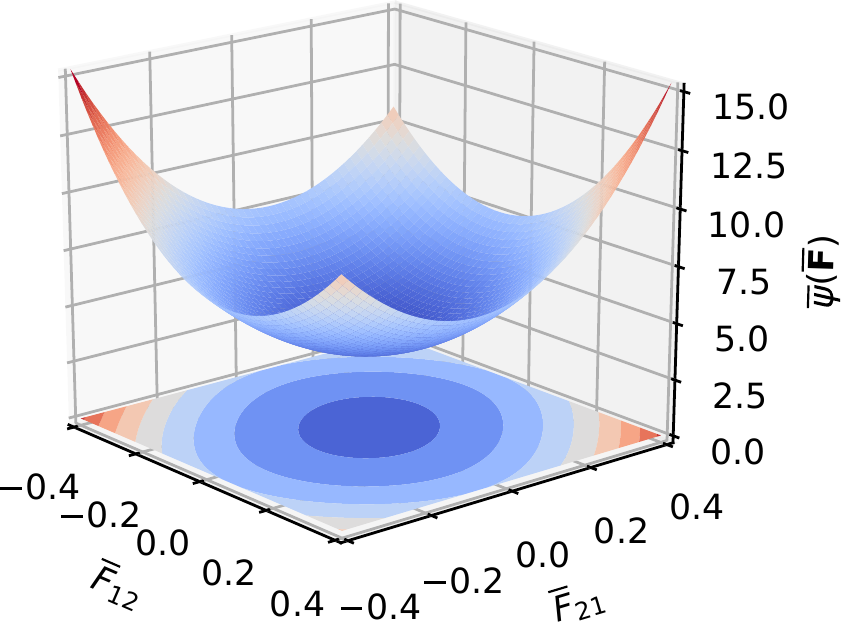} }
			\end{subfigure}\hspace{12pt}
			\begin{subfigure}{0.31\textwidth}
				\centering{ \includegraphics[width=1.05\textwidth]{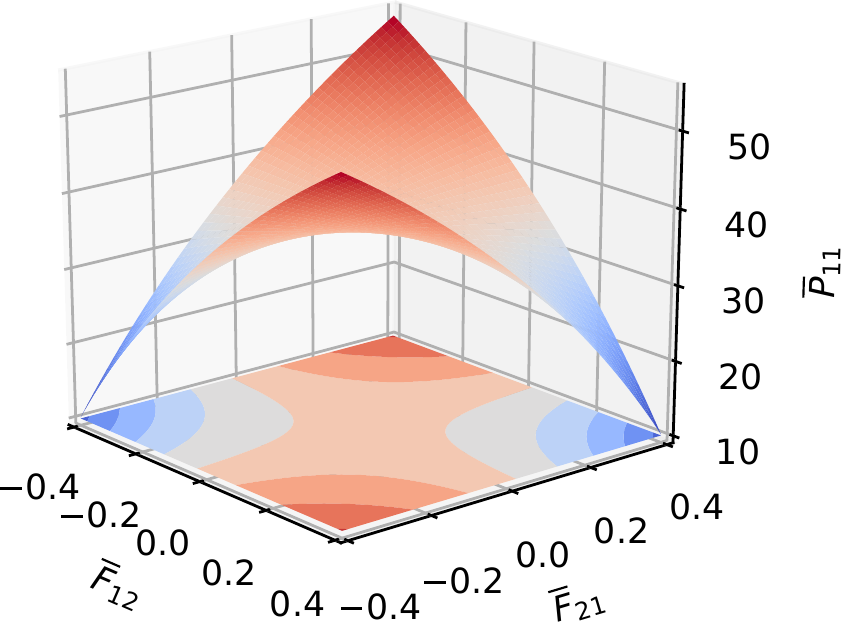} }
			\end{subfigure}\hspace{12pt}
			\begin{subfigure}{0.31\textwidth}
				\centering{ \includegraphics[width=1.05\textwidth]{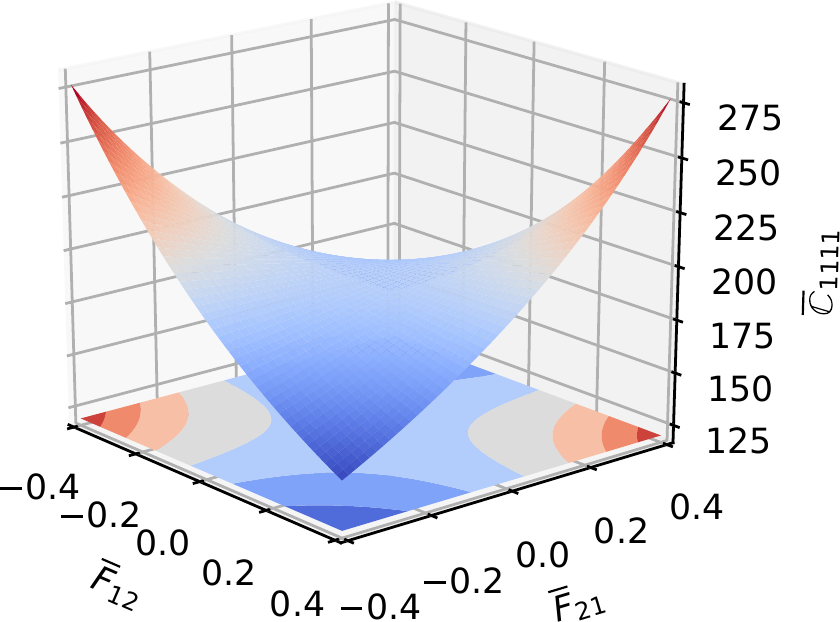} }
			\end{subfigure}
		}
		\caption{\textit{Comparison between surrogate-model computation, FFT solution.} The macro-energy density (\emph{left column}), the stress component $\overline{P}_{11}$ (\emph{middle column}) and the tangent moduli component $\overline{C}_{1111}$ (\emph{right column}) are ordered according to the following navigation: (\emph{top row}) surrogate-model solution, (\emph{bottom row}) high-fidelity solution.}
		\label{Fig:Stress+Tangent-Moduli-Comparison-Circular1}
	\end{figure}
	
	\paragraph{\textbf{\sffamily Numerical results}} In this example, we generate the database of $200\times 10^{3}$ uniformly distributed data points in the following range 
	\begin{equation*}
	\begin{bmatrix}
	\FMacro_{11} & \FMacro_{12} \\
	\FMacro_{21} & \FMacro_{22}
	\end{bmatrix} :: \begin{bmatrix}
	\phantom{-}0.8 \rightarrow 1.2 & -0.5 \rightarrow 0.5 \\
	-0.5 \rightarrow 0.5 & \phantom{-}0.8 \rightarrow 1.2
	\end{bmatrix}
	\end{equation*}
	and use $30\times 10^{3}$ data points out of this database for training process. Since the analytical solution is not available for this RVE problem, we show in Fig.~\ref{Fig:Stress+Tangent-Moduli-Comparison-Circular1} only the numerical results of the macro-energy density $\energyMacro$, stress field $\overline{P}_{11}$, and tangent modulus $\overline{C}_{1111}$ obtained the by both the direct FFT-based and surrogate-model computations. Once again, we have fixed $\FMacro_{11} = \FMacro_{22} = 1.1$ to plot these quantities against the coordinates $\FMacro_{12}, \FMacro_{21} \in [-0.4,0.4]$. It is seen that the neural network function has performed such a good approximation that we hardly see the difference in the macro-quantities produced by surrogate model and two-scale approaches.
	
	\begin{figure}[htb]
		\centering
		\begin{subfigure}{0.49\textwidth}
			\centering
			\includegraphics[width=1.05\textwidth]{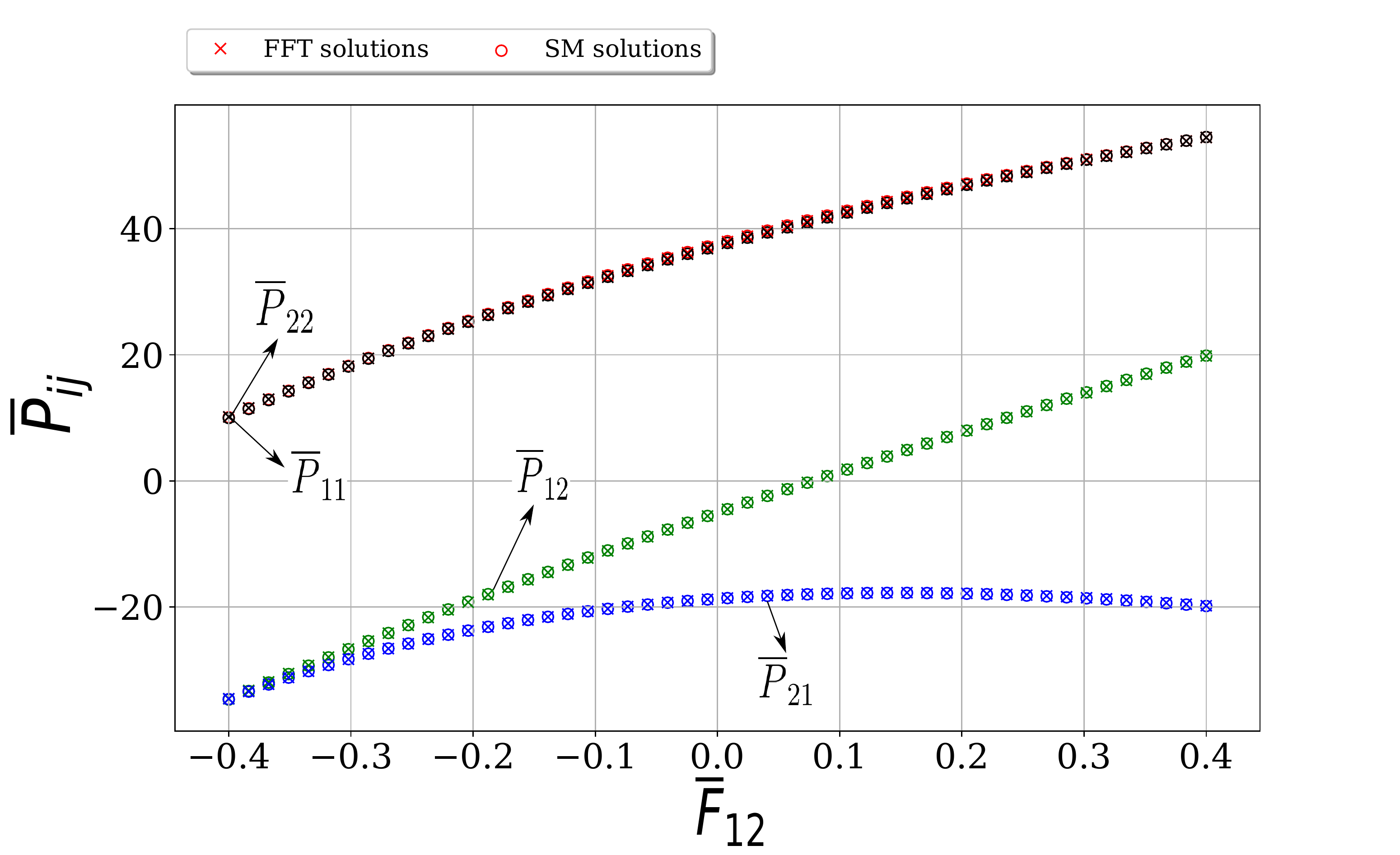}
		\end{subfigure}
		\begin{subfigure}{0.49\textwidth}
			\centering
			\includegraphics[width=1.05\textwidth]{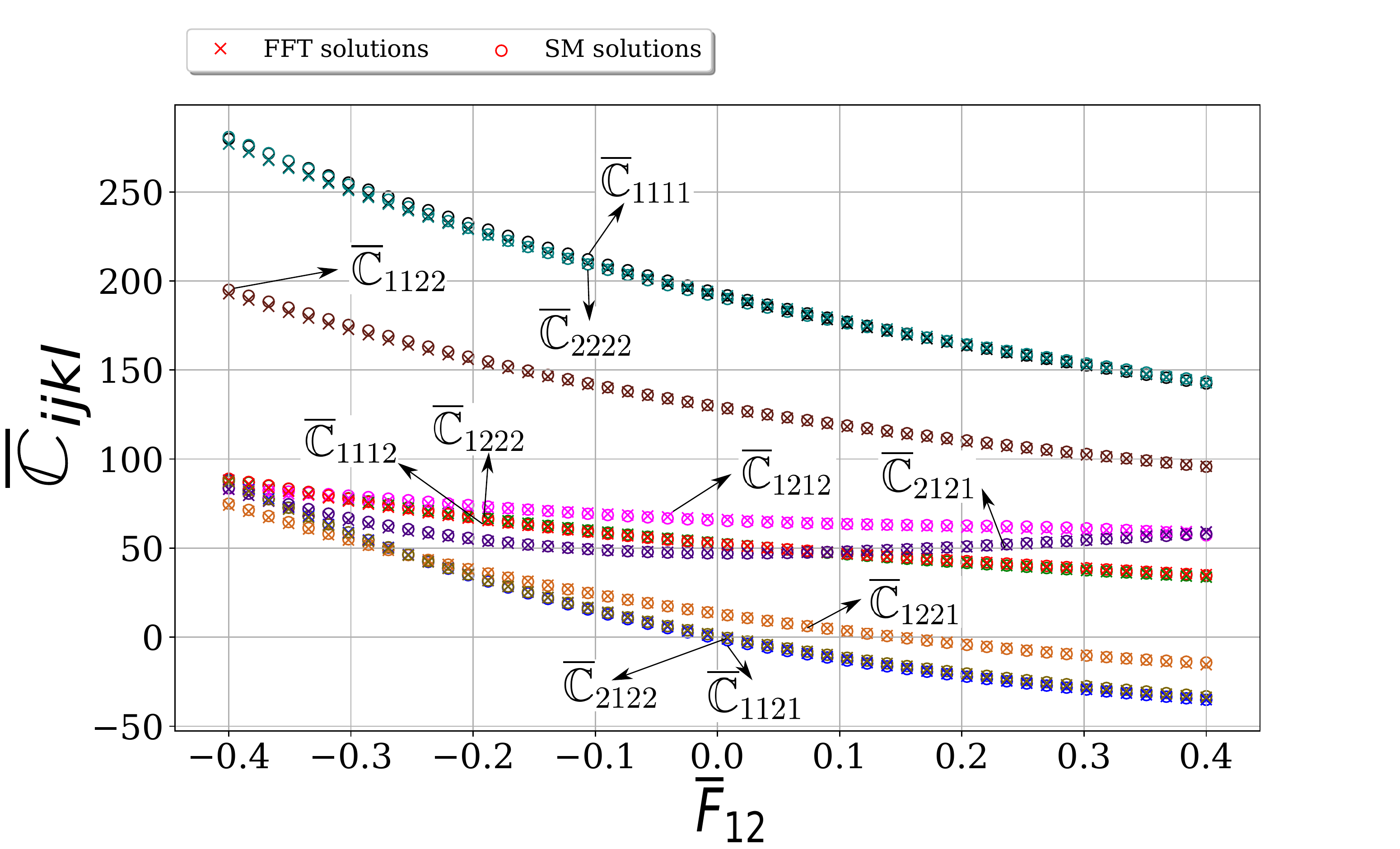}
		\end{subfigure}
		\caption{\textit{Macroscopic stress $\bfPMacro$ and tangent moduli $\bbCMacro$.}}
		\label{Fig:Stress+Tangent-Moduli-Comparison-Circular}
	\end{figure}
	In parallel with the last subsection, we present in Fig.~\ref{Fig:Stress+Tangent-Moduli-Comparison-Circular} the stress tensor $\bfPMacro$ and tangent moduli $\bbCMacro$ for comparison between surrogate-model and FFT solutions. In this numerical experiment, we set $\FMacro_{11} = \FMacro_{22} = 1.1$, $\FMacro_{21} = -0.4$ and let $\FMacro_{12}$ vary in the range $[-0.4,0.4]$.  As expected, the isotropicity of the homogenized material is well-captured in the surrogate model and reflected via the computation of all components $\overline{C}_{ijkl}$.
	
	In Examples 5.4 and 5.5 we examine the two-scale problems with circular-inclusion microstructure depicted in Example 5.3. Thus, we can reuse the trained model in this subsection and save tremendous computation effort paid for both building databases and training the networks. As a consequence, the algorithmic parameters characterizing the neural network architecture in the next two subsections can be consulted from Row \text{Example 5.3} of Table~\ref{Table:Architecture-of-neural-networks}. The great advantage observed herein is that the existing knowledge can be exploited while the improvement of accuracy in the approximate macro-energy density is still possible. 
	
	\subsection{Cook's membrane problem}
	\begin{figure}[H]
		\centering{
			\def\svgwidth{0.65\textwidth}
			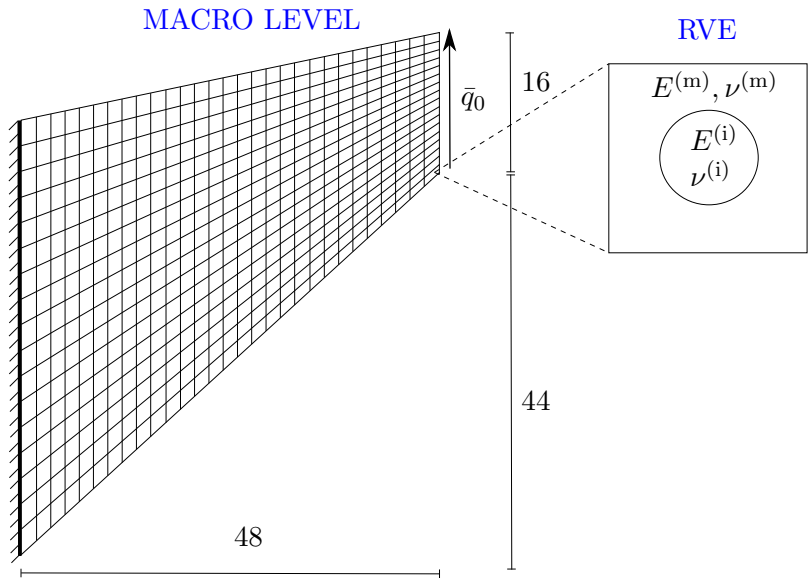
		}
		\caption{\textit{Cook's membrane problem}. The Cook's membrane is described by a two-dimensional trapezoid, clamped along the long left edge and subject to tangential traction $\bar{q}_{0}=4$ along the right edge. The structure is constituted by the materials with microstructure of circular inclusion. The problem is analyzed in the plane strain condition.}
		\label{Fig:Cook-Membrane-problem-setting}
	\end{figure}
	\paragraph{\textbf{\sffamily Problem setting}} This is the first numerical example to show the robustness of the proposed surrogate model using neural networks to approximate the macro-energy density. Particularly, we shall compare the mechanical responses at the macrosale computed by following the surrogate model and two-scale computations. We investigate the well-known Cook's membrane problem whose microstructure is represented by the circular inclusion in Subsection~\ref{Sec:Laminate-RVE-Problem}. This problem is named after the author \textsc{Cook}~\cite{Cook-1974} who first reported it. The geometry of the membrane consists of a trapezoid surface in the $X_1$-$X_2$ plane (see Fig.~\ref{Fig:Cook-Membrane-problem-setting}). The structure is clamped along the left edge and it is loaded by an edge traction load $\bar{q}_{0}=4$ along the right edge in the $X_2$-direction. It is rather thick so that this problem is set in the plane strain condition (see also \textsc{Section 2.1.5, Abaqus Benchmarks Guide}~\cite{Abaqus-simulia}). Note that we do not aim at reproducing the Benchmark results in the aforementioned literature.
	
	\begin{figure}[htb]
		\begin{subfigure}{0.47\textwidth}
			\includegraphics[width=0.9\textwidth]{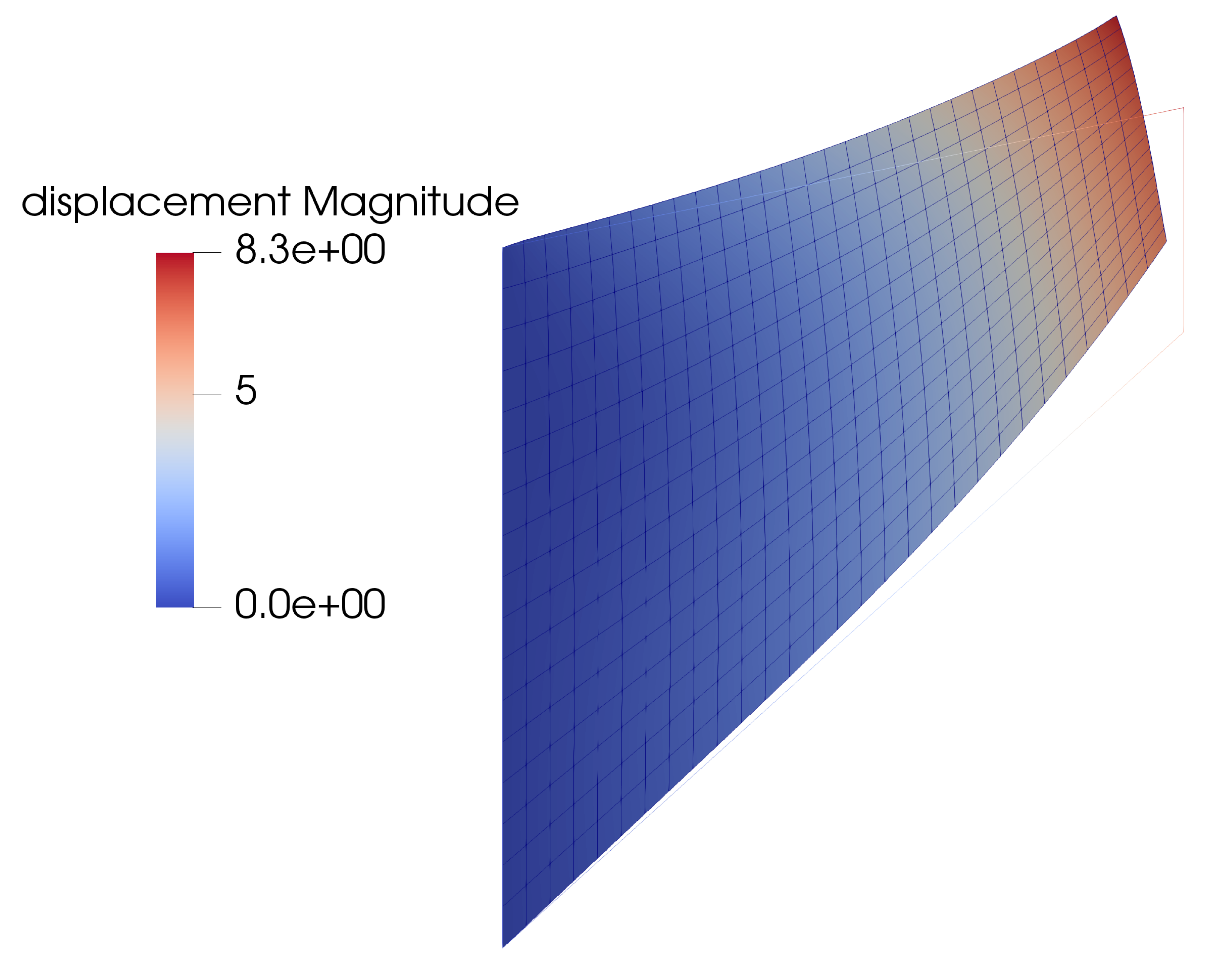}
		\end{subfigure}
		\begin{subfigure}{0.47\textwidth}
			\includegraphics[width=0.9\textwidth]{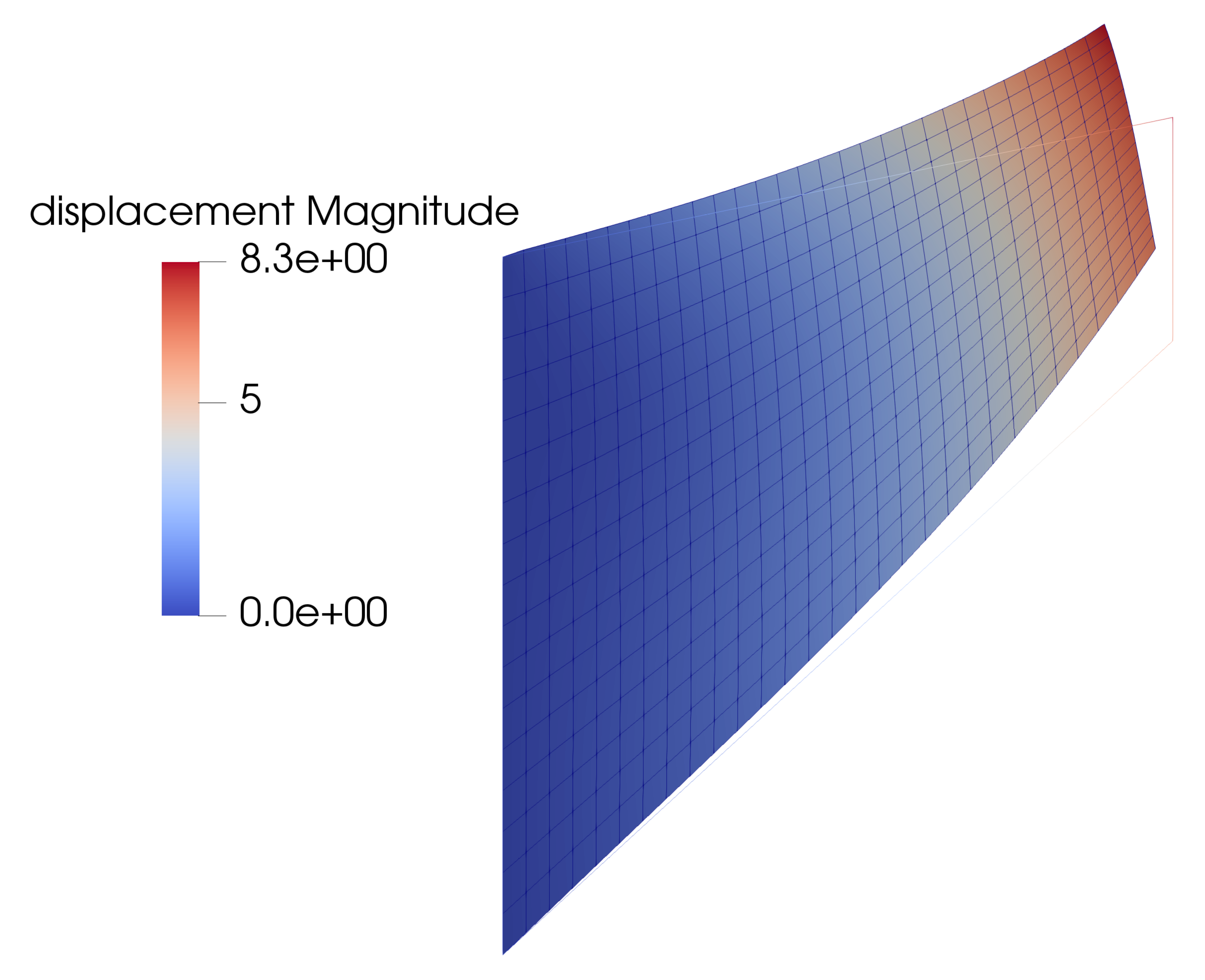}
		\end{subfigure}
		\caption{\textit{Mechanical responses of the Cook's membrane under plane strain condition.} The contour plot shows the vertical displacements computed at the macroscale according to the surrogate model (\emph{left}) and to the two-scale approach (\emph{right}).}
		\label{Fig:Displacement-comparison-Cook-membrane}
	\end{figure}
\begin{figure}[H]
		\centering
		\begin{subfigure}{0.47\textwidth}
			\centering
			\includegraphics[width=1.0\textwidth]{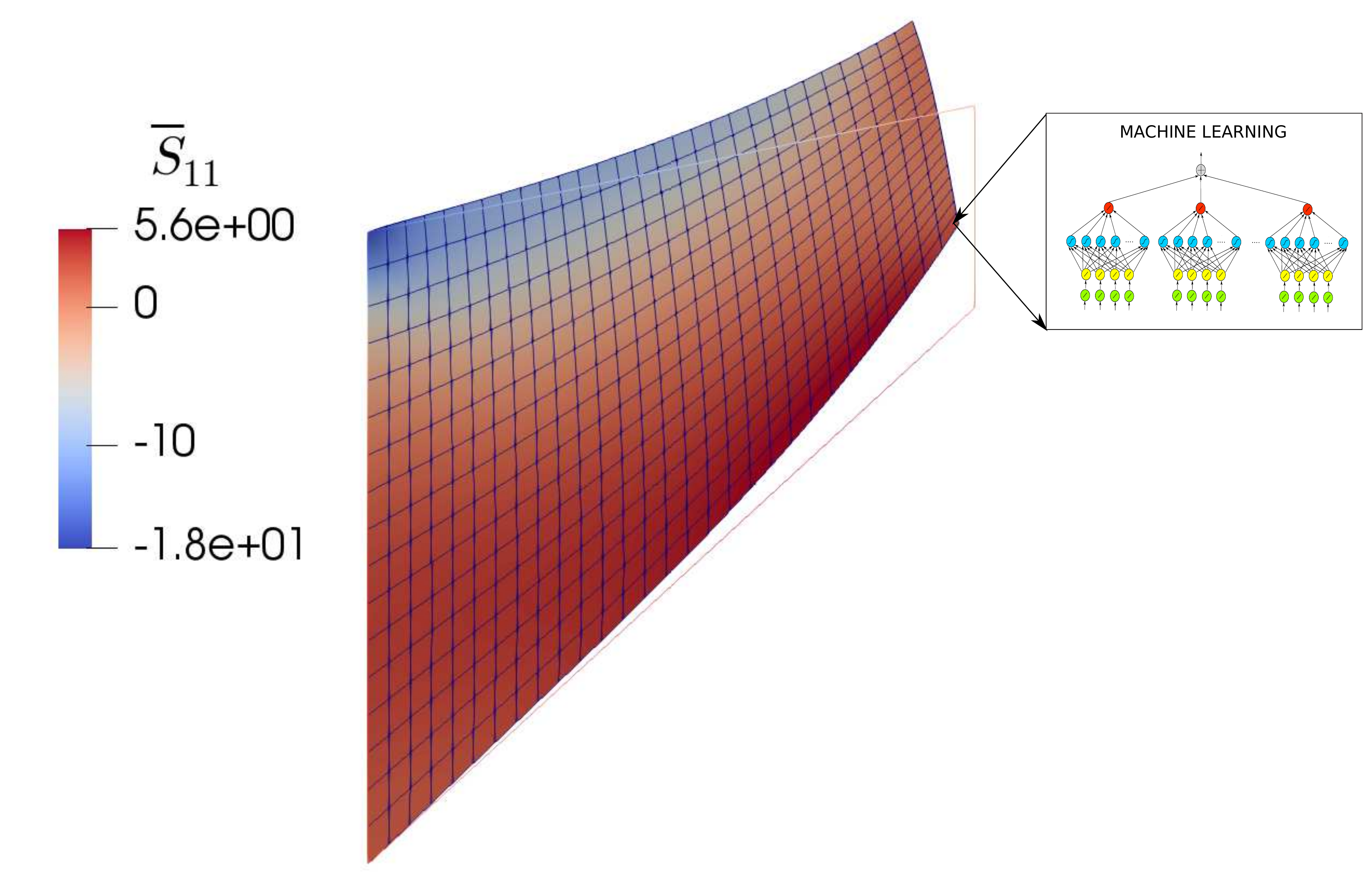}
		\end{subfigure}
		\begin{subfigure}{0.47\textwidth}
			\centering
			\includegraphics[width=1.0\textwidth]{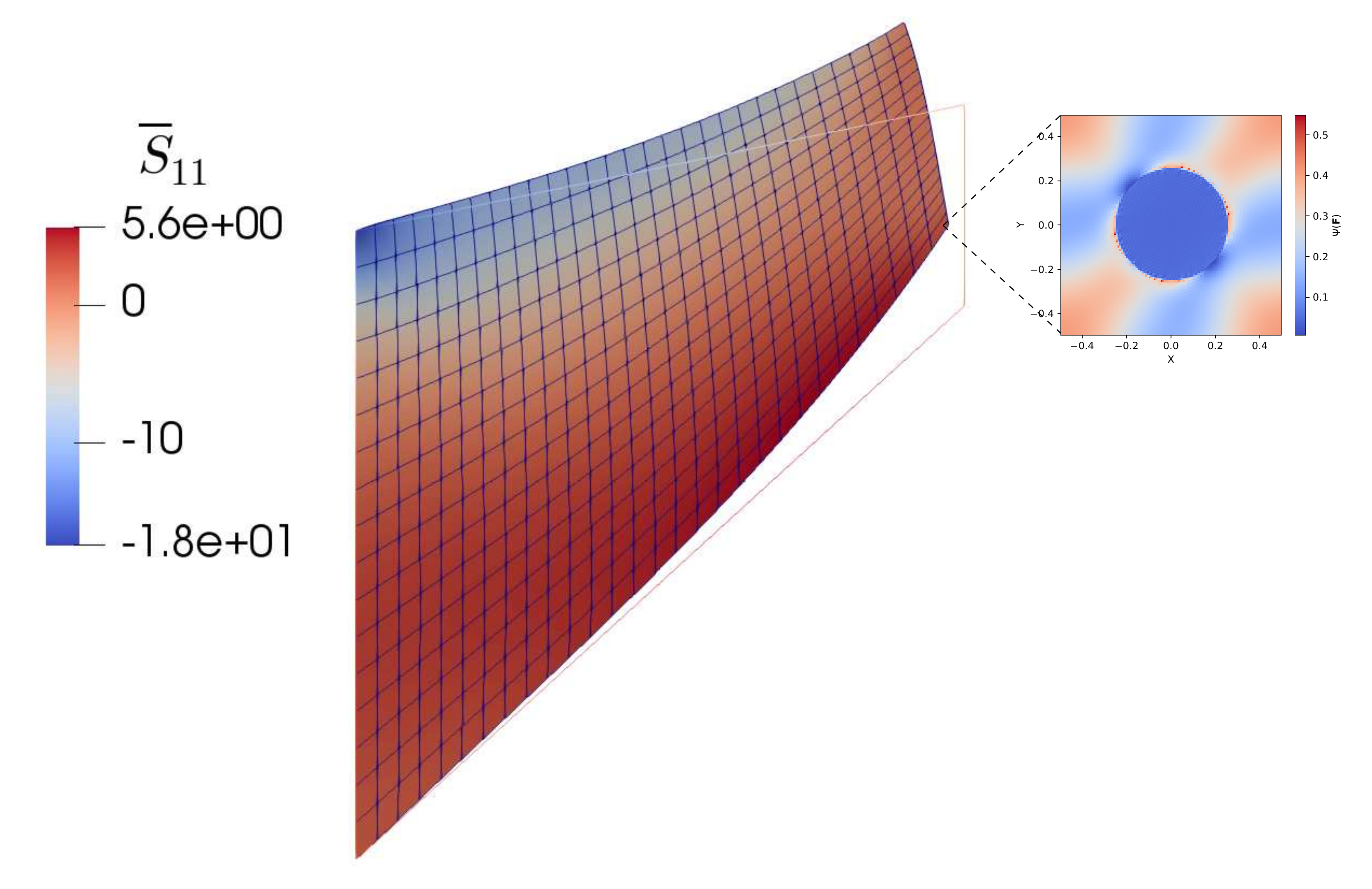}
		\end{subfigure}
		\begin{subfigure}{0.47\textwidth}
			\centering
			\includegraphics[width=1.0\textwidth]{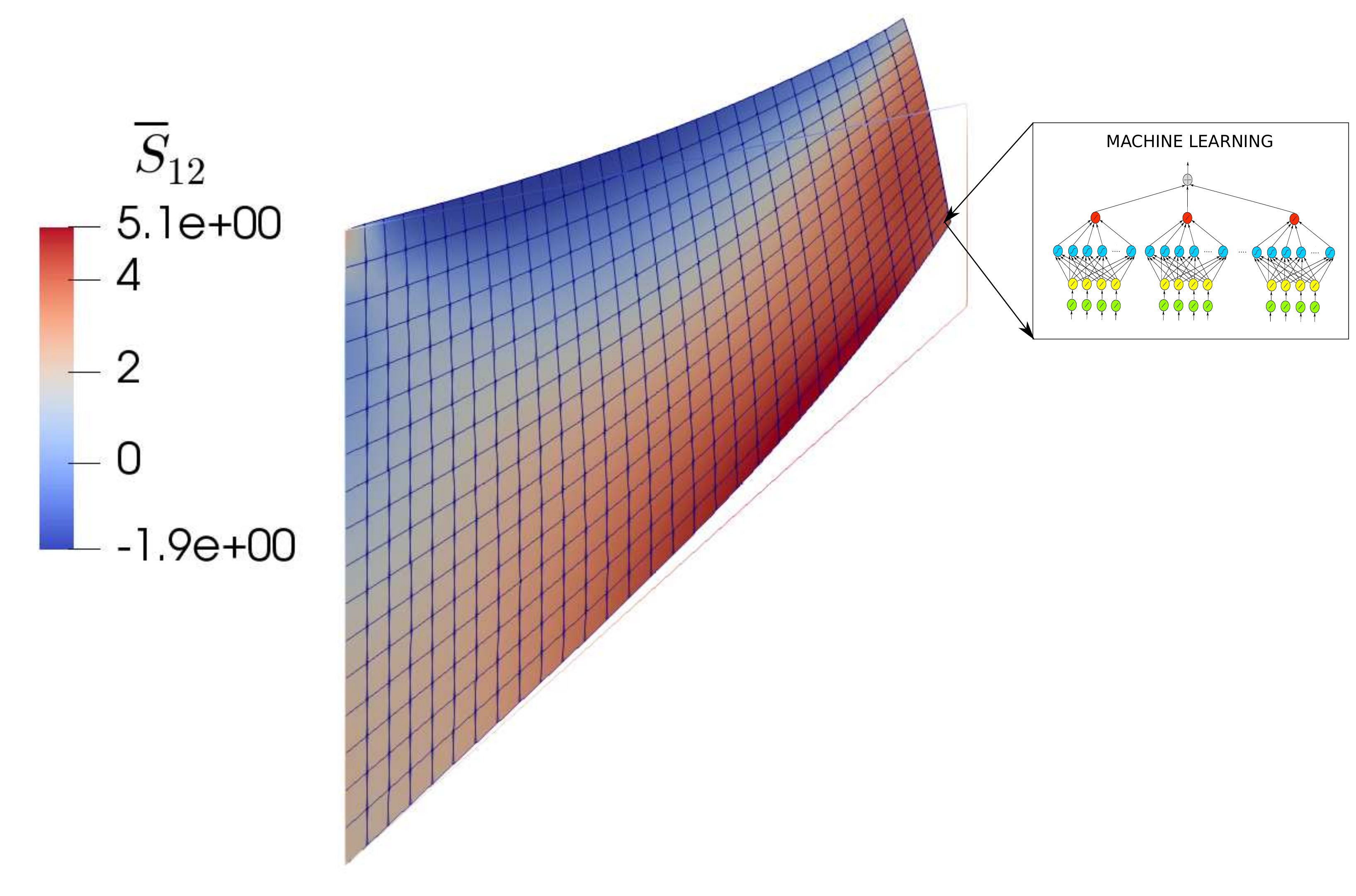}
		\end{subfigure}
		\begin{subfigure}{0.47\textwidth}
			\centering
			\includegraphics[width=1.0\textwidth]{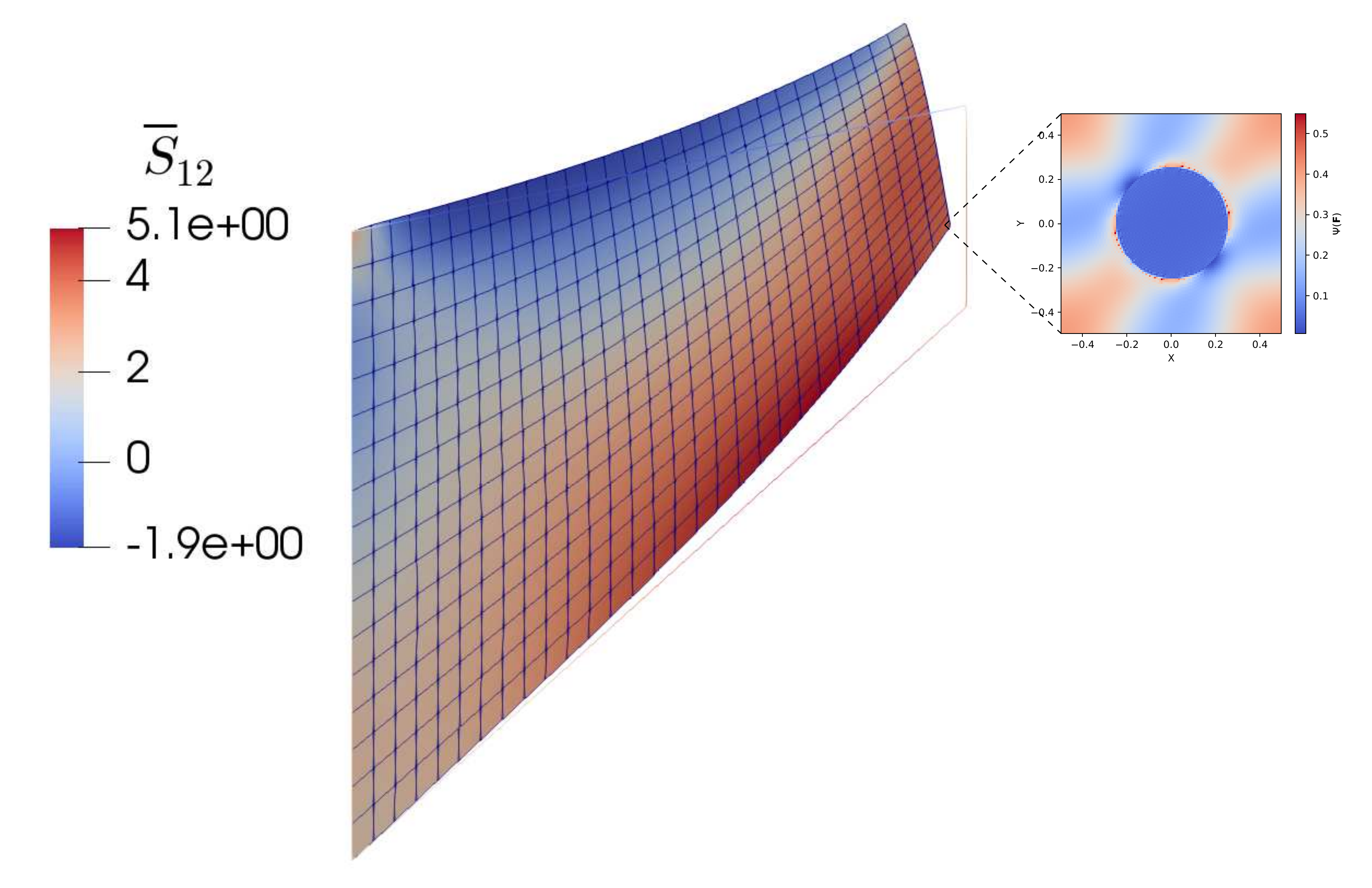}
		\end{subfigure}
		\caption{\textit{Distribution of resultant stress within the Cook's membrane.} The contour plot shows normal stress $\overline{S}_{11}$ (\emph{top}) and $\overline{S}_{12}$ (\emph{bottom}) computed at the macroscale according to the surrogate model (\emph{left}) and to the two-scale approach (\emph{right}), respectively.}
		\label{Fig:Stress-comparison-Cook-membrane}
	\end{figure}

	In order to show that the surrogate model is capable of generating numerical response that is comparable to the two-scale approach (FE-FFT method, FEM solver in macroscale and FFT solver in microscale), we place the results of two approaches adjacent to each other. In Fig.~\ref{Fig:Displacement-comparison-Cook-membrane} and Fig.~\ref{Fig:Stress-comparison-Cook-membrane}, the displacement fields and the resultant stresses corresponding to the surrogate-model and two-scale computations are placed on the \emph{left} and the \emph{right}, respectively.
	
	This example also shows the advantage of our surrogate model after it has been trained because it can re-employ the analysis of the previous example. The computation in two-scale analysis now performs at the macroscale as usual, but at the microscale the neural network will predict the effective stress and tangent moduli in response to the macroscopic deformation gradients at the quadrature points. This approach avoids the nested loop for solving the microscopic BVPs by FFT-based solver.
	
	\subsection{Two-dimensional Cantilever beam under plane strain analysis}
	\begin{figure}[H]
		\begin{subfigure}{1.0\textwidth}
			\centering{
				\def\svgwidth{0.6\textwidth}
				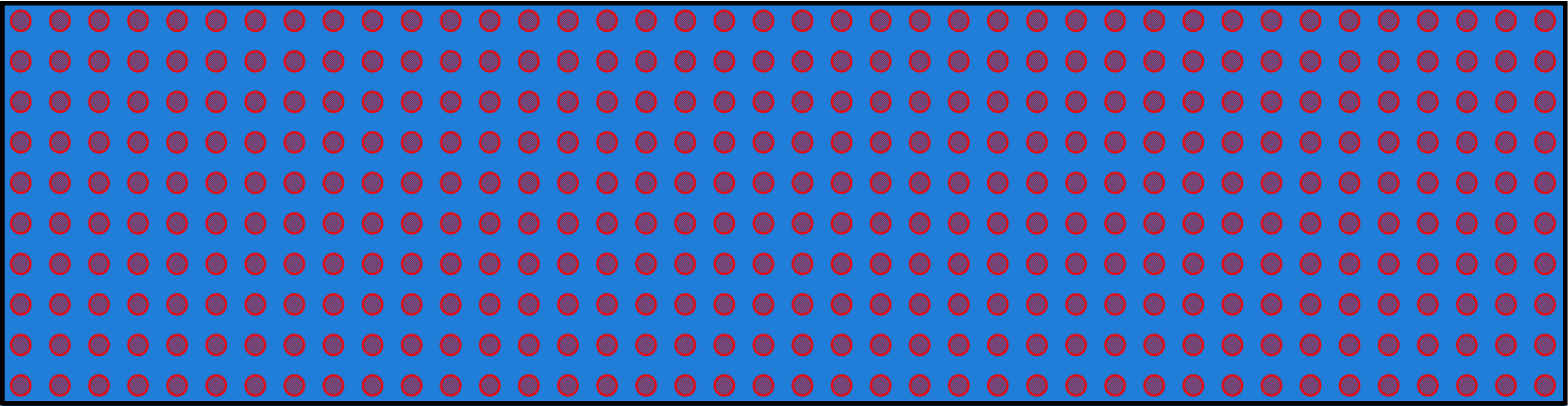
			}
		\end{subfigure}
		\begin{subfigure}{1.0\textwidth}
			\centering{
				\def\svgwidth{0.9\textwidth}
				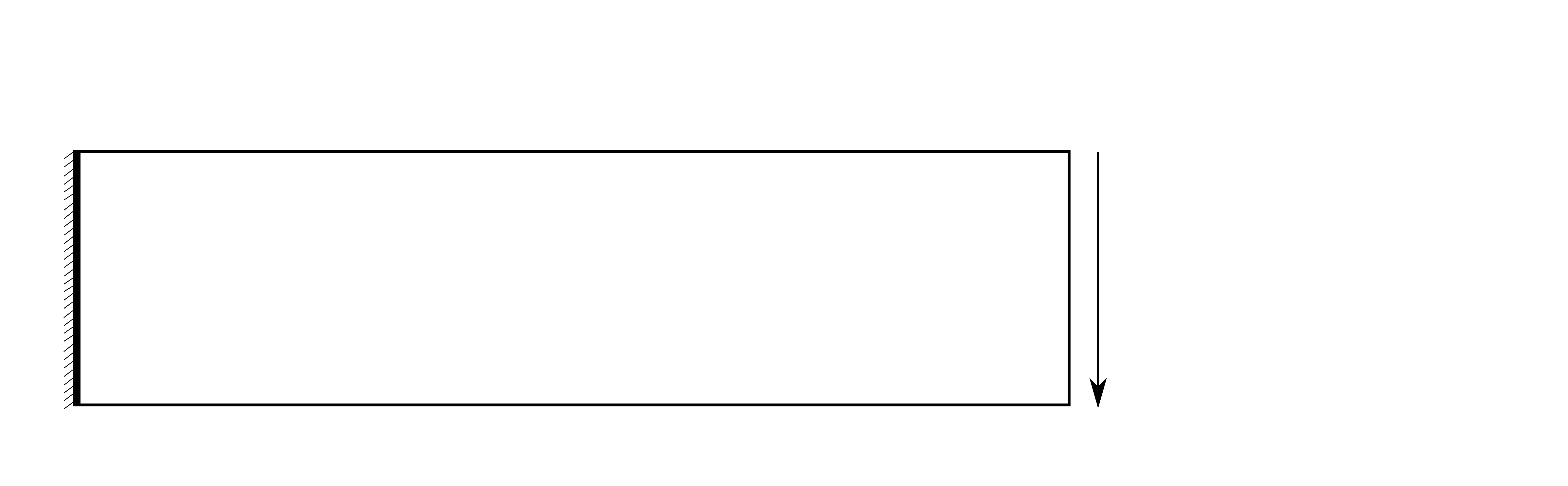
			}
		\end{subfigure}
		\caption{\textit{Cantilever beam of heterogeneous materials under plane strain analysis.} (\emph{top}) The beam is filled with $40\times 10$ circular inclusions of stiffer materials than the matrix's and is subjected to a traction force $\bar{q}_{0}=-0.25$. (\emph{bottom}) When the number of inclusions increases towards infinity, the RVE of circular inclusions represents well the presence of heterogeneity of materials according to the pattern of inclusions in the \emph{top} figure.}
		\label{Fig:Problem-setting-cantilever-beam}
	\end{figure}
	\paragraph{\textbf{\sffamily Problem setting}}
	As the last example, we support the soundness of this computational framework by providing the quantitative comparison between the full-field solution and the homogenized solution achieved by the surrogate modeling. To this end, we investigate the deformation of a two-dimensional cantilever beam, once again, in plane strain analysis. The beam of rectangular geometry is made of including $N_1 \times N_2$ stiffer circular inclusions, where $N_1$ and $N_2$ are the number of inclusions in $X_1$- and $X_2$-direction, respectively. In Fig.~\ref{Fig:Problem-setting-cantilever-beam} (\emph{top}), one typical beam with $(N_1\times N_2) = (40\times 10)$ circular inclusions is shown. For specific mechanical setting, the beam is clamped at its left edge and subject to the constant tangential traction $\bar{q}_{0}=-0.25$ along its right edge, which is visualized in Fig.~\ref{Fig:Problem-setting-cantilever-beam} (\emph{bottom}). The inclusions and matrix are made of materials given in Subsection~\ref{Sec:RVE-with-circular-inclusion} by equation \eqref{matrix-inclusion-material-parameters}. 
	
	\begin{figure}[H]
		\centering
		\begin{subfigure}{0.47\textwidth}
			\centering
			\includegraphics[width=1.0\textwidth]{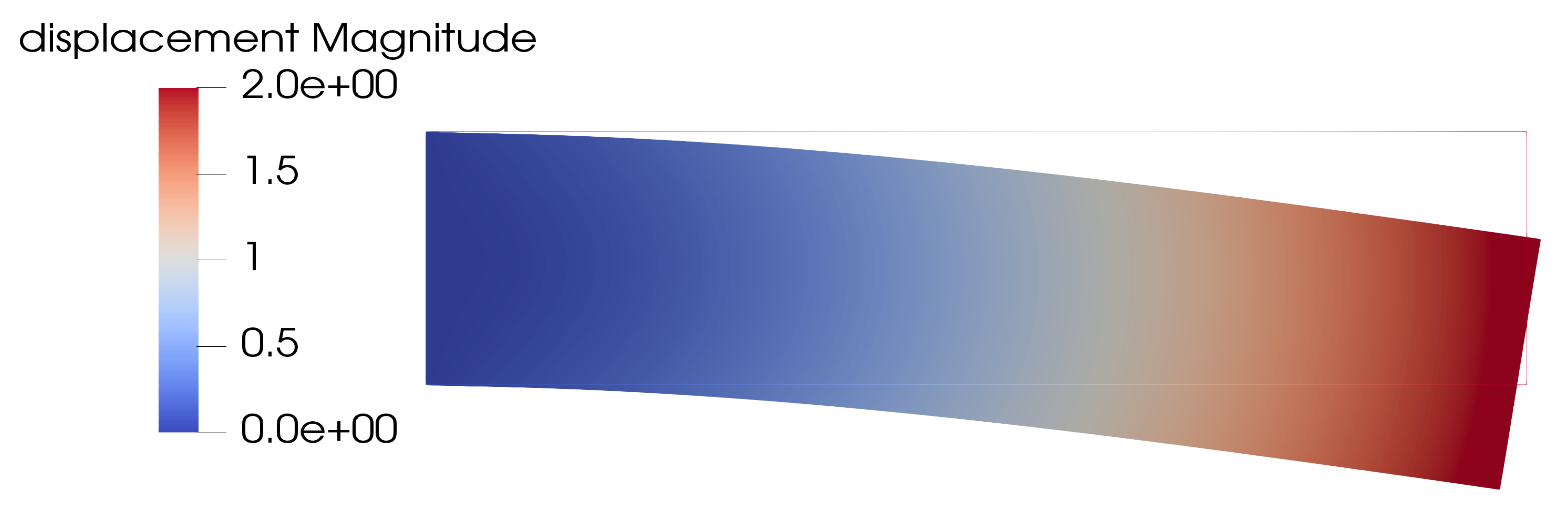}
		\end{subfigure}
		\begin{subfigure}{0.47\textwidth}
			\centering
			\includegraphics[width=1.0\textwidth]{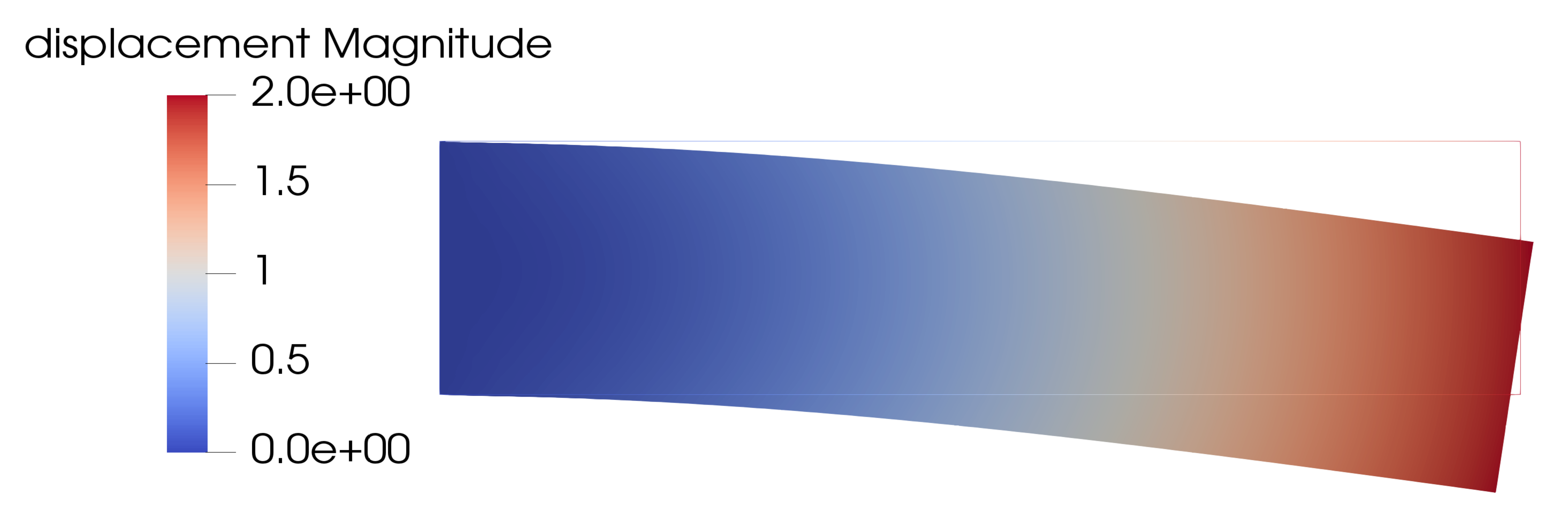}
		\end{subfigure}
		\caption{\textit{Final deformation of the cantilever beam.} The vertical displacement is contour-plotted for both the homogenized solution based on the surrogate modeling (\emph{left}) and the full-field solution based on the structure with $40\times 10$ circular inclusions.}
		\label{Fig:Vertical-displacement-Cantilever-beam}
	\end{figure}
	\paragraph{\textbf{\sffamily Numerical results}}
	The full-field solution using the beam structure of $40\times 10$ circular inclusions (see Fig.~\ref{Fig:Problem-setting-cantilever-beam}) is compared with the homogenized solution based on the surrogate model is presented in Fig.~\ref{Fig:Vertical-displacement-Cantilever-beam}. In this figure, the deformed shape of the beam as well as the contour plot of vertical displacement distributed through the whole domain are shown. The agreement between the two full-field and homogenized solutions regarding the displacements is quite spectacular.
	
\begin{figure}[H]
		\centering
		\begin{subfigure}{0.65\textwidth}
			\centering
			\includegraphics[width=1.0\textwidth]{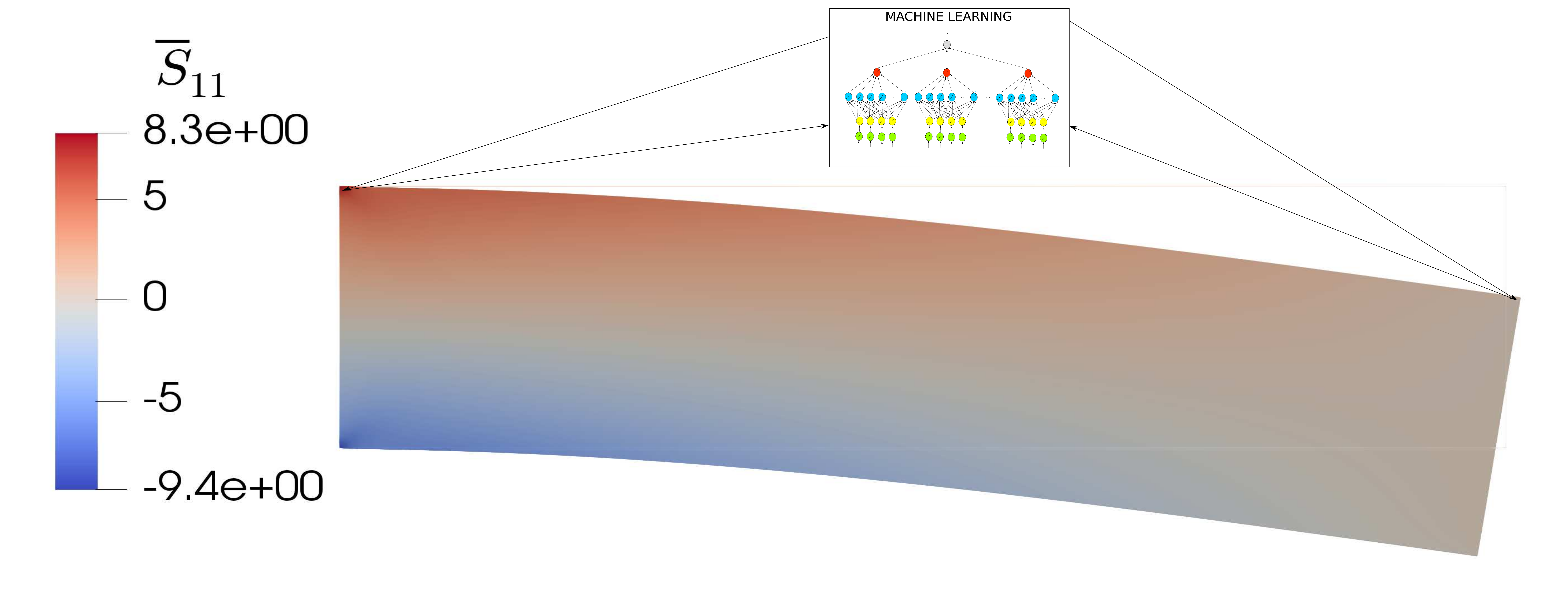}
		\end{subfigure}
		\begin{subfigure}{0.47\textwidth}
			\centering
			\includegraphics[width=1.0\textwidth]{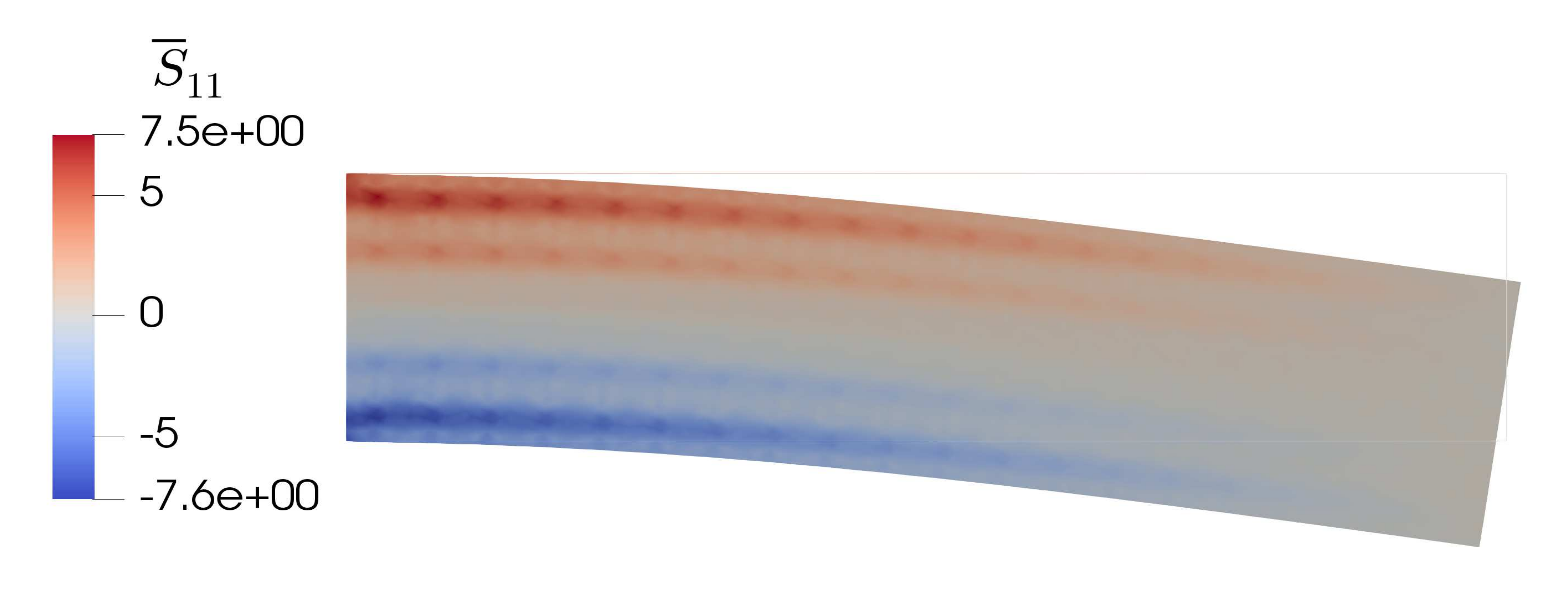}
		\end{subfigure}
		\begin{subfigure}{0.47\textwidth}
			\centering
			\includegraphics[width=1.0\textwidth]{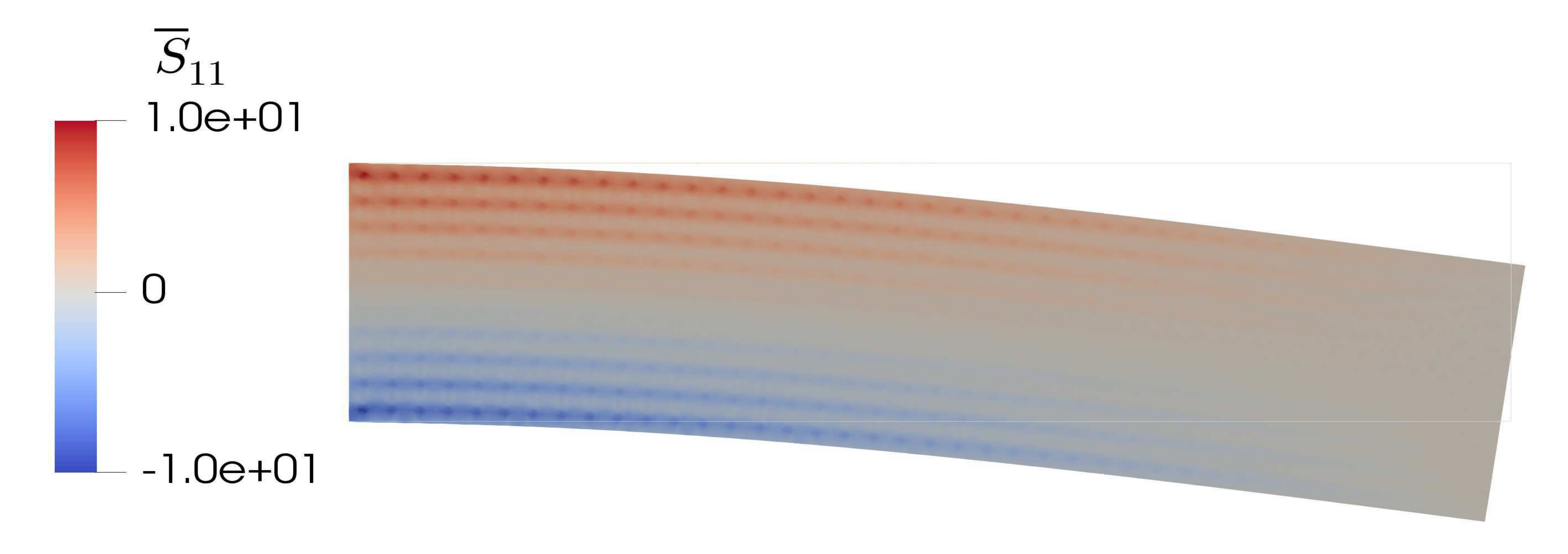}
		\end{subfigure}
		\caption{\textit{Distribution of normal second Piola stress $\overline{S}_{11}$}. As for cantilever beam with real circular inclusions (\emph{bottom}), it is seen that the stress concentrates within the inclusions with higher intensity than that in the matrix phase. Two structures with $20\times 5$ and $40\times 10$ circular inclusions are shown in the \emph{bottom left} and \emph{bottom right} subfigures. As for the homogenized structure (\emph{top}), the stress field appears quite smooth throughout the entire domain, just as expected. }
		\label{Fig:Stress-distribution-Cantilever-beam}
	\end{figure}

	However, it should be expected that we could notice the difference in the stress distribution because the homogenized solution averages out the fluctuations and to capture \emph{overall trend} of the full-field solution. Indeed, this is presented in Fig.~\ref{Fig:Stress-distribution-Cantilever-beam} the distribution of normal second Piola stress $\overline{S}_{11}$ obtained by surrogate modeling and full-field solution.  Nevertheless, it is clearly seen that the overall trend of the homogenized stress is in great compliance with the full-field stress. The ranges of stress values differ from each other as the stress field is concentrated more in the stiffer inclusion and less in the surrounding matrix. The homogenized stress basically \emph{averages out} the fluctuations in stress values across two phases. This is reflected in Fig.~\ref{Fig:Stress-distribution-Cantilever-beam} (\emph{top}) where the perturbations in values (presented by contour plot) totally are filtered out, leaving a smooth transition from one point to the neighbor hood point throughout the entire domain.
	
	These numerical results prove the reliability of our computational framework using HDMR-based neural networks to construct the approximate macro-energy density.
	
	\section{Conclusion}
	This contribution has addressed a surrogate model for two-scale computational homogenization. First, we pointed out that there was a strong connection between the formulation of the Lippmann-Schwinger equation for the microscopic BVP by using the polarization technique and by using the Galerkin-based projection. Indeed, the same result can be arrived at by two different routes of derivation. Second, a surrogate model for computational homogenization of elasticity at finite strains is built based on a neural network architecture that mimics the high-dimensional model representation. Particularly, this black-box function is an approximator of the macroscopic energy density and is trained upon the space of uniformly distributed random data of macroscopic deformation gradients. The database is constructed by solving numerous microscopic problems with the aid of the FFT-based solver to obtain the set of input-target data. The comparison of the numerical results with full-field solution as well two-scale homogenized solution validates both the reliability and robustness of the proposed computational framework.
	
	
\end{document}

%% file: images/Homogenization.pdf_tex
\begingroup%
  \makeatletter%
  \providecommand\color[2][]{%
    \errmessage{(Inkscape) Color is used for the text in Inkscape, but the package 'color.sty' is not loaded}%
    \renewcommand\color[2][]{}%
  }%
  \providecommand\transparent[1]{%
    \errmessage{(Inkscape) Transparency is used (non-zero) for the text in Inkscape, but the package 'transparent.sty' is not loaded}%
    \renewcommand\transparent[1]{}%
  }%
  \providecommand\rotatebox[2]{#2}%
  \ifx\svgwidth\undefined%
    \setlength{\unitlength}{1368.77252053bp}%
    \ifx\svgscale\undefined%
      \relax%
    \else%
      \setlength{\unitlength}{\unitlength * \real{\svgscale}}%
    \fi%
  \else%
    \setlength{\unitlength}{\svgwidth}%
  \fi%
  \global\let\svgwidth\undefined%
  \global\let\svgscale\undefined%
  \makeatother%
  \begin{picture}(1,0.45610262)%
    \put(0,0){\includegraphics[width=\unitlength,page=1]{Homogenization.pdf}}%
    \put(0.16686258,0.40383523){\color[rgb]{0,0,0}\makebox(0,0)[lb]{\smash{\textcolor{blue}{\sffamily MACROSCALE}}}}%
    \put(0.74063049,0.40070417){\color[rgb]{0,0,0}\makebox(0,0)[lb]{\smash{\textcolor{blue}{\sffamily MICROSCALE}}}}%
    \put(0,0){\includegraphics[width=\unitlength,page=2]{Homogenization.pdf}}%
    \put(0.4773183,0.45072812){\color[rgb]{0,0,0}\makebox(0,0)[lb]{\smash{$\overline{\mathbf{t}}$}}}%
    \put(0.32536695,0.33398268){\color[rgb]{0,0,0}\makebox(0,0)[lb]{\smash{$\overline{\mathbf{X}}$}}}%
    \put(0.43131017,0.23403988){\color[rgb]{0,0,0}\makebox(0,0)[lb]{\smash{$\overline{\tsr{x}}$}}}%
    \put(0.06521115,0.31973293){\color[rgb]{0,0,0}\makebox(0,0)[lb]{\smash{$\calB$}}}%
    \put(0.37815682,0.17966854){\color[rgb]{0,0,0}\makebox(0,0)[lb]{\smash{$\calB_t$}}}%
    \put(0.66799769,0.22153095){\color[rgb]{0,0,0}\makebox(0,0)[lb]{\smash{$\calR$}}}%
    \put(0.8837736,0.31368827){\color[rgb]{0,0,0}\makebox(0,0)[lb]{\smash{$\calR_t$}}}%
    \put(0,0){\includegraphics[width=\unitlength,page=3]{Homogenization.pdf}}%
    \put(0.68691346,0.25752429){\color[rgb]{0,0,0}\makebox(0,0)[lb]{\smash{$\mathbf{X}$}}}%
    \put(0.75008157,0.28745362){\color[rgb]{0,0,0}\makebox(0,0)[lb]{\smash{$\tsr{x}$}}}%
    \put(0,0){\includegraphics[width=\unitlength,page=4]{Homogenization.pdf}}%
    \put(0.53218581,0.31059177){\color[rgb]{0,0,0}\makebox(0,0)[lb]{\smash{$\overline{\mathbf{F}}$}}}%
    \put(0.5080394,0.16117462){\color[rgb]{0,0,0}\makebox(0,0)[lb]{\smash{$\overline{\mathbf{P}}$, $\overline{\mathbb{C}}$}}}%
    \put(-0.00068072,0.14912629){\color[rgb]{0,0,0}\makebox(0,0)[lb]{\smash{$\partial\calB$}}}%
    \put(0.24165179,0.12022683){\color[rgb]{0,0,0}\makebox(0,0)[lb]{\smash{$\partial\calB_t$}}}%
    \put(0.75655807,0.04139991){\color[rgb]{0,0,0}\makebox(0,0)[lb]{\smash{$\partial\calR$}}}%
    \put(0.92937544,0.20017722){\color[rgb]{0,0,0}\makebox(0,0)[lb]{\smash{$\partial\calR_t$}}}%
  \end{picture}%
\endgroup%

%% file: images/NeuronNetwork.pdf_tex
\begingroup%
  \makeatletter%
  \providecommand\color[2][]{%
    \errmessage{(Inkscape) Color is used for the text in Inkscape, but the package 'color.sty' is not loaded}%
    \renewcommand\color[2][]{}%
  }%
  \providecommand\transparent[1]{%
    \errmessage{(Inkscape) Transparency is used (non-zero) for the text in Inkscape, but the package 'transparent.sty' is not loaded}%
    \renewcommand\transparent[1]{}%
  }%
  \providecommand\rotatebox[2]{#2}%
  \ifx\svgwidth\undefined%
    \setlength{\unitlength}{6866.55334953bp}%
    \ifx\svgscale\undefined%
      \relax%
    \else%
      \setlength{\unitlength}{\unitlength * \real{\svgscale}}%
    \fi%
  \else%
    \setlength{\unitlength}{\svgwidth}%
  \fi%
  \global\let\svgwidth\undefined%
  \global\let\svgscale\undefined%
  \makeatother%
  \begin{picture}(1,0.42052027)%
    \put(0,0){\includegraphics[width=\unitlength,page=1]{NeuronNetwork.pdf}}%
    \put(0.76703899,-0.65827769){\color[rgb]{0,0,0}\rotatebox{90.33800141}{\makebox(0,0)[lt]{\begin{minipage}{0.09200016\unitlength}\raggedright \end{minipage}}}}%
    \put(0.39015689,-0.72715245){\color[rgb]{0,0,0}\rotatebox{90.33800141}{\makebox(0,0)[lt]{\begin{minipage}{0.29841072\unitlength}\raggedright \end{minipage}}}}%
    \put(0,0){\includegraphics[width=\unitlength,page=2]{NeuronNetwork.pdf}}%
    \put(0.09788576,1.20974353){\color[rgb]{0,0,0}\rotatebox{90.33800141}{\makebox(0,0)[lt]{\begin{minipage}{0.00668345\unitlength}\raggedright \end{minipage}}}}%
    \put(0,0){\includegraphics[width=\unitlength,page=3]{NeuronNetwork.pdf}}%
	\put(0.05092872,-0.02579893){\color[rgb]{0,0,0}\rotatebox{-0.32239775}{\makebox(0,0)[lb]{\smash{$\wb{F}_{11}^{1}$}}}}%
	\put(0.19919484,-0.02579893){\color[rgb]{0,0,0}\rotatebox{-0.32239835}{\makebox(0,0)[lb]{\smash{$\wb{F}_{22}^{1}$}}}}%
	\put(0.10232956,-0.02579893){\color[rgb]{0,0,0}\rotatebox{-0.32239786}{\makebox(0,0)[lb]{\smash{$\wb{F}_{12}^{1}$}}}}%
	\put(0.15257416,-0.02579893){\color[rgb]{0,0,0}\rotatebox{-0.32239799}{\makebox(0,0)[lb]{\smash{$\wb{F}_{21}^{1}$}}}}%
    \put(0,0){\includegraphics[width=\unitlength,page=4]{NeuronNetwork.pdf}}%
    \put(0.89871993,0.03033629){\color[rgb]{0,0,0}\rotatebox{-0.27695307}{\makebox(0,0)[lb]{\smash{\textcolor{green}{\sffamily INPUT LAYER}}}}}%
    \put(0.88544649,0.25445409){\color[rgb]{0,0,0}\rotatebox{-0.31299195}{\makebox(0,0)[lb]{\smash{\sffamily \textcolor{red}{\sffamily OUTPUT LAYER}}}}}%
    \put(0.95644689,0.16907266){\color[rgb]{0,0,0}\rotatebox{-0.30859876}{\makebox(0,0)[lb]{\smash{\textcolor{aqua}{\sffamily HIDDEN LAYER}}}}}%
	\put(0.35414223,-0.02579893){\color[rgb]{0,0,0}\rotatebox{-0.32239775}{\makebox(0,0)[lb]{\smash{$\wb{F}_{11}^{2}$}}}}%
	\put(0.50040823,-0.02579893){\color[rgb]{0,0,0}\rotatebox{-0.32239835}{\makebox(0,0)[lb]{\smash{$\wb{F}_{22}^{2}$}}}}%
	\put(0.40554298,-0.02579893){\color[rgb]{0,0,0}\rotatebox{-0.32239786}{\makebox(0,0)[lb]{\smash{$\wb{F}_{12}^{2}$}}}}%
	\put(0.45578755,-0.02579893){\color[rgb]{0,0,0}\rotatebox{-0.32239799}{\makebox(0,0)[lb]{\smash{$\wb{F}_{21}^{2}$}}}}%
	\put(0.70623688,-0.02579893){\color[rgb]{0,0,0}\rotatebox{-0.32239775}{\makebox(0,0)[lb]{\smash{$\wb{F}_{11}^{L}$}}}}%
	\put(0.85250229,-0.02579893){\color[rgb]{0,0,0}\rotatebox{-0.32239835}{\makebox(0,0)[lb]{\smash{$\wb{F}_{22}^{L}$}}}}%
	\put(0.75763754,-0.02579893){\color[rgb]{0,0,0}\rotatebox{-0.32239786}{\makebox(0,0)[lb]{\smash{$\wb{F}_{12}^{L}$}}}}%
	\put(0.8078816,-0.02579893){\color[rgb]{0,0,0}\rotatebox{-0.32239799}{\makebox(0,0)[lb]{\smash{$\wb{F}_{21}^{L}$}}}}%
	\put(0.09204279,0.29242251){\color[rgb]{0,0,0}\rotatebox{-0.32239818}{\makebox(0,0)[lb]{\smash{$\wb{\psi}^{1}(\wb{\mathbf{F}})$}}}}%
	\put(0.45040036,0.29242249){\color[rgb]{0,0,0}\rotatebox{-0.32239818}{\makebox(0,0)[lb]{\smash{$\wb{\psi}^{2}(\wb{\mathbf{F}})$}}}}%
	\put(0.79930621,0.28488823){\color[rgb]{0,0,0}\rotatebox{-0.32239818}{\makebox(0,0)[lb]{\smash{$\wb{\psi}^{L}(\wb{\mathbf{F}})$}}}}%
	\put(0.42099527,0.41579893){\color[rgb]{0,0,0}\rotatebox{-0.32239818}{\makebox(0,0)[lb]{\smash{$\wb{\psi}^{\mathrm{NN}}(\wb{\mathbf{F}})$}}}}%
  \end{picture}%
\endgroup%

%% file: images/Laminate.pdf_tex
\begingroup%
  \makeatletter%
  \providecommand\color[2][]{%
    \errmessage{(Inkscape) Color is used for the text in Inkscape, but the package 'color.sty' is not loaded}%
    \renewcommand\color[2][]{}%
  }%
  \providecommand\transparent[1]{%
    \errmessage{(Inkscape) Transparency is used (non-zero) for the text in Inkscape, but the package 'transparent.sty' is not loaded}%
    \renewcommand\transparent[1]{}%
  }%
  \providecommand\rotatebox[2]{#2}%
  \ifx\svgwidth\undefined%
    \setlength{\unitlength}{226.88504197bp}%
    \ifx\svgscale\undefined%
      \relax%
    \else%
      \setlength{\unitlength}{\unitlength * \real{\svgscale}}%
    \fi%
  \else%
    \setlength{\unitlength}{\svgwidth}%
  \fi%
  \global\let\svgwidth\undefined%
  \global\let\svgscale\undefined%
  \makeatother%
  \begin{picture}(1,0.99365237)%
    \put(0,0){\includegraphics[width=\unitlength,page=1]{Laminate.pdf}}%
    \put(0.4583005,0.50168359){\color[rgb]{0,0,0}\makebox(0,0)[lb]{\smash{$\mu^{(1)}$}}}%
    \put(0.4583005,0.37168359){\color[rgb]{0,0,0}\makebox(0,0)[lb]{\smash{$\beta^{(1)}$}}}%
    \put(0.09007945,0.50168359){\color[rgb]{0,0,0}\makebox(0,0)[lb]{\smash{$\mu^{(2)}$}}}%
    \put(0.09007945,0.37168359){\color[rgb]{0,0,0}\makebox(0,0)[lb]{\smash{$\beta^{(2)}$}}}%
    \put(0.83138819,0.50168359){\color[rgb]{0,0,0}\makebox(0,0)[lb]{\smash{$\mu^{(2)}$}}}%
    \put(0.83138819,0.37168359){\color[rgb]{0,0,0}\makebox(0,0)[lb]{\smash{$\beta^{(2)}$}}}%
    \put(0,0){\includegraphics[width=\unitlength,page=2]{Laminate.pdf}}%
  \end{picture}%
\endgroup%

%% file: images/Circular.pdf_tex
\begingroup%
  \makeatletter%
  \providecommand\color[2][]{%
    \errmessage{(Inkscape) Color is used for the text in Inkscape, but the package 'color.sty' is not loaded}%
    \renewcommand\color[2][]{}%
  }%
  \providecommand\transparent[1]{%
    \errmessage{(Inkscape) Transparency is used (non-zero) for the text in Inkscape, but the package 'transparent.sty' is not loaded}%
    \renewcommand\transparent[1]{}%
  }%
  \providecommand\rotatebox[2]{#2}%
  \ifx\svgwidth\undefined%
    \setlength{\unitlength}{185.86163859bp}%
    \ifx\svgscale\undefined%
      \relax%
    \else%
      \setlength{\unitlength}{\unitlength * \real{\svgscale}}%
    \fi%
  \else%
    \setlength{\unitlength}{\svgwidth}%
  \fi%
  \global\let\svgwidth\undefined%
  \global\let\svgscale\undefined%
  \makeatother%
  \begin{picture}(1,0.86010489)%
    \put(0,0){\includegraphics[width=\unitlength,page=1]{Circular.pdf}}%
    \put(0.33143918,0.7280266){\color[rgb]{0,0,0}\makebox(0,0)[lb]{\smash{$E^{(\text{m})}, \nu^{(\text{m})}$}}}%
    \put(0.35143918,0.40920836){\color[rgb]{0,0,0}\makebox(0,0)[lb]{\smash{$E^{(\text{i})}, \nu^{(\text{i})}$}}}%
  \end{picture}%
\endgroup%

%% file: images/CookMembrane.pdf_tex
\begingroup%
  \makeatletter%
  \providecommand\color[2][]{%
    \errmessage{(Inkscape) Color is used for the text in Inkscape, but the package 'color.sty' is not loaded}%
    \renewcommand\color[2][]{}%
  }%
  \providecommand\transparent[1]{%
    \errmessage{(Inkscape) Transparency is used (non-zero) for the text in Inkscape, but the package 'transparent.sty' is not loaded}%
    \renewcommand\transparent[1]{}%
  }%
  \providecommand\rotatebox[2]{#2}%
  \ifx\svgwidth\undefined%
    \setlength{\unitlength}{754.72741699bp}
    \ifx\svgscale\undefined%
      \relax%
    \else%
      \setlength{\unitlength}{\unitlength * \real{\svgscale}}%
    \fi%
  \else%
    \setlength{\unitlength}{\svgwidth}%
  \fi%
  \global\let\svgwidth\undefined%
  \global\let\svgscale\undefined%
  \makeatother%
  \begin{picture}(1,0.64271688)%
    \put(0,0){\includegraphics[width=\unitlength,page=1]{CookMembrane.pdf}}%
    \put(0.25128051,0.03550632){\color[rgb]{0,0,0}\makebox(0,0)[lb]{\smash{48}}}%
    \put(0.57178337,0.18775349){\color[rgb]{0,0,0}\makebox(0,0)[lb]{\smash{44}}}%
    \put(0.57178337,0.54268101){\color[rgb]{0,0,0}\makebox(0,0)[lb]{\smash{16}}}%
    \put(0.50398004,0.52730349){\color[rgb]{0,0,0}\makebox(0,0)[lb]{\smash{$\bar{q}_0$}}}%
    \put(0,0){\includegraphics[width=\unitlength,page=2]{CookMembrane.pdf}}%
    \put(0.7160253,0.53597683){\color[rgb]{0,0,0}\makebox(0,0)[lb]{\smash{$E^{(\text{m})}, \nu^{(\text{m})}$}}}%
    \put(0.75968326,0.4772152){\color[rgb]{0,0,0}\makebox(0,0)[lb]{\smash{$E^{(\text{i})}$}}}%
    \put(0.75968326,0.4372152){\color[rgb]{0,0,0}\makebox(0,0)[lb]{\smash{$\nu^{(\text{i})}$}}}%
    \put(0,0){\includegraphics[width=\unitlength,page=3]{CookMembrane.pdf}}%
    \put(0.14956811,0.61251662){\color[rgb]{0,0,0}\makebox(0,0)[lb]{\smash{\textcolor{blue}{MACRO LEVEL}}}}%
    \put(0.72884723,0.63150937){\color[rgb]{0,0,0}\makebox(0,0)[lb]{\smash{}}}%
    \put(0.74546684,0.60064604){\color[rgb]{0,0,0}\makebox(0,0)[lb]{\smash{\textcolor{blue}{RVE}}}}%
  \end{picture}%
\endgroup%

%% file: images/beam_400lo.pdf_tex
\begingroup%
  \makeatletter%
  \providecommand\color[2][]{%
    \errmessage{(Inkscape) Color is used for the text in Inkscape, but the package 'color.sty' is not loaded}%
    \renewcommand\color[2][]{}%
  }%
  \providecommand\transparent[1]{%
    \errmessage{(Inkscape) Transparency is used (non-zero) for the text in Inkscape, but the package 'transparent.sty' is not loaded}%
    \renewcommand\transparent[1]{}%
  }%
  \providecommand\rotatebox[2]{#2}%
  \ifx\svgwidth\undefined%
    \setlength{\unitlength}{987.58465144bp}%
    \ifx\svgscale\undefined%
      \relax%
    \else%
      \setlength{\unitlength}{\unitlength * \real{\svgscale}}%
    \fi%
  \else%
    \setlength{\unitlength}{\svgwidth}%
  \fi%
  \global\let\svgwidth\undefined%
  \global\let\svgscale\undefined%
  \makeatother%
  \begin{picture}(1,0.25890459)%
    \put(0,0){\includegraphics[width=\unitlength,page=1]{beam_400lo.pdf}}%
  \end{picture}%
\endgroup%

%% file: images/beam.pdf_tex
\begingroup%
  \makeatletter%
  \providecommand\color[2][]{%
    \errmessage{(Inkscape) Color is used for the text in Inkscape, but the package 'color.sty' is not loaded}%
    \renewcommand\color[2][]{}%
  }%
  \providecommand\transparent[1]{%
    \errmessage{(Inkscape) Transparency is used (non-zero) for the text in Inkscape, but the package 'transparent.sty' is not loaded}%
    \renewcommand\transparent[1]{}%
  }%
  \providecommand\rotatebox[2]{#2}%
  \ifx\svgwidth\undefined%
    \setlength{\unitlength}{1559.50806625bp}%
    \ifx\svgscale\undefined%
      \relax%
    \else%
      \setlength{\unitlength}{\unitlength * \real{\svgscale}}%
    \fi%
  \else%
    \setlength{\unitlength}{\svgwidth}%
  \fi%
  \global\let\svgwidth\undefined%
  \global\let\svgscale\undefined%
  \makeatother%
  \begin{picture}(1,0.30988858)%
    \put(0,0){\includegraphics[width=\unitlength,page=1]{beam.pdf}}%
    \put(0.71408377,0.13177076){\color[rgb]{0,0,0}\makebox(0,0)[lb]{\smash{$\bar{q}_0$}}}%
    \put(0,0){\includegraphics[width=\unitlength,page=2]{beam.pdf}}%
    \put(0.34858396,0.00027241){\color[rgb]{0,0,0}\makebox(0,0)[lb]{\smash{20}}}%
    \put(0,0){\includegraphics[width=\unitlength,page=3]{beam.pdf}}%
    \put(-0.00148409,0.13209826){\color[rgb]{0,0,0}\makebox(0,0)[lb]{\smash{5}}}%
    \put(0,0){\includegraphics[width=\unitlength,page=4]{beam.pdf}}%
    \put(0.84703875,0.16867588){\color[rgb]{0,0,0}\makebox(0,0)[lb]{\smash{$E^{i}$, $\nu^{i}$}}}%
    \put(0.81239872,0.22409525){\color[rgb]{0,0,0}\makebox(0,0)[lb]{\smash{$E^{m}$, $\nu^{m}$}}}%
    \put(0,0){\includegraphics[width=\unitlength,page=5]{beam.pdf}}%
  \end{picture}%
\endgroup%